\documentclass[pdflatex,sn-basic,trackchanges]{sn-jnl}

\usepackage{graphicx}%
\usepackage{multirow}%
\usepackage{amsmath,amssymb,amsfonts}%
\usepackage{amsthm}%
\usepackage{mathrsfs}%
\usepackage[title]{appendix}%
\usepackage{xcolor}%
\usepackage{textcomp}%
\usepackage{manyfoot}%
\usepackage{booktabs}%
\usepackage{algorithm}%
\usepackage{algorithmicx}%
\usepackage{algpseudocode}%
\usepackage{listings}%
\usepackage{chemfig}%
\usepackage{mhchem}%

\usepackage{siunitx}

\theoremstyle{thmstyleone}%
\theoremstyle{thmstyletwo}%

\theoremstyle{thmstylethree}%

\newcommand{\red}[1]{\textcolor{red}{#1}}

\begin{document}

\title[Article Title]{The Origins \& Reservoirs of Exocomets}





\author[1]{\fnm{Michele} \sur{Bannister}}\email{michele.bannister@canterbury.ac.nz}
\equalcont{These authors contributed equally to this work.}

\author*[2]{\fnm{Susanne} \sur{Pfalzner}}\email{s.pfalzner@fz-juelich.de}

\author[3]{\fnm{Tim} \sur{Pearce}}\email{tim.pearce@warwick.ac.uk}
\equalcont{These authors contributed equally to this work.}

\author[4]{\fnm{Alexander J.} \sur{Mustill}}\email{alexander.mustill@fysik.lu.se}
\equalcont{These authors contributed equally to this work.}

\author[5]{\fnm{Hubert} \sur{Klahr}}\email{klahr@mpia.de}

\author[6]{\fnm{Hideko}\sur{Nomura}}\email{hideko.nomura@nao.ac.jp}

\author[7]{\fnm{Nagayoshi}\sur{Ohashi}}\email{ohashi@asiaa.sinica.edu.tw}

\author[8,9]{\fnm{Rosita} \sur{Kokotanekova}}\email{rkokotanekova@nao-rozhen.org}

\author[10]{\fnm{Sebastian} \sur{Marino}}\email{s.marino-estay@exeter.ac.uk}

\author[11]{\fnm{Dennis} \sur{Bodewits}}\email{dennis@auburn.edu}

\author[12]{\fnm{Raphael}\sur{Marschall}}\email{raphael.marschall@oca.eu}

\author[13,14]{\fnm{Darryl Z.}\sur{Seligman}}\email{dzs@msu.edu}

\author[15,16]{\fnm{Geraint H.} \sur{Jones}}\email{geraint.jones@esa.int}

\author[3,17,18]{\fnm{Dimitri} \sur{Veras}}\email{dimitri.veras@aya.yale.edu}


\affil[1]{\orgdiv{School of Physical and Chemical Sciences -- Te Kura Mat\={u}, \orgname{University of Canterbury}, \orgaddress{\street{Private Bag 4800}, \city{Christchurch}, \postcode{8140}, \country{Aotearoa New Zealand}}}}

\affil[2]{\orgdiv{J\"ulich Supercomputing Centre}, \orgname{Research Centre J\"ulich}, \orgaddress{\street{Wilhelm-Johnen-Strasse}, \city{J\"ulich}, \postcode{52428}, \country{Germany}}}

\affil[3]{\orgdiv{Department of Physics}, \orgname{University of Warwick}, \orgaddress{\street{Gibbet Hill Road}, \city{Coventry}, \postcode{CV4 7AL}, \country{United Kingdom}}}

\affil[4]{\orgdiv{Department of Physics}, \orgname{Lund University}, \orgaddress{\street{Box 118}, \city{Lund}, \postcode{22100}, \country{Sweden}}}

\affil[5]{\orgdiv{}\orgname{Max Planck Institut f\"ur Astronomie}, \orgaddress{\street{K\"onigstuhl 17}, \city{Heidelberg}, \postcode{69117}, \country{Germany}}}

\affil[6]{\orgdiv{Division of Science}, \orgname{National Astronomical Observatory of Japan}, \orgaddress{\street{2-21-1, Osawa, Mitaka}, \city{Tokyo} \postcode{181-8551}, \country{Japan}}}

\affil[7]{\orgname{Academia Sinica Institute of Astronomy \& Astrophysics}, \orgaddress{\street{11F of Astronomy-Mathematics Building, AS/NTU, No.1, Sec. 4, Roosevelt Rd}, \city{Taipei} \postcode{106319}, \country{Taiwan, R.O.C.}}}

\affil[8]{\orgdiv{Institute of Astronomy and NAO}, \orgname{Bulgarian Academy of Sciences}, \orgaddress{\street{72 Tsarigradsko Chaussee Blvd.}, \city{Sofia}, \postcode{1784}, \country{Bulgaria}}}

\affil[9]{\orgname{International Space Science Institute}, \orgaddress{\street{Hallerstrasse 6}, \city{Bern}, \postcode{3012}, \country{Switzerland}}}

\affil[10]{\orgdiv{Department of Physics and Astronomy}, \orgname{University of Exeter}, \orgaddress{\street{Stocker Road}, \city{Exeter}, \postcode{EX4 4QL}, \country{United Kingdom}}}

\affil[11]{\orgdiv{Department of Physics}, \orgname{Auburn University}, \orgaddress{\street{Edmund C. Leach Science Center}, \city{Auburn}, \postcode{36849}, \state{AL} \country{USA}}}

\affil[12]{\orgdiv{Laboratoire J.-L. Lagrange}, \orgname{Observatoire de la C\^ote d'Azur, CNRS}, \orgaddress{\street{CS 34229}, \city{Nice Cedex 4}, \postcode{06304}, \country{France}}}

\affil[13]{\orgdiv{Department of Physics and Astronomy}, \orgname{Michigan State University}, \city{East Lansing}, \postcode{48824}, \state{MI}, \country{USA}}

\affil[14]{\orgdiv{NSF Astronomy and Astrophysics Postdoctoral Fellow}}

\affil[15]{\orgname{UCL Mullard Space Science Laboratory}, \orgaddress{\street{Holmbury St. Mary}, \city{Dorking} \postcode{RH5 6NT}, \country{United Kingdom}}}

\affil[16]{\orgdiv{European Space Technology Centre (ESTEC)}, \orgname{European Space Agency}, \orgaddress{\street{Keplerlaan 1}, \city{2200 AG 
 Noordwijk}, \country{Netherlands}}}

\affil[17]{\orgdiv{Centre for Exoplanets and Habitability}, \orgname{University of Warwick}, \orgaddress{\street{Gibbet Hill Road}, \city{Coventry}, \postcode{CV4 7AL}, \country{United Kingdom}}}

\affil[18]{\orgdiv{Centre for Space Domain Awareness}, \orgname{University of Warwick}, \orgaddress{\street{Gibbet Hill Road}, \city{Coventry}, \postcode{CV4 7AL}, \country{United Kingdom}}}


\abstract{

Small bodies exist in distinct populations within their planetary systems. 
These reservoir populations hold a range of compositions, which to first order are dependent on formation location relative to their star. 
We provide a general overview of the nature of the reservoirs that source exocomets, from the influence of the stellar environment through planetesimal formation to comparisons with Solar System populations.
Once transitioned from a young protoplanetary disc to a debris disc, a star can expect to be rained with exocomets.
While exocomets are predominantly detected to date at A-type stars, planetesimals plausibly exist across a range of stellar masses, based on exoplanet abundance, debris disc occurrence and white dwarf infall.



}

\keywords{}



\maketitle

\section{Introduction}
\label{sec:intro}

The small-body populations of a planetary system typically exist in discrete dynamical groups: \textit{reservoirs}.
These are remnant populations of small worlds (often for generality termed `bodies' or `objects') orbiting their star after the dispersal of the gas disc. 
As every exocomet is sourced from some point within a planetesimal disc, understanding the reservoirs of a system is necessary to develop an understanding of the potential origin, composition, and thermal history of an object that is only seen in the late brief flare of a demising exocomet.

\begin{figure}
    \centering
    \includegraphics[width=0.99\textwidth]{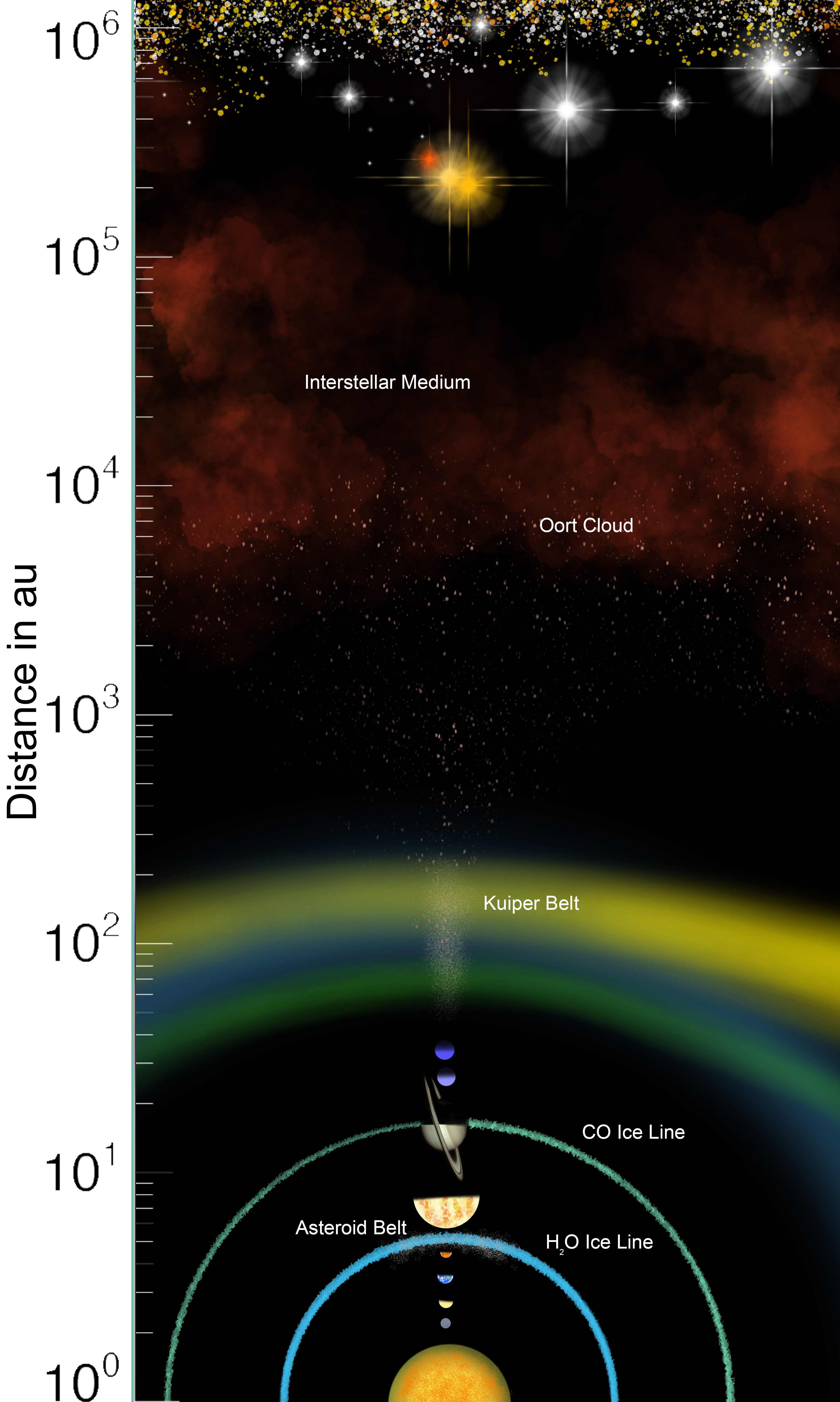}
    \caption{Schematic picture illustrating the approximate location of the different small-body and comet reservoirs in the Solar System. }
    \label{fig:Location_reservoirs}
\end{figure}

In the Solar System, observational surveys over many decades have refined our understanding of the small-body reservoirs that exist today across a wide range of semi-major axes (Figure~\ref{fig:Location_reservoirs}).
Smallest in semimajor axis are the sparse population of `Ayló`chaxnim asteroids, found entirely interior to Venus's orbit (0.72 au) \citep{Bottke:2002,Bolin:2023}; no asteroids have been found within the orbit of Mercury, despite two centuries of searches for such ``Vulcanoids'' \citep{Perrine1902,Steffl2013}.
Other populations of near-Earth asteroids are within or cross Earth's orbit (Atiras, Atens, Apollos and Amors respectively), with some thirty thousand greater than 100 m diameter \citep{Nesvorny:2024,Deienno:2025}.

The above asteroids are sourced by dynamical erosion from the larger reservoir of the main belt of asteroids with \mbox{$2 \lesssim a \lesssim 3.5$ au}, where $a$ is the semi-major axis, between the orbits of Mars and Jupiter.
A complex structure, the main belt exhibits compositional mixing of largely volatile-poor objects rich in silicates and some organics in the inner main belt, with an overall gradient of more water-rich objects toward the outside of the belt \citep{DeMeo:2014}. 
The main belt is sculpted by dynamical resonances with Jupiter to leave the Kirkwood gaps.
Asteroids displaying cometary or non-gravitational activity are now observationally well established in this region \citep{Hsieh2006}, though only recently confirmed using JWST to be due to volatile outgassing \citep{Kelley2023,Hsieh2025}.
Various co-orbital asteroids are expected (and found) for most planets from Venus outwards \citep{Pan:2025}, either as highly transient 1:1 resonances in horseshoe or quasi-satellite orbits, or as more permanent Trojan populations, depending on the exact orbital stability regions for each planet \citep{Morais:2006, Carruba:2025, Granvik:2012, Fedorets:2017, Tabachnik:1999, Greenstreet:2024,Alexandersen:2021}\footnote{No co-orbitals are known for Mercury. Note that Trojans are not stable for Saturn or Uranus.}. 

At larger semimajor axes, resonant populations become more and more numerous.
The Hildas are outside the main belt in a 3:2 resonance with Jupiter, numbering some six thousand, but the Jupiter Trojans are a far larger population.
While no objects are currently known on stable orbits between Uranus and Neptune, they are theoretically possible \citep{Holman:1997,Zhang2022}, though the population must be small \citep[of order 80 to $H_r < 10$;, where $H_r$ is the absolute magnitude,][]{Dorsey:2023}.
The clouds of irregular satellites orbiting each giant planet appear to originate from the same reservoir as their Trojans, and were captured at some point \citep{Jewitt2007}. 

The massive population of the Neptune Trojans \citep{Parker2015,Lin2021} is a harbinger of the true scale of small-body populations: they are far and away the province of the outer Solar System.
As the outermost known planet, Neptune's gravitational influence sculpts the whole trans-Neptunian region; for a detailed review, see \citet{Gladman2021}.
Here exterior resonances at larger semimajor axis than that of Neptune do not remove objects as in the Kirkwood gaps, but instead hold stable populations.
Two main trans-Neptunian populations are present.
With \mbox{$42.5 < a < $ 47.5 au,} the cold classical Kuiper belt is a dynamically quiescent group that appears to have been retained in situ \citep{Batygin:2011, Gladman2021}.
Overlapping this population are several populations, which may be grouped together as dynamically excited: objects in a wide variety of mean-motion resonances with Neptune, those in the dynamically hot classical Kuiper belt, the scattered disc that are actively changing semimajor axis due to interactions with Neptune, the fraction of scattering TNOs that are weakly sticking to distant resonances of Neptune while evolving in $a$, and more distantly in perihelia, the detached disc and the less-understood population of extreme trans-Neptunian objects.
Together, these populations are both more massive and more numerous (estimated at $>10^6$ objects down to an absolute magnitude $H_r<12$; \citealt{Lawler2018}) than all inner populations; for a detailed population analysis see \citet{Crompvoets:2022,Beaudoin:2023}. 

Almost all trans-Neptunian objects have predominantly icy surface compositions, with a range of volatiles. The main species detected are \ce{H2O}, \ce{CO2} (and its isotopologue \ce{^{13}CO2}, CO,  \ce{CH4}, and \ce{CH3OH} and complex organics and molecules containing aliphatic C–H, \ce{C#N}, O–H and N–H bonds
\citep{Brown:2012,Pinilla-Alonso2025}.
As in the inner system, a small fraction of bodies excited from the scattering disc will transition into dynamically temporary populations.
These populations, the Centaurs and the short-period or Jupiter family comets, cross the orbits of inward giant planets \citep{Fraser2022}.
We consider the orbital structure and compositions of the most distant and most numerous outer Solar System reservoir, the Oort cloud, in more detail later in the chapter. Comet surveys imply it comprises ${\sim10^{11}}$ objects larger than ${0.3 \; \rm km}$ in radius, with semimajor axes above ${\sim1000 \; \rm au}$, though these numbers are highly dependent on assumptions about the comet infall rate, flux-radius relation, and orbital distribution \citep{Kaib:2024}.

In this chapter, we discuss the process of crafting an exocomet, from its star's birth environment to the formation of a planetesimal in the protoplanetary disc, and the ways this process may vary across the range of stellar environments.
We describe the relationship of formation theory and Solar System analogues seen in the present day in the reservoirs in our system, used as empirical tests.
As discs evolve, they move from the protoplanetary stage of planetesimal formation to a stage with collisional evolution, termed debris discs.
The mature stage of our main-sequence Solar System seen today comprises several main reservoirs of bodies considered comets, such as the Oort cloud.
Common dynamical processes suggest these reservoirs are likely to exist in other systems, though the level of observational evidence varies.

Once formed, reservoirs of small bodies are not static populations: they evolve with time, both through dynamical processes and through a broad range of physical processes that modify the interior and surfaces of individual bodies.
These evolutionary effects are considered in Mustill et al. (submitted).

\section{Star formation and its environment}


Stars are formed in dense molecular clouds, also often called dense cores, with a scale of 0.1~pc through their gravitational collapse \citep[e.g.,][]{Myers1983,Shu1987}. Protostars deeply embedded in dense cores are optically invisible, whereas they are identified as infrared sources \citep[e.g.,][]{Beichman1986}. Some dense cores, however, have no infrared sources, and such dense cores are considered to be in the stage of pre-star formation \citep[e.g.,][]{Bergin2007ARA&A}. Dense cores forming stars not only gravitationally collapse but also often slowly rotate \citep[e.g.,][]{Goodman1993}, and infalling materials conserve their specific angular momenta \citep[e.g.,][]{Ohashi1997}, forming discs around central protostars \citep[e.g.,][]{Ohashi2014}. Discs around protostars are often called protostellar discs. They are also sometimes called embedded discs because they are still embedded in infalling envelopes. Discs are considered to gradually increase their masses and sizes as central protostars grow \citep[e.g.,][]{Terebey1984}, although the details of such evolution are still under debate \citep[e.g.,][]{Sheehan2022, Yen2024}. When materials in dense cores infall, a fraction of material is removed as outflows or more collimated jets, which carry away angular momentum of the infalling materials. Optically invisible protostars eventually become optically visible pre-main-sequence stars when infalling materials are consumed, both through mass infall to the central star-disc systems and also through mass outflows.

Young stars are categorized into four stages based on their evolution and surrounding matter: class 0, class I, class II, and class III \citep{Williams2011}. This classification relies on the slope of the spectral energy distribution (SED) in the 2 - 20 $\mu$m range. Class 0 stars are strongly embedded and lack complete emission in this range. Class I shows a blackbody component from the central object, with most energy from a dusty cocoon. Class II, or young T Tauri stars, primarily emit from the central object, with some contribution from an optically thick disk. In Class III, the disk is optically thin and significantly less massive.

These discs around young stellar objects are considered to be the site of planet formation. 
Indeed, recent direct imaging of protoplanets in the disc around T Tauri star PDS 70 \citep{Benisty2021} strongly support this notion. In addition, recent ALMA observations at high angular resolution reveal that substructures, such as rings and gaps, are ubiquitous in discs around Class II sources \citep[e.g.,][]{Andrews2018, Cieza2019}. Although the origin of these substructures is still under debate \citep[e.g.,][]{Okuzumi2016, Takahashi2018}, these substructures strongly suggest that planet formation actually takes place in discs around Class II sources \citep[e.g.,][]{Dodson-Robinson2011, Dipierro2016}. Because of the importance of discs around Class II stars for planet formation, they have been extensively observed in the past decades, allowing us to do demographics of discs. Such studies provide us with overall physical conditions of discs, such as disc mass and radius \citep[e.g.,][see more details in the next section]{Manara2023}. 

In contrast to Class II discs, discs around Class 0/I protostars are still undergoing strong development.  They are much more massive than discs around Class II stars and may play an important role in planet formation. Recent studies suggest that dust masses of Class II discs are insufficient to account for the solid masses of most observed exoplanets \citep[e.g.,][]{Manara2018}, while Class 0/I discs do seem to have enough mass for exoplanets \citep[e.g.,][]{Tychoniec2020}. A recent imaging survey for Class 0/I discs at a high angular resolution revealed that they have fewer distinctive substructures in marked contrast to Class II discs, suggesting that substructures could be rapidly developed when central stars evolve from protostars to Class II sources \citep{Ohashi2023}. The survey also revealed that Class 0/I discs are more spatially thick than Class II discs \citep{Takakuwa2024}. Interestingly, their disc sizes increase as a function of the central stellar mass \citep{Yen2024}. It is still not clear whether this trend is due to disc evolution or spectral type dependence.  

Most stars are born within a cluster of stars \citep{Lada2003}, not in isolation. 
The star cluster environment may affect the protoplanetary disc surrounding the young stars and, therefore, also the resulting exocomet populations. The environment can affect the discs in two ways -- by close stellar flybys and by intense radiation \citep{Winter2018,Concha2023}. 
Both mechanisms can lead to disc truncation and even disc destruction, thus they may reduce the amount of disc material available for planetesimal formation. How strongly the environment affects the discs depends strongly on the cluster density, which can vary over seven orders of magnitude \citep{Pfalzner2013}.

During the early disc phases ($<5$~Myr), it depends on the cluster type whether external photo-evaporation or stellar flybys are the dominant effect on the protoplanetary discs. 
The radiation field is only efficient in removing discs if the cluster contains rare, massive, O and B stars. 
This external photo-evaporation dominates in massive long-lived clusters, where the number of stars in the clusters $N_{\rm stars}$ is large and stellar density $n$ high, ($N_{\rm stars} >$ 1000, $n >$ 1000/pc$^3$) \citep{Winter2018,Concha2023}. 
By contrast, flybys dominate in low-mass short-lived clusters \citep{Pfalzner:2021}. 

As external photo-evaporation only affects the gas in the disc, it is no longer efficient in the debris disc phase (debris discs contain much less detectable gas than protoplanetary discs; e.g. \citealt{Kral2017}). By contrast, flybys may affect discs throughout the star's lifetime. However, close stellar flybys are most common during the early phases. The frequency of close stellar flybys differs greatly depending on the type of star cluster \citep{Adams2010}. The underlying reason for this diversity is that star clusters can be divided into two major types -- long-lived and short-lived, sometimes referred to as clusters and associations \citep{Pfalzner2009}. While both cluster types expand by about a factor of 5--10 during the first 10 Myr of their existence, their stellar density differs by a factor of 100--1000 at any given age. Thus, close encounters can happen in both environments, but are more likely in long-lived clusters. The likelihood of a close stellar flyby also depends on the star's mass. High-mass (A-type and above) and M-type stars are more likely to undergo a close stellar flyby than intermediate-mass stars, as high-mass stars function as gravitational foci \citep{Vincke:2018}. 


The environment can strongly influence the properties of the disc from which the planetesimals may form.  It also may lead to matter transport within the disc, potentially affecting the chemistry within the disc. The question is what role these discs' physical and chemical properties play in forming the planetesimals. 

\section{Initial chemical composition and physical conditions of protoplanetary discs} 




Planetesimals, and then planets, are believed to be formed in protoplanetary discs, where the number density of dust particles is high. 
The dust-to-gas mass ratio in these discs is commonly much higher than the typical ISM value of 0.01. 
This high dust density allows the dust to grow to larger sizes through processes of collisional sticking \citep[e.g.,][]{Nakagawa1986}. 
Processes such as trapping in gas pressure bumps and the sedimentation of dust particles lead to 
locally high dust-to-gas ratios, further accelerating the formation of planetesimals.
Recent ALMA-survey observations of protoplanetary discs show a large variety of dust masses ranging from $\sim 10^{3}$M$_{\oplus}$ to $\sim 0.1$M$_{\oplus}$, depending on the age of the discs \citep{Bae2023}. The uncertainty in measuring the gas mass in the disc is quite large, since neither the gas-to-dust mass ratio nor the CO-to-H$_2$ gas mass ratio are uniform in the discs, which is different from the case of molecular clouds. A recently proposed method to measure the mass of the disc gas using the pressure broadening of the CO transition lines shows that a protoplanetary disc has gas of $\sim 7$ Jupiter mass inside the disc radius of $\sim 5$ au \citep{Yoshida2022}. The measured sizes of the discs are very different, depending on the observed tracers. For example, the size traced by CO transition lines or infrared dust-scattered light is more than five times larger than that traced by millimeter dust continuum emission in a protoplanetary disc, which could be explained by drift of pebbles (mm-dm grains) towards the central star. ALMA observations can resolve the dust continuum emission, showing that the majority are compact (the disc radii of $\leq 50$ au), while some discs have larger disc radii of up to $\sim 200$ au \citep[e.g.,][]{Miotello2023}.

The composition of the reservoirs is significantly controlled by the thermal conditions. 
The materials in the discs are heated by stellar irradiation, and the temperature profile depends on the stellar luminosity and the distance from the central star. 
Turbulent viscosity could also heat the materials near the midplane of the inner discs, depending on the strength of the turbulence. 
In addition, the properties of the dust grains, such as the size distribution and spatial distribution, affect the temperature profile of the discs through the opacity of the dust \citep[e.g.,][]{Oka2011, Harsono2015}. 
Dust grains near the midplane of the outer discs are cold because they are shielded from the direct irradiation of the star, so that gas-phase species are frozen out on dust grains and subsequent surface reactions proceed until dust grains are locked into planetesimals and larger objects, the disc gases disperse, or the grains drift past a snow line.

In the hot inner discs, water and organic molecules thermally vaporise into the gas phase. 
The boundary outside of which a certain molecule exists as a solid is called its ‘snowline’, and the location of the snowline depends on how easily each molecule sublimates. For example, carbon monoxide (CO) sublimates at a lower temperature than water. The sublimation temperatures of CO and water are around 20~K and 150~K, respectively, with the location of the CO snowline being around 20–30 au, while the location of the water snowline is around 1–2 au in protoplanetary discs around solar-mass stars. The location depends on the luminosity of the central star, and for more luminous stars, the snowlines appear at much greater distances from the disc. Because water is the main component of ices, the composition of planetesimals (icy/rocky) created in the discs varies significantly inside and beyond the water snowline; the mass of ice is significantly reduced inside the snowline, where only refractory organic matter and water contained in hydrous minerals, such as some silicates, can exist as solids. 

Recent ALMA observations and model calculations suggest that the compositions of gas and ice are not simply controlled by the molecular snowlines. 
Observations have revealed CO gas depletion in the outer discs as well as even inside the CO snowline \citep[e.g.,][]{Zhang2019, Yoshida2022}. 
The CO gas depletion is likely to proceed as the discs evolve from Class 0/I to Class II objects with a timescale of approximately one million years \citep[e.g.,][]{Bergner2020, Zhang2020}. 
It is explained by theoretical models through destruction of CO in gas-phase by UV radiation and/or cosmic-rays and the subsequent freeze-out of molecules on dust grains following grain surface reactions, which is stimulated by turbulent mixing of gas and dust in the discs \citep[e.g.,][]{Krijt2020, Furuya2022}. 
Anomalies in elemental abundances in carbon, oxygen, and nitrogen in gas-phase in protoplanetary discs have been suggested by recent ALMA observations. The carbon-to-oxygen elemental abundance ratio is higher than one (C$/$O$>1$), in contradiction to the local interstellar medium and our Solar System, and even the nitrogen elemental abundance is larger than carbon and oxygen elemental abundances in gas-phase in some discs \citep[e.g.,][]{Oberg2023, Krijt2023}. Such anomalies are also explained by the above-mentioned model with freeze-out of molecules on dust grains with different tendency depending on the elements; oxygen, carbon, and nitrogen are more likely incorporated into ice in this order. The composition of icy planetesimals must depend on where and when they form in the discs.

The type of star is one of the keys to determine the location of the snowlines.
The more luminous the star, the more distant the snowlines are situated. 
As a result, the ice lines of more massive stars are positioned at significantly greater distances than those around solar-type stars. In the next section, we explore the impact of the mass of the host star on planetesimal formation and, consequently, the creation of exocomet reservoirs. Notably, exocomets have been predominantly/exclusively discovered in the vicinity of A stars.

\section{Dependence on host star mass}

The existence of exocomets requires both a source population (a disc or Oort-like cloud) and one or more massive bodies in the system (planets or stars) to drive the cometary nuclei down to small pericentres. The formation of these populations depends on conditions of star and planet formation, which in turn depend on the mass of the primary (proto-)star. Meanwhile, our observations are differentially sensitive to bodies orbiting stars of different masses.

Observationally, we have a good understanding of stellar multiplicity as a function of stellar mass, with the fraction of stars in binaries (or higher multiples) rising with mass, from around 1 in 4 for M dwarfs, 1 in 2 for Sun-like stars, to over 60\% for B stars \citep{Duchene2013}. 
Detection of lower-mass (planetary) companions becomes strongly affected by detection biases for many regions of parameter space, while precise occurrence rates for a particular class of planet (such as ``Jupiter-like'') can be sensitive to the cuts in semimajor axis (or period) and planetary mass $m$ (or $m\sin I$, where $I$ is inclination). 
We content ourselves here therefore with a reasonably qualitative description: exoplanet demographics is a complex field, and many reviews exist \citep[e.g.,][]{Zhu2021}.

In the context of exocomets, we are primarily interested in wide-orbit planets beyond a few au, where planets are close to the source reservoir. 
Here, the most prolific detection method (planetary transit) fails. 
Direct imaging is most powerful at wide separations, albeit only to gas-giant (Jovian or super-Jovian) planets at present. 
Roughly 10\% of B stars host such planets detectable by direct imaging \citep{Janson2021,Delorme2024}. This is roughly comparable to the occurrence rate of wide-orbit giant planets around Sun-like stars, although other imaging surveys reveal a positive correlation with stellar mass \citep{Nielsen2019,Vigan2021}. White dwarf studies show that about 10\% of white dwarfs whose main-sequence progenitor masses were above $3.5\mathrm{\,M_{\odot}}$ currently host, and therefore once formed, planetary systems \citep{OuldRouis2024}.
One of the highest of these masses is $4.8\mathrm{\,M_{\odot}}$, for the white dwarf WD~J2317+1830 \citep{Hollands2021}. Radial-velocity surveys again reveal roughly similar rates, with a peak in giant planet occurrence at around $2\mathrm{\,M}_\odot$ \citep{Reffert2015}; note that this relies on the study of \emph{evolved} former A stars, as main-sequence A stars have too broad spectral lines for precision RV measurements. For reasons discussed in Mustill et al. (submitted), lower-mass planets may be preferred as dynamical drivers of exocometary activity. Unfortunately, sub-Jupiter planets on wide orbits are essentially invisible with current technology, although some are picked up in gravitational microlensing surveys, which suggest occurrence rates of at least tens of percent for planets roughly Neptune mass or larger located around or beyond the snowline \citep{Shvartzvald2016,Poleski2021}. The host stars probed by these surveys tend to be low mass (M dwarfs). The overall picture, then, is a positive correlation between stellar mass and the presence of a wide-orbit Jupiter-like planet, at least up to $\sim2\mathrm{\,M}_\odot$, while lower-mass planets are considerably more numerous but too hard to probe to yet say whether their presence on wide orbits depends on stellar mass. 

The occurrence rates of close-in
planets ($\lesssim1$\,au) have been provided by transit surveys \citep[e.g.,][]{Dressing2013,Mulders2015,Hardegree-Ullman2019,Yang2020,He2021}, which have 
relatively simple biases, large sample sizes, and sensitivity to small (down to Earth-sized) planets. There appears to be  
a decrease in close-in planet occurrence rate between M and F-type stars, although it is unclear how this can be extrapolated to the wider-orbit planets presumably responsible for driving exocomet dynamics. These observations have motivated many studies of the dependence of planet
formation as a function of stellar mass for roughly
Solar-mass stars and smaller \citep[e.g.,][]{Liu2019,Burn2021,Mulders2021}. 
Less interest has attached to A stars, 
possibly because, by virtue of their smaller number, 
they are less common in transit surveys. Some work here
has been motivated by observations of post-main sequence planetary systems: see \cite{Veras2020HighMass,Kunitomo2021,Johnston2024}. 

The observed prevalence of planet formation around stars of different masses indicates that exocomets should not be formed exclusively, or even predominantly, around A-type stars, where their
detection is favoured (see Korth et al., submitted, for discussion).  
Debris discs are probably the primary source of exocomets, and debris discs are more commonly detected around younger stars and, for stars of a similar age, earlier spectral types (\mbox{Figure \ref{fig:DebrisDiscMassAndLocationVsAge}}, and e.g., \citealt{Su2006, Krivov2010, Wyatt2008, Matthews2014, Moor2016, Hughes2018, Sibthorpe2018, Marino2022, Pearce2024Review}). Note however that it is easier to detect a disc of a given temperature and fractional luminosity around earlier-type stars, as the disc emission is more clearly separated from the stellar blackbody \citep[e.g.,][]{Wyatt2008}.
In reality, all main-sequence stars probably have some amount of debris, and thus potentially exocomets. 
The phenomenon of exozodiacal dust within $\sim1$\,au of the host star, detectable as extended near- or mid-IR emission via interferometric observations \citep{Ertel2014,Ertel2020}, may be related to cometary activity (\citealt{Sezestre2019, Pearce2022Comets}; though also see \citealt{Pearce2020}). It correlates strongly with the presence of a detected cold debris disc in the system, and detected presence declines somewhat with lower stellar mass (comparing A--early F with late F--GK stars), but the latter can be explained by increased detection sensitivity for earlier spectral types \citep{Ertel2020}.




\section{Early stages of planetesimal formation}
\label{sec:early_planetesimal}

\begin{figure}
    \centering
    \includegraphics[width=0.99\textwidth]{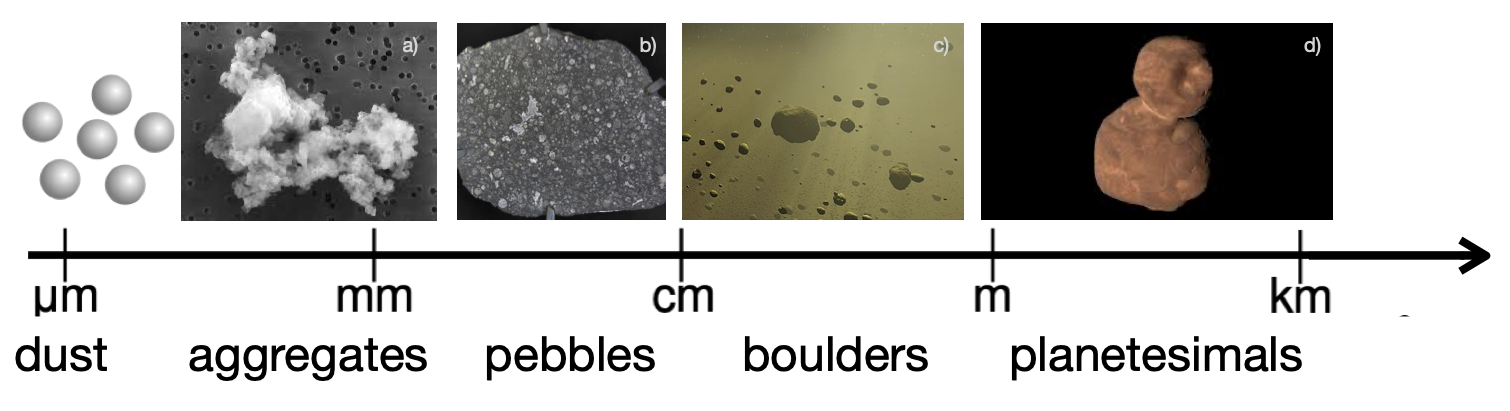}
    \caption{Schematic picture illustrating the growth from dust to planetesimal sizes. 
    `Dust' of silicate- and organics-rich grains are shown schematically; these grow to fractal dust structures via sticking collisions. 
    After compactifying and further growth they reach planetesimal sizes. 
    Credits: a) Amara/Wikipedia (CC), b) via Wikimedia Commons, c) adapted from Wikipedia Commons, and d) cold classical Arrokoth (NASA).}
    \label{fig:Dust2planetesimals}
\end{figure}

It is generally assumed that dust in protoplanetary discs grows from initial sizes of $\sim $ 0.1 -- 1 $\mu$m dust grains to planetesimals of sizes in the range $\sim $1–1000 km, a process shown schematically in Fig.~\ref{fig:Dust2planetesimals}. 
Mutual collisions among protoplanetary dust particles are important for understanding the buildup of these larger bodies. 
Pioneering theoretical work on dust growth \citep{Weidenschilling1977,Weidenschilling1993} was followed in the last two decades by complementary laboratory experiments \citep[for a review, see][]{Blum2018}. 
For dust grains or small aggregates smaller than $\sim10-100\,\mu$m, the collisions are mainly driven by Brownian motion \citep{Kempf:1999,Dominik2007}, with typical velocities around 1 mm/s. 
Larger aggregates ($>$ 10–100\,$\mu$m) experience a systematic drift relative to the gas, due to the sub-Keplerian motion of the pressure-supported gas. 
The result is a net inward force on the dust, leading to an increased radial drift velocity with larger aggregate size, peaking at around 60 m/s for 1 m diameter bodies; this limits their lifetimes to merely $\sim$~years at 1 au  \citep{Blum2018}. 

Differences in how dust and gas stick together cause variations in the speed of drift among aggregates.
Systematic drift of dust particles relative to the gas also occurs, due to sedimentary motion towards the disc midplane. 
Larger dust aggregates have a greater surface area-to-mass ratio. 
Thus, they sediment faster than smaller aggregates, resulting in collisions among aggregates of different sizes. 
Additionally, gas turbulence significantly influences dust aggregate evolution by causing collisions, especially among larger aggregates --- leading to stochastic particle motion, regardless of their aerodynamic properties.

A key problem in aggregate growth was identified experimentally: when dust aggregates of $\mu$m- or mm-sized silicate grains collide, they bounce or fragment, rather than sticking \citep{Blum:1993}. 
Since this early work, the parameter space has been expanded to dust sizes between $\sim$ 1 $\mu$m and 10 cm, impact velocities between $\sim 10^{-3}$ m/s and $\sim100$ m/s, and agglomerates of water ice \citep{Gundlach2015}, CO$_2$ ice and CO$_2$--H$_2$O ice mixtures \citep{Musiolik2016} have been tested.
Monte-Carlo simulations of the aggregation process of silicate dust grains show that after an initial process of fractal growth, sticking and bouncing collisions lead to the formation of compact mm- to cm-sized aggregates, which have relatively low volume filling factors of $\sim 0.36$ \citep{Zsom2010}. 
Further growth is inhibited by the bouncing barrier, as no further mass transfer can be achieved \citep{Wurm2021}.  Such a bouncing barrier may not exist for icy particles \citep[see, for example, ][]{Okuzumi2012}. 

Aggregate growth has some dependence on both location in the protoplanetary disc, and the disc parameter termed dust-to-ice ratio: the ratio of silicate- and organics-rich particles to those of volatile ices.
The maximum aggregate size that can form at a given point in the disc depends on both the distance to the star and the composition of the dust \citep{Lorek:2018}. 
Increasing distances to the star, and increasing dust-to-ice ratios, result in smaller maximum aggregate sizes. 

After reaching the bouncing barrier \citep{Dominik2024}, the gap to fragmentation, mass transfer or cratering may be overcome, due to the velocity distribution of the dust aggregates. 
The destructive processes eroding aggregates will form new, smaller aggregates.
Once these projectile aggregates develop, their mass transfer allows some dust aggregates to grow to planetesimal sizes \citep{Windmark2012}.

Although the time scales for growth to the 100-metre level are reasonably short at distances of 1~au, they are much longer further out in the disc, and the maximum aggregate sizes are thus considerably smaller \citep{Birnstiel2024}. 
\citet{Garaud2013} showed that at 30~au, even after 6 $\times$ 10$^5$ years, the maximum aggregate size is only a few metres. 
Typically, the growth time scales are so long that outside dust traps in the disc, where the radial pressure gradient of the gas vanishes, halting dust drift, radial drift limits the maximum size achievable. The formation of planetesimals likely proceeds through alternative pathways, as described in the next Section.



\section{Late stages of planetesimal formation}
\label{sec:late_planetesimal}

\begin{figure}
    \centering
    \includegraphics[width=0.89\textwidth]{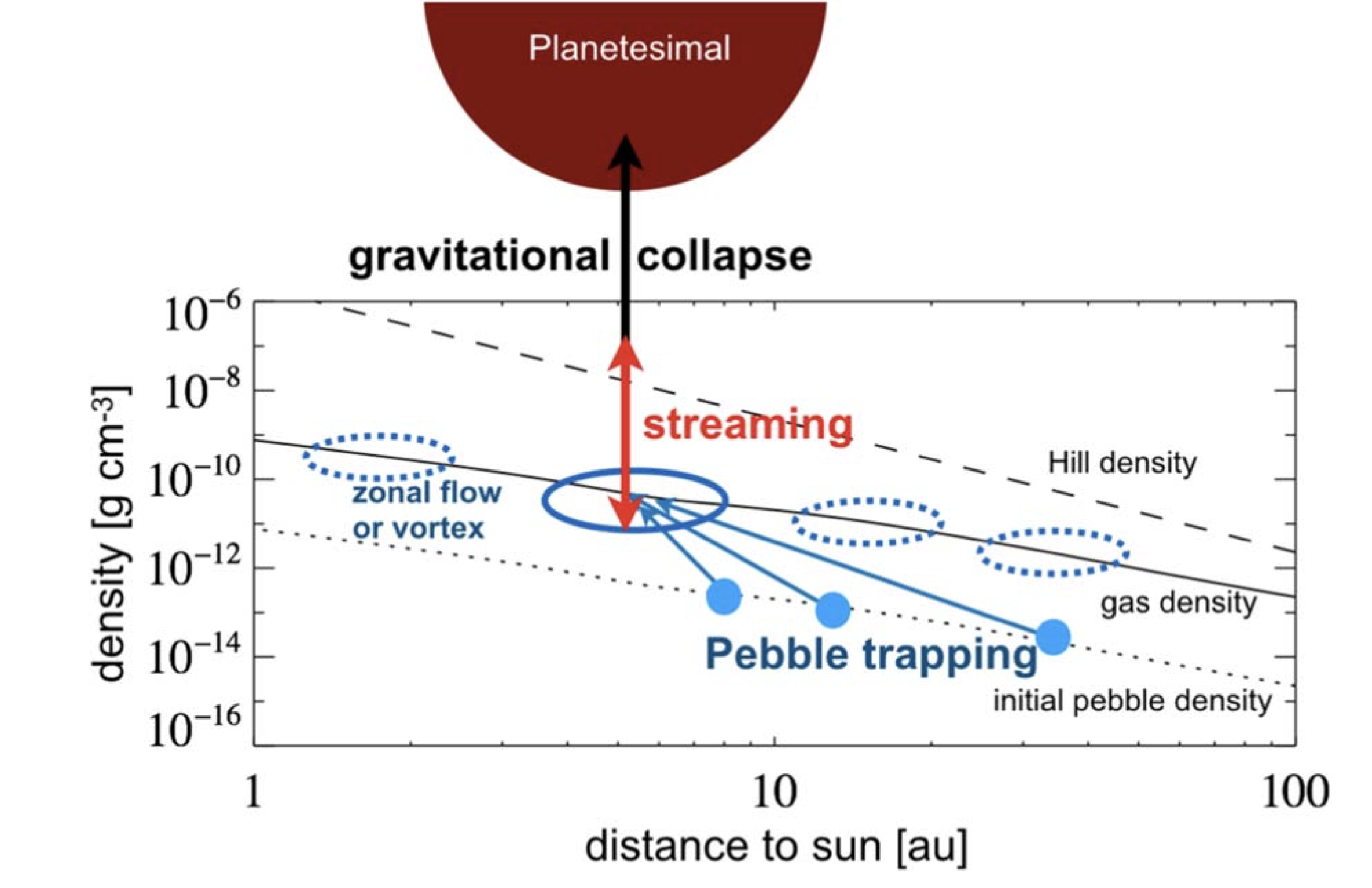}
    \caption{Drift-trap-stream-collapse (schematic representation of densities). 
    In the current paradigm the sequence of planetesimal formation starts with particles growing to pebble size; they then sediment to the midplane and drift toward the star to get temporarily trapped in a zonal flow or vortex \citep{2019ApJ...874...36L}. 
    Here the density of pebbles in the midplane is eventually sufficient to trigger the streaming instability. 
    This instability concentrates and diffuses pebbles likewise, leading to clumps that reach the Hill density at which tidal forces from the star can no longer shear the clumps away. 
    If the turbulent diffusion is now also weak enough to let the pebble cloud collapse, planetesimal formation will occur.}
    \label{fig:PlanetesimalFormation}
\end{figure}

\subsection{The challenge of planetesimal formation}

Planetesimals form most likely through the gravitational collapse of clouds of pebbles (mm--dm grains) where these can locally accumulate in the disc, reaching densities exceeding the Hill density where self-gravity overcomes the Keplerian shear that tears the cloud apart.
The process was initially described in the foundational work by \citet{Safronov1972} and \citet{Goldreich1973}.
This original idea was then largely abandoned for several decades, as \citet{Weidenschilling1980} pointed out that at the necessary high dust-to-gas ratios for gravitational collapse in the midplane, strong turbulence will be triggered by vertical shear in the disc. 
That is, the dust-dominated midplane will rotate at a Keplerian value, whereas the gas-dominated layers above and below will rotate at sub-Keplerian value, due to the radial pressure gradient in the gas. 
This results in the development of the Kelvin--Helmholtz Instability \citep[MRI;][]{Chandrasekhar1961}, and so the dust concentration should be quickly diffused by turbulence, before a gravitational collapse can happen.

The conclusion by \citet{Weidenschilling1980} was that possibly coagulation, i.e.\ non-elastic and sticking collisions of dust grains, leads to growth up to the sizes of comets and asteroids. Yet as discussed in Sec~\ref{sec:early_planetesimal}, extended experimental data on collision velocities that would lead to bouncing and fragmentation rather than sticking showed that this pathway to planetesimal formation is not possible \citep{Blum2018}.

Much attention has therefore focused on identifying conditions that might permit sufficiently dense clouds of pebbles to accumulate and collapse. An early suggestion was concentration by Kolmogorov turbulence in the gas disc: particles can stochastically be driven to high densities between turbulent eddies \citep{Cuzzi1993,Pan2011,Hartlep2020}.
However, the stochastic nature and short lifetime of these eddies means that the probability of attaining the high density required may be low.

Many disc processes, however, give rise to large-scale structures that can trap particles. Particle trapping is systematic in slowly-rotating, long-lived vortices  \citep{BargeSommeria1995,Klahr1997}, which do not appear in the Kolmogorov turbulence invoked in \citet{Cuzzi1993}. However, in both hydrodynamical \citep{Raettig2015,Manger2018} and magnetohydrodynamical simulations \citep{Johansen2005,Johansen2006,Lesur2023}, the formation of such particle-trapping vortices, as well as pressure bumps, is a common phenomenon. These can be driven, for example, by the magnetorotational instability \citep[MRI; ][]{Balbus1991}, and early simulations of turbulent discs driven by  the MRI and including the self gravity of pebbles showed rapid formation of planetesimals in zonal flows \citep{2007Natur.448.1022J}.

Particle concentration is aided by vertical settling, which \cite{Weidenschilling1980} argued should be prevented by KHI. However, the first direct numerical simulations of the KHI \citep{Johansen2006b} of pebbles sedimenting to the midplane of the disc showed that while on average pebbles diffused away from the midplane, strong local clustering in fact occurred. Such simulations of turbulence caused by sedimentation of particles have often been interpreted as the Streaming Instability, discussed next.

\subsection{The streaming instability}

The streaming instability (SI) arises as a result of the back-reaction of the pebbles on the gas: pebbles orbit at the Keplerian velocity, while the gas disc's azimuthal velocity is affected by the radial pressure gradient. The drag force on the solids also causes the gas to be accelerated towards the solid velocity, which process becomes more significant as the solid:gas ratio increases. Particles embedded in the gas drift radially inward due to their lack of pressure support to balance the diminished centrifugal force caused by the sub-Keplerian gas motion. This drift results in local accumulation of solid particles, as follows.

The streaming instability relies on a fundamental oscillation mode of the disc gas, in which the gas moves diagonally (radial--vertical) while conserving its angular momentum. 
This is a generalization of the familiar epicyclic oscillations of particles around a circular orbit.  
In the part of the gas bands where the gas is at its furthest outward position during the oscillation, the gas has lower angular momentum than its surroundings. 
As a result, the particles drift faster than average, while at the opposite point in the oscillation cycle, the particles drift slower. 
Consequently, the particles accumulate at the neutral point of the oscillation, when the gas moves outward, giving the gas a ``kick'' towards the Keplerian value. 
In contrast, the inward-moving gas experiences a deficiency of dust, and this results in an inward ``kick'' on the gas. 
If the particles in the gas move radially inward at the same velocity as the phase velocity of the gas oscillations, a resonant amplification of the gas oscillation occurs (from this, SI can also be classified as a resonant drag instability; \citealt{Squire2018SI}). 
As a result, SI can amplify the local dust-to-gas ratio by an order of magnitude before nonlinear effects kick in. 
As the radial drift velocity of the pebbles depends on their stopping time, the wavelength, oscillation frequency and thus the phase velocity of the amplified oscillation are strongly dependent on the particle size \citep{Youdin2005}. This has been argued as an obstacle for SI if one has a wide spectrum of particle sizes \citep{Krapp2019}, though problems may not arise at sufficiently high dust-to-gas ratios \citep{Schaffer2021,ZhuYang2021,Yang2021}.

The streaming instability relies on a significant back-reaction from the dust on the gas, thus requiring a sufficiently high dust-to-gas ratio. The initial condition to trigger SI is a local dust-to-gas ratio of at least order unity, similar to the condition for KHI. This can be attained through vertical sedimentation; however, a super-Solar overall metallicity (about 2--3 times Solar) is required to overcome turbulent diffusion and trigger gravitational collapse \citep{Johansen2009b,Carrera2017,Yang2017,Gerbig2020,Klahr2020,Klahr2021,Li2021,Lim2024}.
The required high levels of solid content can be potentially caused by photo-evaporation in late-stage discs or early trapping in vortices and zonal flows \citep{Carrera2017,Lenz_2019}.
The key condition for planetesimal formation may not merely be reaching a midplane dust-to-gas ratio of unity, 
but achieving sufficient local mass to reach Hill density. 
Overcoming this threshold minimizes turbulence effects and facilitates gravitational
collapse \citep{2007Natur.448.1022J}.

It has now become common to refer to the simulations of turbulence by sedimenting pebbles as `planetesimal formation via the streaming instability'.
However, we want to stress that the role of SI in the generation of the high pebble-to-gas ratios, beyond the typical one order of magnitude, has not yet been explained \citep{Li2018}. 
Yet for the conclusions from the related numerical work with and without the inclusion of extra turbulence by other mechanisms than the particle feedback, it is not relevant to check for the role of SI.
It may become important in the future once we better know parameters such as the actual pebble size distribution, the actual level of turbulence in the gas nebula, and finally its global mass content of solid material.

\subsection{Initial mass distribution}

Simulations of planetesimal formation in the so-called `SI scenario' reveal size distributions with shallow power laws, with mass distributions such as $dN/dM_{\rm P} \propto M_{\rm P}^{-p}$ ($p \approx 1.6$ or $p \approx 1.3$ for truncated distributions) favoring large planetesimals \citep{Johansen2015,Simon2016,Schaefer2017,Abod_2019, Ormel2025}. 
High-resolution studies suggest a critical collapse mass may explain the deficiency of smaller planetesimals, emphasizing the role of turbulent diffusivity \citep{Klahr2021}. The initial size of planetesimals depends on the size distribution of pebbles and the level of turbulence in the gas nebula. 

Many planetesimals from the initial disc merged into planets, but the remnant populations, such as asteroids and trans-Neptunian objects, and in other systems exo-comets, are diagnostic of the original populations. 
Studying these bodies provides key insights into planetesimal formation and planetary evolution, though reconstructing their initial size distribution requires accounting for the system's lifetime of collisional evolution.

In the Solar System, when using the parameters of \citet{Lenz2020}, one finds that both asteroids and cold classical KBOs should form with a diameter of $\approx 100$ km.
Observations reveal that some asteroids, predominantly $\sim$100 km in size, are not collisional fragments but primordial objects, supporting this ``born big" hypothesis \citep{Morbidelli2009,Delbo2017}. 
Additional evidence includes the bi-lobed structure of Arrokoth and the high binary fraction among cold classical TNOs, consistent with formation via gravitational collapse \citep{Stern2019,Nesvorny2019}.
Numerical simulations of gas and dust that trace the gravitational collapse in streaming instability have demonstrated that it provides a qualitative agreement with the observed distributions of the orbital inclinations of Kuiper Belt binaries \citep{Nesvorny2019}. 
The observed heterogeneity of the reflectance spectra, colors, and surface composition of trans-Neptunian binaries further supports formation via gravitational collapse of a pebble concentration \citep{Marsset2020}, as does the size-dependent bulk density of TNOs \citep{Brown2013,WahlbergJansson2014}.

How may planetesimal formation proceed around other stars?
Under the assumption that the formation of exo-comets happens analogously to the formation of minor bodies in the Solar System, we can translate the predicted mass for planetesimals $m_c$ from that for Solar-type stars.
\cite{Klahr2021} predict $m_c\propto M$, where $M$ is the stellar mass.
Therefore, the radius of these icy and rocky objects will not depend strongly on the stellar type.
As in the Solar System, planetesimals will obtain typical diameters of about 100 km at the time of their formation. The size may also depend on formation location in the disc, with a mass predicted to increase roughly linearly with distance from the star \citep{Li2019,Liu2020}.

\section{From protoplanetary disc to debris disc}

\begin{figure}
    \centering
    \includegraphics[width=0.99\textwidth]{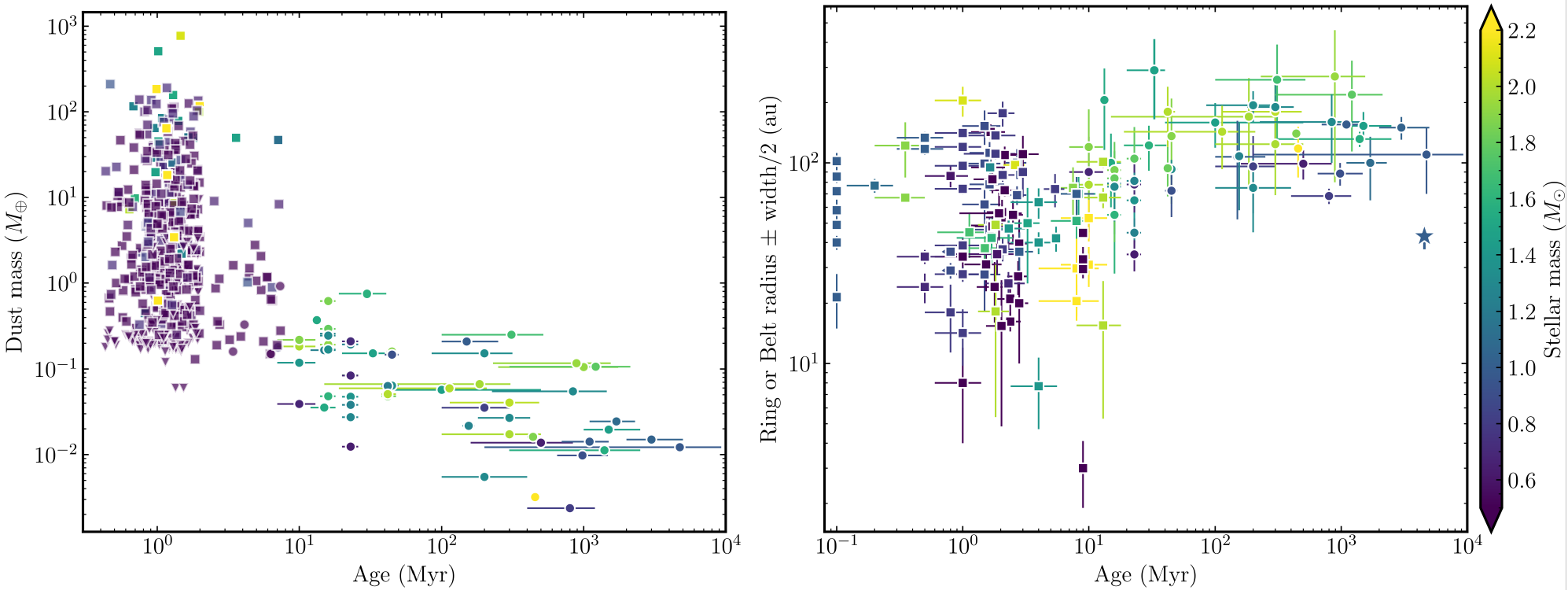}
    \caption{Dust mass (left) and radius (right) of protoplanetary and debris discs as a function of system age. Protoplanetary discs are represented with squares and debris or evolved discs with circles. The Kuiper belt is represented with a star on the right panel. Debris data from \cite{Matra2025}, and protoplanetary disc data from \citet{Bae2023}. The colour scale indicates the stellar mass in both panels.
    }
    \label{fig:DebrisDiscMassAndLocationVsAge}
\end{figure}

Looking now at the entire extent of a protoplanetary disc, the dust growth, pebble and planetesimal formation happen with different efficiencies, and may not be simultaneous everywhere in the disc. 
While early dust growth is observable, the planetesimal growth process itself is not directly observable. 
However, we can consider how long it takes for protoplanetary discs to turn into debris discs. 
The lifetime of a protoplanetary disc is defined as the time it takes to disperse most of its gas. 
At this point, gas giants can no longer be formed. 
Initially, it was assumed that by 6 Myr, all stars would have lost their discs \citep{Haisch2001}, with typical disc half-life times of 1--3 Myr. 
However, it turned out that such short disc lifetimes are only characteristic of stars of at least 0.8 solar masses. 
The lifetime of protoplanetary discs is a strong function of the mass/stellar type of the host star \citep{Ribas2015}. 
A-type stars, where exocomets have most typically been observed to date, typically have relatively short proto-planetary disc lifetimes (1-4 Myr), while the smaller M- and K-type stars have mean disc lifetimes of 5--10 Myr \citep{Michel2021, Pfalzner:2022}. The distribution of the protoplanetary disc lifetime is very broad, so that some M-types stars are still surrounded by a protoplanetary disc at an age of 20 Myr, with some rare examples still having a disc at 40 Myr and older  \citep[e.g.,][]{Silverberg:2020}.



Once the protoplanetary disc disperses, we are left with a disc of remnant solid material that did not make it into planets. This is the debris disc. A sizeable potential reservoir of exocomets in nearby stars is their cold debris discs, which are located in the outer regions of systems and are detected around 20\% of nearby AFGK stars \citep{Wyatt2008, Hughes2018}. 
These dusty discs are brighter analogues of the Solar System's classical Kuiper belt, and are thought to have compositions similar to those of TNOs (i.e. rich in water and hypervolatiles; \citealt{Matra2017}). 
These discs are found to be very diverse, some of them being as narrow as the present-day classical Kuiper belt (cf. 30--50~au, though at different radii) and a majority being much wider and extending beyond 100~au \citep{Marino2022, Pearce2024Review, Matra2025}. 
Some show morphological features such as warps, gaps, eccentric shapes, clumps and sharp edges \citep{Golimowski2006, Marino2018_HD107146, Faramaz2019, Booth2023, ImazBlanco2023}; these may be indicative of planetary interactions, which could drive material into the inner system to become exocomets (e.g. \citealt{Mouillet1997, Pearce2014, Pearce2015, Sefilian2021, Friebe2022, Booth2023, Pearce2024Edges}). 
The specific dynamical mechanisms by which material could be driven into the inner system are described in Mustill et al. (submitted).

How massive these discs are is an open question. Observations in the infrared and millimetre wavelengths are only sensitive to the $\mu$m- to cm-sized grains which are most likely a small percentage of the total disc mass. These observations constrain the dust disc masses (up to cm-sized grains) to $10^{-3}-1$~$M_{\oplus}$ \citep{Matra2025}, the lower end being limited by the current sensitivity of observatories. We expect much larger bodies to be present, whose collisions replenish the observed dust levels in a collisional cascade. If we extrapolate their dust masses to total mass assuming the largest exocomets/planetesimals are 1~km in size and a size distribution exponent of -3.5, we expect these discs have masses of 0.3-300~$M_{\oplus}$. It is possible that there are even larger bodies present, in which case disc masses could be even larger, but it is very unlikely that disc masses are larger than ${\sim}$1000~$M_{\oplus}$ given the available solid mass in protoplanetary discs \citep{Krivov2021}. However, since the dust masses observed in many extrasolar debris discs are considerably greater than in our own Solar System (e.g. \citealt{Pearce2024Review}), it is plausible that there is significantly more material in those exocometary reservoirs than in our own.

The material in debris discs depletes over time, as the bodies grind down through mutual collisions. The resulting dust, when small enough, is removed from the system via radiation forces (see Mustill et al., submitted). This is evidenced by dust mass decreasing with age in observed debris discs, as shown on the left panel of \mbox{Figure \ref{fig:DebrisDiscMassAndLocationVsAge}}.  This means that older debris discs have less material available to become comets than younger ones, so we may expect exocomets to be preferentially detected in younger systems.

This collisional grinding implies that planetesimals in debris discs have non-zero eccentricities and/or inclinations, to enable frequent and violent collisions that destroy bodies to release dust \citep{Wyatt2008}. There is some observational evidence of this; ALMA observations show that some debris discs have smooth edges, as expected from a population of eccentric bodies \citep{Marino2021, ImazBlanco2023, Rafikov2023}, as opposed to the sharp edges expected if discs are sculpted by planets \citep{Pearce2024Edges}.  However, it is thought that planetesimals form in protoplanetary discs on near-circular orbits, then some processes excite them up to moderate eccentricities to enable destructive collisions in debris discs. Similarly, since debris discs are the presumed reservoirs of (exo)comets, some process must change the orbits of some planetesimals significantly to drive them onto highly eccentric cometary orbits. There are several mechanisms that can do this, as described in Mustill et al. (submitted).


\section{Observational evidence from the Solar System}
\label{sec:solar_system}


In both the Solar System and beyond, we can learn about a system's formation by studying extant planetesimal populations or debris discs. 
The locations of such discs tell us where planetesimal formation succeeded but planet formation failed. 
This could be due to planet-formation timescales being too long in the outer region of systems, or because already formed planets perturb nearby planetesimals and prevent further formation (e.g. \citealt{Petit2001, Eriksson2021}), or because newly formed planets create pressure bumps in the protoplanetary disc which can facilitate planetesimal formation (e.g. \citealt{Miller2021}). 
Furthermore, debris discs contain clues about where planets are and how they migrated; the past migration of Neptune is imprinted on the orbits of resonant trans-Neptunian objects \citep{Malhotra1993}, and the shapes, features and locations of extrasolar debris discs may be telling us about planetary interactions in those systems too (e.g. \citealt{Mouillet1997, Sefilian2021, Friebe2022, Pearce2022ISPY, Booth2023, Pearce2024Edges}).

While exocomets and debris discs around other stars have only been studied for a few decades, the small-body populations in the Solar System have been the subject of astrophysical and spacecraft exploration for much longer. 
Alongside remote and in-situ observations, laboratory sample investigations have also greatly contributed to our understanding of the composition and material properties of not only meteorites, but also of samples from asteroids by Hayabusa \citep{Tsuchiyama2011}, Hayabusa2 \citep{Oba2023}, OSIRIS-REx \citep{Lauretta2024} and comet Wild 2 by Stardust \citep{Brownlee2006}. 
One of the main drivers of these research efforts has been to constrain the processes of planetesimal formation across the Solar System's protoplanetary disc, as well as the subsequent early dynamical evolution which has shaped the architecture of the Solar System. 
However, one fundamental challenge in using the properties of the current small bodies as evidence for the early stages of the Solar System is these bodies' subsequent evolution. 
The small-body populations we observe in the Solar System today have undergone various levels of dynamical, physical and chemical processing (see Mustill et al., submitted). 
Therefore, linking the observed properties of today's populations to their early progenitors relies on a careful consideration of the variety of processes that have affected them since formation. 

With this caveat in mind, in this Section, we highlight some of the observational evidence that \red{has} been used to address key questions about Solar System formation. These examples are selected to illustrate the diversity of research questions, lines of evidence, and approaches currently being pursued by Solar System observers.

Of all available observables, the orbital distribution of the populations is probably the most widely accessible property. 
The orbits and relative populations of the main small-body reservoirs and their subpopulations are becoming well characterized (e.g. see \citet{Jedicke2015} for asteroids and \citet{Gladman2008,Kavelaars2020} for TNOs), with the exception of the Oort Cloud, which retains substantial uncertainties (see Section~\ref{sec:Oort}). 
The orbital distribution of TNOs is especially diagnostic of the late disc evolution in the Solar System.
The dynamical properties of the objects in the trans-Neptunian region, the irregular satellites of the giant planets and the Trojan populations are used to constrain the dynamical evolution of the early Solar System \citep[e.g.][]{Nesvorny2018, Pirani2019}.
They are also used to quantify the influence of the stellar environment through the effects of close stellar encounters \citep[][]{Nesvorny2023b,Pfalzner2024,Pfalzner2024b}. 


The detailed characterization of the dynamical links between the outer Solar System populations reveals that comets, TNOs, irregular satellites and Trojans share a common origin in the primordial distant disc \citep[see][]{Nesvorny2018,Fraser2022}. 
All of these populations can therefore be used to discern the evolutionary properties from the primordial signatures of the conditions in their original reservoir in the outer regions of the protoplanetary disc. 
While faint, it is easiest to study these distant bodies' surface properties, and they have been explored for evidence of differences in composition. 
Multiband photometric studies have provided strong evidence that the surface colours of small TNOs, Centaurs, Jupiter Trojans and possibly Neptune Trojans are bimodal \citep[e.g.,][and references therein]{Szabo2007,Peixinho2015,Tegler2016,Wong2017,Marsset2019,Fraser2023,Markwardt2023}. 
While the details of the different color classifications are outside the scope of this chapter, we want to highlight the possibility of using them as probes of the primordial planetesimal disc composition and its dependence on heliocentric distance. 
For example, \citet{Wong2016} hypothesized that the different surface colors of Jupiter Trojans reflect the formation region of the objects with respect to the region in the primordial disc where H${_2}$S could be retained. 
Additionally, studies of the surface colors in combination with the dynamical properties of TNOs \citep[e.g.][]{Nesvorny2020,Marsset2023} have been used to probe for a compositional gradient in the primordial disc and sublimation-driven surface depletion in the early Solar System, which leads to today's differences in surface colors. 
The launch of \textit{JWST} has enabled us to obtain near-IR spectra of distant objects with unprecedented sensitivity. 
The first \textit{JWST} programs are already beginning to reveal significant differences in the surface composition among the populations \citep[e.g][]{Licandro2025,Belyakov2024,Wong2024,Harrington2023,DePra2025}. 
The first large \textit{JWST} program targeting medium-sized TNOs (DiSCo-TNOs) identified three distinct spectral groups among TNOs and Centaurs, giving rise to the hypothesis that the objects from these populations were formed in different locations with respect to the sublimation radius of CO$_2$ and several other molecules such as CH$_3$OH, C$_2$H$_6$ and HCN \citep{Pinilla-Alonso2025}. 

Before JWST, most of the progress on compositional studies of small bodies came from active objects. 
Comets and active Centaurs experience outgassing, which allows us to probe the composition of their subsurface, less altered material. 

Ongoing observational surveys aim to connect the chemical abundances of comets to their formation and thus to the properties of the protoplanetary disc \citep{Eistrup2019}. 
These include studies of easily detectable fragment species such as OH, CN, C$_2$, C$_3$, and NH in hundreds of comets \citep{AHearn1995, Cochran2012,Fink2009,Schleicher2007}; analyses of multiple minor species directly released from comet nuclei for 30 comets \citep{Mumma2011,DelloRusso2016,Saki2024}; and measurements of the relative abundance of the main ices in a sample of 25 comets and Centaurs, i.e. H$_2$O, CO$_2$, and CO \citep{AHearn2012,Pinto2022}. 
However, limitations in the sample size and observational biases towards bright Oort Cloud comets in the latter two datasets make it difficult to connect these. 
Additionally, the chemical origins of CN and C$_2$ are not understood \citep{Feldman2005}, which hampers connecting the largest dataset to the other two, or to Solar System formation. 
In addition, most observations capture only brief snapshots of cometary activity.
Long-term studies suggest that the composition of the coma can vary significantly due to evolutionary and seasonal effects \citep{Combi2020, Bodewits2014}.

\newpage

Without doubt, in-situ missions provide the most detailed compositional analysis. For example, the JAXA Hayabusa2 mission recently returned samples from the carbonaceous asteroid (162173) Ryugu that showed organic molecules \citep{Naraoka2023,Oba2023}, in contrast to the samples returned earlier from the volatile depleted stony asteroid (25143) Itokawa, which were organic poor \citep{Tsuchiyama2011}\footnote{For a recent review of the composition of (162173) Ryugu sample compositions, see \citet{Yokoyama2025}.}. Similarly, the Rosetta mission, and more specifically its Rosina instrument, provided a detailed inventory of the volatiles in the coma of comet 67P/Churyumov–Gerasimenko \citep{LeRoy2015}. 
One of the most surprising findings 
was the elevated D/H isotope ratio of 67P with respect to other Jupiter-family comets \citep{Altwegg2015}; the D/H ratio is believed to be a strong clue to formation location \citep[see][]{Jacquet2013}. 
However, results from the past decade reveal that interpreting the measured D/H ratio remains challenging. 
It cannot be ruled out that the D/H ratio estimates for a single comet vary along its orbit \citep{Paganini2017}, while the activity level of a comet seems to be linked to the measured isotope ratio \citep{Lis2019}, raising the possibility that the activity mechanisms releasing the observed gas can result in differences in the observed D/H isotope ratio. 
Even the ROSINA results have been challenged by the reanalysis of the archival data by \cite{Mandt2024}, which identify a much lower D/H ratio than originally derived from the original Rosetta results. 
The rich datasets from Rosetta still reveal new key evidence. 
\citet{Marschall2025} were able to constrain the approximate temperature and location in the disc where comet 67P has formed, estimated to lie around 25--35\,au. Other ratios also encode information about formation location and environment; for example, the volatile carbon-to-oxygen ratio can provide a zeroth-order insight into formation distance \citep{Seligman2022}.

\section{Oort clouds}
\label{sec:Oort}

In the Solar System, the Oort cloud is the source of the long-period comets. 
When looking at reservoirs for the production of exocomets, we have so far considered formation of the closer reservoirs, and the general state of debris disc structures. 
Here, we look more closely at the formation mechanisms of the Oort cloud and their likelihood for other stars. 
For a detailed review, see \citet{Kaib:2024}.
There are two regions distinguished in the Oort cloud --- the inner Oort cloud is located between 2000 and 10$^4$ au, beyond which the outer Oort cloud starts. While not directly observable, models predict the inner cloud to be the much denser of the two, having tens or hundreds of times as many cometary nuclei as the outer cloud.
Simulations of the ejection of planetesimals from the vicinity of stars show that a fraction of $\sim$~1--3\% of the ejected planetesimals remain bound to their parent star. 
The orbits of these planetesimals isotropize and circularize due to the Galactic tidal field, and eventually form an Oort cloud between $\sim$ 10$^4$ and \mbox{$\sim$ 2 $\times$ 10$^5$ au} \citep{Portegies2021}.

While long-period comets are routinely detected by sky surveys such as PanSTARRS \citep[e.g.][]{Boe:2019}, the Solar System's Oort cloud has not been observed in situ. 
The existence of a distant cloud of cometary objects that orbit the Sun is based on a spike in the reciprocal orbital separation at  \mbox{1/$a_0$ $\leq$ 10$^{-4}$ au$^{-1}$ }  \citep{Oort1950}, where $a_0$ is the original semi-major axis. One promising approach for detecting Oort cloud objects in situ is by searching for their sub-second occultations of distant stars \citep{Schlichting2012}. 
Using $\approx$1 s integration by \mbox{$>$1 m telescopes} at the optimal region near the quadrature points will be  marginally dominated by Oort cloud objects rather than Kuiper belt objects \citep{Ofek2024A}.

The formation timing of our Oort cloud is somewhat open: it may have formed together with the Solar System, potentially during a giant-planet instability, or it might have been acquired in part later on, from other stars in the Sun's birth cluster. 
It remains also an open question to what extent the Oort cloud is still in its primordial dynamical state.  
If the Oort cloud is a pristine relic of Solar System formation, then structural mapping of the Oort cloud may provide information regarding the stellar environment in which the Sun was born. 
It would also provide a picture of the planetesimal population during the outer planets' formation phase. 

There are many indications that the Solar System was born in a stellar cluster. 
During this phase, the Sun would have been exposed to closer stellar flybys than in the Galactic field. 
Minor planets with a semi-major axis between $\sim$100 au and several 10$^3$ au may still bear the signatures of the Sun being born in a $\gtrsim 1000 \mathrm{\,M_\odot\,pc}^{-3}$ star cluster \citep{Brown2004,Schwamb2010}. These signatures of the environment include the steep decline in material beyond 30 au, the Fe-60 and Al-26 content attributed to a closeby supernova and possibly the orbits and colours distributions of the trans-Neptunian objects \citep{Pfalzner:2024,Pfalzner:2025}. 
During the first 10--100 Myr after its formation, the Solar System's birth cluster either dissolved or the Sun was ejected from it. 
Either most of the outer Oort cloud formed after the Solar System was ejected \citep{Portegies2021}, or it was severely modified during that stage \citep{Pfalzner2024}.

The bulk of the Oort cloud originates from the region outside Jupiter's orbit and are potentially the least thermally processed objects \citep{Gkotsinas:2024}. 
However, the Oort cloud may also contain planetesimals that originally formed around other stars and are non-native to the Solar System. 
These objects would have been captured during the cluster phase, or some could be interstellar objects approaching with low relative velocity \citep{Levison2010,Dehnen:2022,Penarubbia2023}. 
The relative fraction of non-native members is still an open question \citep{Hands2020,Napier2021}, though none are currently known from their dynamics \citep{Morbidelli2020}.

Naturally, the cluster environment would affect potentially existing exo-Oort clouds. 
Oort clouds are not necessarily entirely destroyed by a close stellar flyby, but their planetesimal content is considerably reduced \citep{Pfalzner2024}. 
However, the orbits of the planetesimals are strongly affected \citep{Portegies2021}, and as a consequence \red{an} intense inward stream of exocomets is released. 
Thus, the cluster environment simultaneously increases the long-period exocomet rate on a short time scale, but reduces it on longer time scales, due to the reduced Oort cloud reservoir. 

The existence of exo-Oort clouds around other stars is therefore theoretically possible, but lacks observational proof at present. 
Generally, there is the question of how comets are launched from the exo-Oort cloud region to the regions close to the host star. 
\citet{Wyatt2017} find that three principles maximise the cometary influx from exo-Kuiper belts: a chain of closely separated planets interior to the belt, none of which is a Jupiter-like ejector; planet masses not increasing strongly with distance and ongoing replenishment of comets, possibly by embedded low-mass planets. 
Thus, the structure of the planetary system strongly influences the degree of long-period exocomet influx.

\section{Open questions} 



We close this chapter with a summary of questions that remain largely unanswered and thus `open' for consideration.
One of the most exciting aspects of astronomy as a modern science is the degree to which our growing understanding both encompasses satisfactory answers to older questions, and grows a host of new questions in their place.
As authors, we hope some of the readers of this chapter --- perhaps even the earliest-career among us --- will themselves bring answers in their turn.
 
\begin{itemize}
    \item What is the role of the natal prestellar environment for the subsequently emerging star-disc system?
    \item How important is it for the planetesimals if the host star is binary (with the disc in either a circumstellar or circumbinary configuration)?
    \item Which steps are needed to make a self-consistent model all the way from disc composition to evolved planetesimal in a reservoir?
    \item How well do we understand the potential differences of planetesimal formation around the Sun vs. other stars?
    \item Which small-body reservoir of a system is its dominant source of exocomets? Can a distinctive signature be found?
    \item Are planets required to produce high-eccentricity comets?
    \item How large is the fraction of non-local objects (captured from other stars) in an Oort cloud? Are they compositionally distinguishable?
    \item How massive are debris discs, which are the likely reservoirs for exocomets?
    \item What sets where debris discs form?
    \item Is the diversity seen in debris-disc shapes due to different formation conditions, or the action of planets? How do these link to exocomet production?
    \item How far can the analogy of debris discs being classical Kuiper belt analogues be pushed?
    \item What will the Vera C.\ Rubin Observatory's LSST reveal about the structure of the Oort Cloud?
    \item Are planetesimals primarily implanted into Oort-cloud analogues during the protoplanetary disc stage or afterwards?
\end{itemize}

\section{Conclusions}

Reservoirs of small bodies suitable for sourcing exocomets are a probable outcome of the planetesimal-formation process at any young star. 
While exocomets are often detected at present around A-type stars, planet formation happens at stars of a wide variety of masses. 
The dust grains and dust-to-gas ratio throughout a disc can vary substantially. 
While the planetesimal formation pathway has a range of remaining uncertainties, turbulence and oscillation within the dust-rich disc gas permit gravitational collapse and allow the growth of planetesimals. 
These form in huge numbers in a typical disc. 

The formation history of a given system will be influenced by its stellar environment. 
To first order, the thermal conditions and resulting snowlines within a disc will strongly influence the composition of the planetesimals that form at that location. 
The mass of the star affects the distance of these snowlines, but complex additional processes can affect the available gas and ice compositions as the discs evolve with time. 
The presence of planets will set the dynamical architecture of the system; however, our understanding of these systems may still be affected by the current observational biases of the sample of planets around stars of different masses. 
The remnant solid material not incorporated into planets is visible as debris discs in the early stages, and over subsequent Myr, collisional grinding slowly depletes the populations of small bodies into dust.

Thus, at almost any main-sequence star, there will be small-body reservoirs. They present a complex mix of possibilities for exocomets to emerge. The local planetary architecture will permit a particular range of stable reservoirs, with transient populations feeding between them and swapping material out into active exocomets.
The compositions of the small bodies within each reservoir will vary widely depending on their source region in the disc and the mixing history of that system.

The detailed knowledge about the small-body population of the Solar System illustrates the wealth of small-body reservoirs that can be present in a single system with relatively straightforward mass partitioning between the planets.
The sheer nuance of some of these small-body populations would not be looked for in other systems if examples were absent in our system.
Equally, it is always worth considering how much our detailed familiarity with this system, ``the fields we know", may lead to expectations that influence our assumptions about other systems.
In general, for any planetary system, resonant relationships of the planets will sculpt and dominate the orbital stability of the small bodies, from their primordial formation locations to their eventual orbits.
A given system's current planetary architecture is only a snapshot of the range of states it may have existed in over time, and its small-body reservoirs reflect that.
Several of the exocomet reservoirs described here require processes that change their orbits to turn them into an exocomet; this transformation is discussed in Mustill et al. (submitted).
The home of any exocomet may therefore be a transient thing.

Solar System objects provide observational constraints to the models of planetesimal formation that can be used to understand the diversity of objects around stars beyond the sun. 
Exploring the properties of Solar System objects with ever-more-sensitive instruments allows us to address challenging questions which test our understanding of formation processes.
Even though there are many unanswered questions, the combination of the details we can observe in the Solar System with the diversity of populations around other stars will lead to an improved understanding of the processes leading to the formation of small-body reservoirs across the Galaxy.

The range of open questions for exocomet reservoirs suggests many of these details will be resolved over the next decade, as further observation by JWST and ALMA take place, and a wider range of stars are studied for exocomet observation. Furthermore, the community can look forward to new facilities as the Vera C.\ Rubin Observatory, NASA's SPHEREx (Spectro-Photometer for the History of the Universe, Epoch of Reionization and Ices Explorer) and Nancy Grace Roman Space Telescope, and the Extremely Large Telescope and Giant Magellan Telescope.

\backmatter



\bmhead{Acknowledgments}

We gratefully acknowledge support by the International Space Science Insitute, ISSI, Bern, for supporting  and hosting the workshop on ``Exocomets: Bridging our Understanding of Minor Bodies in Solar and Exoplanetary Systems'', during which this work was initiated in July 2024.
We would like to thank the reviewers for their helpful suggestions for improving the manuscript.


M.T.B. appreciates support by the Rutherford Discovery Fellowships from New Zealand Government funding, administered by the Royal Society Te Ap\={a}rangi. 
A.J.M. acknowledges support from the Swedish Research Council (Project Grant 2022-04043) and the Swedish National Space Agency (Career Grant 2023-00146). 
T.D.P. is supported by a UKRI Stephen Hawking Fellowship and a Warwick Prize Fellowship, the latter made possible by a generous philanthropic donation. 
SM was supported by a Royal Society University Research Fellowship (URF- R1-221669).
R.K. would like to acknowledge the support from “L’Oreal UNESCO For Women in Science” National program for Bulgaria.
H.K.\ acknowledges funding from the European Research Council (ERC) via the ERC Advanced Grant “TiPPi - Turbulence, Pebbles and Planetesimals: The Origin of Minor Bodies in the Solar System" (project ID 855130) and from the
Heidelberg Cluster of Excellence (EXC 2181 - 390900948) “STRUCTURES: A unifying approach to emergent phenomena in the physical world, mathematics, and complex data,” funded by the German Excellence Strategy. H.N. is supported by JSPS and MEXT Grants-in-Aid for Scientific
Research, 18H05441, 19K03910, and 20H00182.
N.O. is supported by the National Science and Technology Council (NSTC) in Taiwan (NSTC 113-2112-M-001-037).

This work benefited from discussions within ISSI International Team project 504 “The Life Cycle of Comets” funded by the International Space Science Institute (ISSI) in Bern. D.Z.S. is supported by an NSF Astronomy and Astrophysics Postdoctoral Fellowship under award AST-2303553. This research award is partially funded by a generous gift of Charles Simonyi to the NSF Division of Astronomical Sciences.  The award is made in recognition of significant contributions to Rubin Observatory’s Legacy Survey of Space and Time. 

\bmhead{Competing interests}

None declared.

\bibliography{sn-bibliography}

\begin{thebibliography}{272}
\providecommand{\natexlab}[1]{#1}
\providecommand{\url}[1]{{#1}}
\providecommand{\urlprefix}{URL }
\providecommand{\doi}[1]{\url{https://doi.org/#1}}
\providecommand{\eprint}[2][]{\url{#2}}
 \bibcommenthead

\bibitem[{{Abod} et~al(2019){Abod}, {Simon}, {Li}, {Armitage}, {Youdin}, and {Kretke}}]{Abod_2019}
{Abod} CP, {Simon} JB, {Li} R, et~al (2019) {The Mass and Size Distribution of Planetesimals Formed by the Streaming Instability. II. The Effect of the Radial Gas Pressure Gradient}. \apj 883(2):192. \doi{10.3847/1538-4357/ab40a3}, {\href{https://arxiv.org/abs/1810.10018}{{arXiv:1810.10018}}} {[astro-ph.EP]}

\bibitem[{{Adams}(2010)}]{Adams2010}
{Adams} FC (2010) {The Birth Environment of the Solar System}. \araa 48:47--85. \doi{10.1146/annurev-astro-081309-130830}, {\href{https://arxiv.org/abs/1001.5444}{{arXiv:1001.5444}}} {[astro-ph.SR]}

\bibitem[{A'Hearn et~al(1995)A'Hearn, Millis, Schleicher, Osip, and Birch}]{AHearn1995}
A'Hearn MF, Millis RL, Schleicher DG, et~al (1995) {The ensemble properties of comets: Results from narrowband photometry of 85 comets, 1976-1992.} Icarus 118:223. \doi{10.1006/icar.1995.1190}, \urlprefix\url{http://adsabs.harvard.edu/cgi-bin/nph-data\_query?bibcode=1995Icar..118..223A\&link\_type=EJOURNAL}

\bibitem[{A'Hearn et~al(2012)A'Hearn, Feaga, Keller, Kawakita, Hampton, Kissel, Klaasen, McFadden, Meech, and Schultz}]{AHearn2012}
A'Hearn MF, Feaga LM, Keller HU, et~al (2012) {Cometary Volatiles and the Origin of Comets}. Astrophysical Journal 758(1):29. \doi{10.1088/0004-637x/758/1/29}

\bibitem[{{Alexandersen} et~al(2021){Alexandersen}, {Greenstreet}, {Gladman}, {Bannister}, {Chen}, {Gwyn}, {Kavelaars}, {Petit}, {Volk}, {Lehner}, and {Wang}}]{Alexandersen:2021}
{Alexandersen} M, {Greenstreet} S, {Gladman} BJ, et~al (2021) {OSSOS. XXIII. 2013 VZ$_{70}$ and the Temporary Coorbitals of the Giant Planets}. \psj 2(5):212. \doi{10.3847/PSJ/ac1c6b}, {\href{https://arxiv.org/abs/2110.09627}{{arXiv:2110.09627}}} {[astro-ph.EP]}

\bibitem[{{Altwegg} et~al(2015){Altwegg}, {Balsiger}, {Bar-Nun}, {Berthelier}, {Bieler}, {Bochsler}, {Briois}, {Calmonte}, {Combi}, {De Keyser}, {Eberhardt}, {Fiethe}, {Fuselier}, {Gasc}, {Gombosi}, {Hansen}, {H{\"a}ssig}, {J{\"a}ckel}, {Kopp}, {Korth}, {LeRoy}, {Mall}, {Marty}, {Mousis}, {Neefs}, {Owen}, {R{\`e}me}, {Rubin}, {S{\'e}mon}, {Tzou}, {Waite}, and {Wurz}}]{Altwegg2015}
{Altwegg} K, {Balsiger} H, {Bar-Nun} A, et~al (2015) {67P/Churyumov-Gerasimenko, a Jupiter family comet with a high D/H ratio}. Science 347(6220):1261952. \doi{10.1126/science.1261952}

\bibitem[{{Andrews} et~al(2018){Andrews}, {Huang}, {P{\'e}rez}, {Isella}, {Dullemond}, {Kurtovic}, {Guzm{\'a}n}, {Carpenter}, {Wilner}, {Zhang}, {Zhu}, {Birnstiel}, {Bai}, {Benisty}, {Hughes}, {{\"O}berg}, and {Ricci}}]{Andrews2018}
{Andrews} SM, {Huang} J, {P{\'e}rez} LM, et~al (2018) {The Disk Substructures at High Angular Resolution Project (DSHARP). I. Motivation, Sample, Calibration, and Overview}. \apjl 869(2):L41. \doi{10.3847/2041-8213/aaf741}, {\href{https://arxiv.org/abs/1812.04040}{{arXiv:1812.04040}}} {[astro-ph.SR]}

\bibitem[{{Bae} et~al(2023){Bae}, {Isella}, {Zhu}, {Martin}, {Okuzumi}, and {Suriano}}]{Bae2023}
{Bae} J, {Isella} A, {Zhu} Z, et~al (2023) {Structured Distributions of Gas and Solids in Protoplanetary Disks}. In: {Inutsuka} S, {Aikawa} Y, {Muto} T, et~al (eds) Protostars and Planets VII, p 423, \doi{10.48550/arXiv.2210.13314}, \eprint{2210.13314}

\bibitem[{{Balbus} and {Hawley}(1991)}]{Balbus1991}
{Balbus} SA, {Hawley} JF (1991) {A Powerful Local Shear Instability in Weakly Magnetized Disks. I. Linear Analysis}. \apj 376:214. \doi{10.1086/170270}

\bibitem[{Barge and Sommeria(1995)}]{BargeSommeria1995}
Barge P, Sommeria J (1995) {Did planet formation begin inside persistent gaseous vortices?} A{\&}A 295(1):L1--L4. {\href{https://arxiv.org/abs/9501050}{{arXiv:9501050}}} {[astro-ph]}

\bibitem[{{Batygin} et~al(2011){Batygin}, {Brown}, and {Fraser}}]{Batygin:2011}
{Batygin} K, {Brown} ME, {Fraser} WC (2011) {Retention of a Primordial Cold Classical Kuiper Belt in an Instability-Driven Model of Solar System Formation}. \apj 738(1):13. \doi{10.1088/0004-637X/738/1/13}, {\href{https://arxiv.org/abs/1106.0937}{{arXiv:1106.0937}}} {[astro-ph.EP]}

\bibitem[{{Beaudoin} et~al(2023){Beaudoin}, {Gladman}, {Huang}, {Bannister}, {Kavelaars}, {Petit}, and {Volk}}]{Beaudoin:2023}
{Beaudoin} M, {Gladman} B, {Huang} Y, et~al (2023) {OSSOS. XXIX. The Population and Perihelion Distribution of the Detached Kuiper Belt}. \psj 4(8):145. \doi{10.3847/PSJ/ace88d}, {\href{https://arxiv.org/abs/2306.12847}{{arXiv:2306.12847}}} {[astro-ph.EP]}

\bibitem[{{Beichman} et~al(1986){Beichman}, {Myers}, {Emerson}, {Harris}, {Mathieu}, {Benson}, and {Jennings}}]{Beichman1986}
{Beichman} CA, {Myers} PC, {Emerson} JP, et~al (1986) {Candidate Solar-Type Protostars in Nearby Molecular Cloud Cores}. \apj 307:337. \doi{10.1086/164421}

\bibitem[{{Belyakov} et~al(2024){Belyakov}, {Davis}, {Milby}, {Wong}, and {Brown}}]{Belyakov2024}
{Belyakov} M, {Davis} MR, {Milby} Z, et~al (2024) {JWST Spectrophotometry of the Small Satellites of Uranus and Neptune}. \psj 5(5):119. \doi{10.3847/PSJ/ad3d55}, {\href{https://arxiv.org/abs/2404.06660}{{arXiv:2404.06660}}} {[astro-ph.EP]}

\bibitem[{{Benisty} et~al(2021){Benisty}, {Bae}, {Facchini}, {Keppler}, {Teague}, {Isella}, {Kurtovic}, {P{\'e}rez}, {Sierra}, {Andrews}, {Carpenter}, {Czekala}, {Dominik}, {Henning}, {Menard}, {Pinilla}, and {Zurlo}}]{Benisty2021}
{Benisty} M, {Bae} J, {Facchini} S, et~al (2021) {A Circumplanetary Disk around PDS70c}. \apjl 916(1):L2. \doi{10.3847/2041-8213/ac0f83}, {\href{https://arxiv.org/abs/2108.07123}{{arXiv:2108.07123}}} {[astro-ph.EP]}

\bibitem[{{Bergin} and {Tafalla}(2007)}]{Bergin2007ARA&A}
{Bergin} EA, {Tafalla} M (2007) {Cold Dark Clouds: The Initial Conditions for Star Formation}. \araa 45(1):339--396. \doi{10.1146/annurev.astro.45.071206.100404}, {\href{https://arxiv.org/abs/0705.3765}{{arXiv:0705.3765}}} {[astro-ph]}

\bibitem[{{Bergner} et~al(2020){Bergner}, {{\"O}berg}, {Bergin}, {Andrews}, {Blake}, {Carpenter}, {Cleeves}, {Guzm{\'a}n}, {Huang}, {J{\o}rgensen}, {Qi}, {Schwarz}, {Williams}, and {Wilner}}]{Bergner2020}
{Bergner} JB, {{\"O}berg} KI, {Bergin} EA, et~al (2020) {An Evolutionary Study of Volatile Chemistry in Protoplanetary Disks}. \apj 898(2):97. \doi{10.3847/1538-4357/ab9e71}, {\href{https://arxiv.org/abs/2006.12584}{{arXiv:2006.12584}}} {[astro-ph.SR]}

\bibitem[{{Birnstiel}(2024)}]{Birnstiel2024}
{Birnstiel} T (2024) {Dust Growth and Evolution in Protoplanetary Disks}. \araa 62(1):157--202. \doi{10.1146/annurev-astro-071221-052705}, {\href{https://arxiv.org/abs/2312.13287}{{arXiv:2312.13287}}} {[astro-ph.EP]}

\bibitem[{{Blum}(2018)}]{Blum2018}
{Blum} J (2018) {Dust Evolution in Protoplanetary Discs and the Formation of Planetesimals. What Have We Learned from Laboratory Experiments?} \ssr 214(2):52. \doi{10.1007/s11214-018-0486-5}, {\href{https://arxiv.org/abs/1802.00221}{{arXiv:1802.00221}}} {[astro-ph.EP]}

\bibitem[{{Blum} and {M{\"u}nch}(1993)}]{Blum:1993}
{Blum} J, {M{\"u}nch} M (1993) {Experimental Investigations on Aggregate-Aggregate Collisions in the Early Solar Nebula}. \icarus 106(1):151--167. \doi{10.1006/icar.1993.1163}

\bibitem[{Bodewits et~al(2014)Bodewits, Farnham, A'Hearn, Feaga, McKay, Schleicher, and Sunshine}]{Bodewits2014}
Bodewits D, Farnham TL, A'Hearn MF, et~al (2014) {The Evolving Activity of the Dynamically Young Comet C/2009 P1 (Garradd)}. Astrophysical Journal 786(1):48. \doi{10.1088/0004-637x/786/1/48}

\bibitem[{{Boe} et~al(2019){Boe}, {Jedicke}, {Meech}, {Wiegert}, {Weryk}, {Chambers}, {Denneau}, {Kaiser}, {Kudritzki}, {Magnier}, {Wainscoat}, and {Waters}}]{Boe:2019}
{Boe} B, {Jedicke} R, {Meech} KJ, et~al (2019) {The orbit and size-frequency distribution of long period comets observed by Pan-STARRS1}. \icarus 333:252--272. \doi{10.1016/j.icarus.2019.05.034}, {\href{https://arxiv.org/abs/1905.13458}{{arXiv:1905.13458}}} {[astro-ph.EP]}

\bibitem[{{Bolin} et~al(2023){Bolin}, {Ahumada}, {Dokkum}, {Fremling}, {Hardegree-Ullman}, {Purdum}, {Serabyn}, and {Southworth}}]{Bolin:2023}
{Bolin} BT, {Ahumada} T, {Dokkum} Pv, et~al (2023) {Preliminary estimates of the Zwicky Transient Facility 'Ayl{\'o}'chaxnim asteroid population completeness}. \icarus 394:115442. \doi{10.1016/j.icarus.2023.115442}, {\href{https://arxiv.org/abs/2009.04125}{{arXiv:2009.04125}}} {[astro-ph.EP]}

\bibitem[{{Booth} et~al(2023){Booth}, {Pearce}, {Krivov}, {Wyatt}, {Dent}, {Hales}, {Lestrade}, {Cruz-S{\'a}enz de Miera}, {Faramaz}, {L{\"o}hne}, and {Chavez-Dagostino}}]{Booth2023}
{Booth} M, {Pearce} TD, {Krivov} AV, et~al (2023) {The clumpy structure of $\epsilon$ Eridani's debris disc revisited by ALMA}. \mnras 521(4):6180--6194. \doi{10.1093/mnras/stad938}, {\href{https://arxiv.org/abs/2303.13584}{{arXiv:2303.13584}}} {[astro-ph.EP]}

\bibitem[{{Bottke} et~al(2002){Bottke}, {Morbidelli}, {Jedicke}, {Petit}, {Levison}, {Michel}, and {Metcalfe}}]{Bottke:2002}
{Bottke} WF, {Morbidelli} A, {Jedicke} R, et~al (2002) {Debiased Orbital and Absolute Magnitude Distribution of the Near-Earth Objects}. \icarus 156(2):399--433. \doi{10.1006/icar.2001.6788}

\bibitem[{{Brown}(2012)}]{Brown:2012}
{Brown} ME (2012) {The Compositions of Kuiper Belt Objects}. Annual Review of Earth and Planetary Sciences 40(1):467--494. \doi{10.1146/annurev-earth-042711-105352}, {\href{https://arxiv.org/abs/1112.2764}{{arXiv:1112.2764}}} {[astro-ph.EP]}

\bibitem[{{Brown}(2013)}]{Brown2013}
{Brown} ME (2013) {The Density of Mid-sized Kuiper Belt Object 2002 UX25 and the Formation of the Dwarf Planets}. \apjl 778(2):L34. \doi{10.1088/2041-8205/778/2/L34}, {\href{https://arxiv.org/abs/1311.0553}{{arXiv:1311.0553}}} {[astro-ph.EP]}

\bibitem[{{Brown} et~al(2004){Brown}, {Trujillo}, and {Rabinowitz}}]{Brown2004}
{Brown} ME, {Trujillo} C, {Rabinowitz} D (2004) {Discovery of a Candidate Inner Oort Cloud Planetoid}. \apj 617(1):645--649. \doi{10.1086/422095}, {\href{https://arxiv.org/abs/astro-ph/0404456}{{arXiv:astro-ph/0404456}}} {[astro-ph]}

\bibitem[{{Brownlee} et~al(2006){Brownlee}, {Tsou}, {Al{\'e}on}, {Alexander}, {Araki}, {Bajt}, {Baratta}, {Bastien}, {Bland}, {Bleuet}, {Borg}, {Bradley}, {Brearley}, {Brenker}, {Brennan}, {Bridges}, {Browning}, {Brucato}, {Bullock}, {Burchell}, {Busemann}, {Butterworth}, {Chaussidon}, {Cheuvront}, {Chi}, {Cintala}, {Clark}, {Clemett}, {Cody}, {Colangeli}, {Cooper}, {Cordier}, {Daghlian}, {Dai}, {D'Hendecourt}, {Djouadi}, {Dominguez}, {Duxbury}, {Dworkin}, {Ebel}, {Economou}, {Fakra}, {Fairey}, {Fallon}, {Ferrini}, {Ferroir}, {Fleckenstein}, {Floss}, {Flynn}, {Franchi}, {Fries}, {Gainsforth}, {Gallien}, {Genge}, {Gilles}, {Gillet}, {Gilmour}, {Glavin}, {Gounelle}, {Grady}, {Graham}, {Grant}, {Green}, {Grossemy}, {Grossman}, {Grossman}, {Guan}, {Hagiya}, {Harvey}, {Heck}, {Herzog}, {Hoppe}, {H{\"o}rz}, {Huth}, {Hutcheon}, {Ignatyev}, {Ishii}, {Ito}, {Jacob}, {Jacobsen}, {Jacobsen}, {Jones}, {Joswiak}, {Jurewicz}, {Kearsley}, {Keller}, {Khodja}, {Kilcoyne}, {Kissel}, {Krot}, {Langenhorst}, {Lanzirotti}, {Le},
  {Leshin}, {Leitner}, {Lemelle}, {Leroux}, {Liu}, {Luening}, {Lyon}, {MacPherson}, {Marcus}, {Marhas}, {Marty}, {Matrajt}, {McKeegan}, {Meibom}, {Mennella}, {Messenger}, {Messenger}, {Mikouchi}, {Mostefaoui}, {Nakamura}, {Nakano}, {Newville}, {Nittler}, {Ohnishi}, {Ohsumi}, {Okudaira}, {Papanastassiou}, {Palma}, {Palumbo}, {Pepin}, {Perkins}, {Perronnet}, {Pianetta}, {Rao}, {Rietmeijer}, {Robert}, {Rost}, {Rotundi}, {Ryan}, {Sandford}, {Schwandt}, {See}, {Schlutter}, {Sheffield-Parker}, {Simionovici}, {Simon}, {Sitnitsky}, {Snead}, {Spencer}, {Stadermann}, {Steele}, {Stephan}, {Stroud}, {Susini}, {Sutton}, {Suzuki}, {Taheri}, {Taylor}, {Teslich}, {Tomeoka}, {Tomioka}, {Toppani}, {Trigo-Rodr{\'\i}guez}, {Troadec}, {Tsuchiyama}, {Tuzzolino}, {Tyliszczak}, {Uesugi}, {Velbel}, {Vellenga}, {Vicenzi}, {Vincze}, {Warren}, {Weber}, {Weisberg}, {Westphal}, {Wirick}, {Wooden}, {Wopenka}, {Wozniakiewicz}, {Wright}, {Yabuta}, {Yano}, {Young}, {Zare}, {Zega}, {Ziegler}, {Zimmerman}, {Zinner}, and
  {Zolensky}}]{Brownlee2006}
{Brownlee} D, {Tsou} P, {Al{\'e}on} J, et~al (2006) {Comet 81P/Wild 2 Under a Microscope}. Science 314(5806):1711. \doi{10.1126/science.1135840}

\bibitem[{{Burn} et~al(2021){Burn}, {Schlecker}, {Mordasini}, {Emsenhuber}, {Alibert}, {Henning}, {Klahr}, and {Benz}}]{Burn2021}
{Burn} R, {Schlecker} M, {Mordasini} C, et~al (2021) {The New Generation Planetary Population Synthesis (NGPPS). IV. Planetary systems around low-mass stars}. \aap 656:A72. \doi{10.1051/0004-6361/202140390}, {\href{https://arxiv.org/abs/2105.04596}{{arXiv:2105.04596}}} {[astro-ph.EP]}

\bibitem[{{Carrera} et~al(2017){Carrera}, {Gorti}, {Johansen}, and {Davies}}]{Carrera2017}
{Carrera} D, {Gorti} U, {Johansen} A, et~al (2017) {Planetesimal Formation by the Streaming Instability in a Photoevaporating Disk}. \apj 839(1):16. \doi{10.3847/1538-4357/aa6932}, {\href{https://arxiv.org/abs/1703.07895}{{arXiv:1703.07895}}} {[astro-ph.EP]}

\bibitem[{{Carruba} et~al(2025){Carruba}, {Di Ruzza}, {Carit{\'a}}, {Aljbaae}, {Domingos}, {Huaman}, {Araujo}, {Mour{\~a}o}, {Alves}, {Delfino}, and {Silva}}]{Carruba:2025}
{Carruba} V, {Di Ruzza} S, {Carit{\'a}} G, et~al (2025) {Time scales for Co-orbital Cycles of Venus Trojans Asteroids}. \icarus 433:116508. \doi{10.1016/j.icarus.2025.116508}

\bibitem[{{Chandrasekhar}(1961)}]{Chandrasekhar1961}
{Chandrasekhar} S (1961) {Hydrodynamic and hydromagnetic stability}

\bibitem[{{Cieza} et~al(2019){Cieza}, {Ru{\'\i}z-Rodr{\'\i}guez}, {Hales}, {Casassus}, {P{\'e}rez}, {Gonzalez-Ruilova}, {C{\'a}novas}, {Williams}, {Zurlo}, {Ansdell}, {Avenhaus}, {Bayo}, {Bertrang}, {Christiaens}, {Dent}, {Ferrero}, {Gamen}, {Olofsson}, {Orcajo}, {Pe{\~n}a Ram{\'\i}rez}, {Principe}, {Schreiber}, and {van der Plas}}]{Cieza2019}
{Cieza} LA, {Ru{\'\i}z-Rodr{\'\i}guez} D, {Hales} A, et~al (2019) {The Ophiuchus DIsc Survey Employing ALMA (ODISEA) - I: project description and continuum images at 28 au resolution}. \mnras 482(1):698--714. \doi{10.1093/mnras/sty2653}, {\href{https://arxiv.org/abs/1809.08844}{{arXiv:1809.08844}}} {[astro-ph.EP]}

\bibitem[{Cochran et~al(2012)Cochran, Barker, and Gray}]{Cochran2012}
Cochran AL, Barker ES, Gray CL (2012) {Thirty years of cometary spectroscopy from McDonald Observatory}. Icarus 218(1):144 -- 168. \doi{10.1016/j.icarus.2011.12.010}

\bibitem[{Combi et~al(2020)Combi, Shou, Fougere, Tenishev, and 2020}]{Combi2020}
Combi MR, Shou Y, Fougere N, et~al (2020) {The surface distributions of the production of the major volatile species, H2O, CO2, CO and O2, from the nucleus of comet 67P/Churyumov-Gerasimenko …}. Icarus 335:113421. \doi{10.1016/j.icarus.2019.113421}

\bibitem[{{Concha-Ram{\'\i}rez} et~al(2023){Concha-Ram{\'\i}rez}, {Wilhelm}, and {Portegies Zwart}}]{Concha2023}
{Concha-Ram{\'\i}rez} F, {Wilhelm} MJC, {Portegies Zwart} S (2023) {Evolution of circumstellar discs in young star-forming regions}. \mnras 520(4):6159--6172. \doi{10.1093/mnras/stac1733}, {\href{https://arxiv.org/abs/2101.07826}{{arXiv:2101.07826}}} {[astro-ph.GA]}

\bibitem[{{Crompvoets} et~al(2022){Crompvoets}, {Lawler}, {Volk}, {Chen}, {Gladman}, {Peltier}, {Alexandersen}, {Bannister}, {Gwyn}, {Kavelaars}, and {Petit}}]{Crompvoets:2022}
{Crompvoets} BL, {Lawler} SM, {Volk} K, et~al (2022) {OSSOS XXV: Large Populations and Scattering-Sticking in the Distant Trans-Neptunian Resonances}. \psj 3(5):113. \doi{10.3847/PSJ/ac67e0}, {\href{https://arxiv.org/abs/2204.09139}{{arXiv:2204.09139}}} {[astro-ph.EP]}

\bibitem[{{Cuzzi} et~al(1993){Cuzzi}, {Dobrovolskis}, and {Champney}}]{Cuzzi1993}
{Cuzzi} JN, {Dobrovolskis} AR, {Champney} JM (1993) {Particle-Gas Dynamics in the Midplane of a Protoplanetary Nebula}. \icarus 106(1):102--134. \doi{10.1006/icar.1993.1161}

\bibitem[{{De Pr{\'a}} et~al(2025){De Pr{\'a}}, {H{\'e}nault}, {Pinilla-Alonso}, {Holler}, {Brunetto}, {Stansberry}, {de Souza Feliciano}, {Carvano}, {Harvison}, {Licandro}, {M{\"u}ller}, {Peixinho}, {Lorenzi}, {Guilbert-Lepoutre}, {Bannister}, {Pendleton}, {Cruikshank}, {Schambeau}, {McClure}, and {Emery}}]{DePra2025}
{De Pr{\'a}} MN, {H{\'e}nault} E, {Pinilla-Alonso} N, et~al (2025) {Widespread CO$_{2}$ and CO ices in the trans-Neptunian population revealed by JWST/DiSCo-TNOs}. Nature Astronomy 9:252--261. \doi{10.1038/s41550-024-02276-x}

\bibitem[{{Dehnen} et~al(2022){Dehnen}, {Hands}, and {Sch{\"o}nrich}}]{Dehnen:2022}
{Dehnen} W, {Hands} TO, {Sch{\"o}nrich} R (2022) {Capture of interstellar objects - II. By the Solar system}. \mnras 512(3):4078--4085. \doi{10.1093/mnras/stab3666}, {\href{https://arxiv.org/abs/2112.07486}{{arXiv:2112.07486}}} {[astro-ph.EP]}

\bibitem[{{Deienno} et~al(2025){Deienno}, {Denneau}, {Nesvorn{\'y}}, {Vokrouhlick{\'y}}, {Bottke}, {Jedicke}, {Naidu}, {Chesley}, {Farnocchia}, and {Chodas}}]{Deienno:2025}
{Deienno} R, {Denneau} L, {Nesvorn{\'y}} D, et~al (2025) {The debiased Near-Earth object population from ATLAS telescopes}. \icarus 425:116316. \doi{10.1016/j.icarus.2024.116316}, {\href{https://arxiv.org/abs/2409.10453}{{arXiv:2409.10453}}} {[astro-ph.EP]}

\bibitem[{{Delbo'} et~al(2017){Delbo'}, {Walsh}, {Bolin}, {Avdellidou}, and {Morbidelli}}]{Delbo2017}
{Delbo'} M, {Walsh} K, {Bolin} B, et~al (2017) {Identification of a primordial asteroid family constrains the original planetesimal population}. Science 357(6355):1026--1029. \doi{10.1126/science.aam6036}

\bibitem[{{Delorme} et~al(2024){Delorme}, {Chomez}, {Squicciarini}, {Janson}, {Flasseur}, {Schib}, {Gratton}, {Lagrange}, {Langlois}, {Mayer}, {Helled}, {Reffert}, {Kiefer}, {Biller}, {Chauvin}, {Fontanive}, {Henning}, {Kenworthy}, {Marleau}, {Mesa}, {Meyer}, {Mordasini}, {Ringqvist}, {Samland}, {Vigan}, and {Viswanath}}]{Delorme2024}
{Delorme} P, {Chomez} A, {Squicciarini} V, et~al (2024) {Population of giant planets around B stars from the first part of the BEAST survey}. \aap 692:A263. \doi{10.1051/0004-6361/202451461}, {\href{https://arxiv.org/abs/2409.18793}{{arXiv:2409.18793}}} {[astro-ph.EP]}

\bibitem[{{DeMeo} and {Carry}(2014)}]{DeMeo:2014}
{DeMeo} FE, {Carry} B (2014) {Solar System evolution from compositional mapping of the asteroid belt}. \nat 505(7485):629--634. \doi{10.1038/nature12908}, {\href{https://arxiv.org/abs/1408.2787}{{arXiv:1408.2787}}} {[astro-ph.EP]}

\bibitem[{{Dipierro} et~al(2016){Dipierro}, {Laibe}, {Price}, and {Lodato}}]{Dipierro2016}
{Dipierro} G, {Laibe} G, {Price} DJ, et~al (2016) {Two mechanisms for dust gap opening in protoplanetary discs}. \mnras 459(1):L1--L5. \doi{10.1093/mnrasl/slw032}, {\href{https://arxiv.org/abs/1602.07457}{{arXiv:1602.07457}}} {[astro-ph.EP]}

\bibitem[{{Dodson-Robinson} and {Salyk}(2011)}]{Dodson-Robinson2011}
{Dodson-Robinson} SE, {Salyk} C (2011) {Transitional Disks as Signposts of Young, Multiplanet Systems}. \apj 738(2):131. \doi{10.1088/0004-637X/738/2/131}, {\href{https://arxiv.org/abs/1106.4824}{{arXiv:1106.4824}}} {[astro-ph.EP]}

\bibitem[{{Dominik} and {Dullemond}(2024)}]{Dominik2024}
{Dominik} C, {Dullemond} CP (2024) {The bouncing barrier revisited: Impact on key planet formation processes and observational signatures}. \aap 682:A144. \doi{10.1051/0004-6361/202347716}, {\href{https://arxiv.org/abs/2312.06000}{{arXiv:2312.06000}}} {[astro-ph.EP]}

\bibitem[{{Dominik} et~al(2007){Dominik}, {Blum}, {Cuzzi}, and {Wurm}}]{Dominik2007}
{Dominik} C, {Blum} J, {Cuzzi} JN, et~al (2007) {Growth of Dust as the Initial Step Toward Planet Formation}. In: {Reipurth} B, {Jewitt} D, {Keil} K (eds) Protostars and Planets V, p 783, \doi{10.48550/arXiv.astro-ph/0602617}, \eprint{astro-ph/0602617}

\bibitem[{{Dorsey} et~al(2023){Dorsey}, {Bannister}, {Lawler}, and {Parker}}]{Dorsey:2023}
{Dorsey} RC, {Bannister} MT, {Lawler} SM, et~al (2023) {OSSOS. XXVII. Population Estimates for Theoretically Stable Centaurs between Uranus and Neptune}. \psj 4(6):110. \doi{10.3847/PSJ/acd771}, {\href{https://arxiv.org/abs/2305.11412}{{arXiv:2305.11412}}} {[astro-ph.EP]}

\bibitem[{{Dressing} and {Charbonneau}(2013)}]{Dressing2013}
{Dressing} CD, {Charbonneau} D (2013) {The Occurrence Rate of Small Planets around Small Stars}. \apj 767(1):95. \doi{10.1088/0004-637X/767/1/95}, {\href{https://arxiv.org/abs/1302.1647}{{arXiv:1302.1647}}} {[astro-ph.EP]}

\bibitem[{{Duch{\^e}ne} and {Kraus}(2013)}]{Duchene2013}
{Duch{\^e}ne} G, {Kraus} A (2013) {Stellar Multiplicity}. \araa 51(1):269--310. \doi{10.1146/annurev-astro-081710-102602}, {\href{https://arxiv.org/abs/1303.3028}{{arXiv:1303.3028}}} {[astro-ph.SR]}

\bibitem[{Eistrup et~al(2019)Eistrup, Walsh, and Dishoeck}]{Eistrup2019}
Eistrup C, Walsh C, Dishoeck EFv (2019) {Cometary compositions compared with protoplanetary disk midplane chemical evolution - An emerging chemical evolution taxonomy for comets}. Astronomy and Astrophysics Supplement Series 629:A84. \doi{10.1051/0004-6361/201935812}

\bibitem[{{Eriksson} et~al(2021){Eriksson}, {Ronnet}, and {Johansen}}]{Eriksson2021}
{Eriksson} LEJ, {Ronnet} T, {Johansen} A (2021) {The fate of planetesimals formed at planetary gap edges}. \aap 648:A112. \doi{10.1051/0004-6361/202039889}, {\href{https://arxiv.org/abs/2011.05769}{{arXiv:2011.05769}}} {[astro-ph.EP]}

\bibitem[{{Ertel} et~al(2014){Ertel}, {Absil}, {Defr{\`e}re}, {Le Bouquin}, {Augereau}, {Marion}, {Blind}, {Bonsor}, {Bryden}, {Lebreton}, and {Milli}}]{Ertel2014}
{Ertel} S, {Absil} O, {Defr{\`e}re} D, et~al (2014) {A near-infrared interferometric survey of debris-disk stars. IV. An unbiased sample of 92 southern stars observed in H band with VLTI/PIONIER}. \aap 570:A128. \doi{10.1051/0004-6361/201424438}, {\href{https://arxiv.org/abs/1409.6143}{{arXiv:1409.6143}}} {[astro-ph.EP]}

\bibitem[{{Ertel} et~al(2020){Ertel}, {Defr{\`e}re}, {Hinz}, {Mennesson}, {Kennedy}, {Danchi}, {Gelino}, {Hill}, {Hoffmann}, {Mazoyer}, {Rieke}, {Shannon}, {Stapelfeldt}, {Spalding}, {Stone}, {Vaz}, {Weinberger}, {Willems}, {Absil}, {Arbo}, {Bailey}, {Beichman}, {Bryden}, {Downey}, {Durney}, {Esposito}, {Gaspar}, {Grenz}, {Haniff}, {Leisenring}, {Marion}, {McMahon}, {Millan-Gabet}, {Montoya}, {Morzinski}, {Perera}, {Pinna}, {Pott}, {Power}, {Puglisi}, {Roberge}, {Serabyn}, {Skemer}, {Su}, {Vaitheeswaran}, and {Wyatt}}]{Ertel2020}
{Ertel} S, {Defr{\`e}re} D, {Hinz} P, et~al (2020) {The HOSTS Survey for Exozodiacal Dust: Observational Results from the Complete Survey}. \aj 159(4):177. \doi{10.3847/1538-3881/ab7817}, {\href{https://arxiv.org/abs/2003.03499}{{arXiv:2003.03499}}} {[astro-ph.SR]}

\bibitem[{{Faramaz} et~al(2019){Faramaz}, {Krist}, {Stapelfeldt}, {Bryden}, {Mamajek}, {Matr{\`a}}, {Booth}, {Flaherty}, {Hales}, {Hughes}, {Bayo}, {Casassus}, {Cuadra}, {Olofsson}, {Su}, and {Wilner}}]{Faramaz2019}
{Faramaz} V, {Krist} J, {Stapelfeldt} KR, et~al (2019) {From Scattered-light to Millimeter Emission: A Comprehensive View of the Gigayear-old System of HD 202628 and its Eccentric Debris Ring}. \aj 158(4):162. \doi{10.3847/1538-3881/ab3ec1}, {\href{https://arxiv.org/abs/1909.04162}{{arXiv:1909.04162}}} {[astro-ph.EP]}

\bibitem[{{Fedorets} et~al(2017){Fedorets}, {Granvik}, and {Jedicke}}]{Fedorets:2017}
{Fedorets} G, {Granvik} M, {Jedicke} R (2017) {Orbit and size distributions for asteroids temporarily captured by the Earth-Moon system}. \icarus 285:83--94. \doi{10.1016/j.icarus.2016.12.022}

\bibitem[{Feldman et~al(2004)Feldman, Cochran, and Combi}]{Feldman2005}
Feldman pD, Cochran AL, Combi MR (2004) {Spectroscopic Investigations of Fragment Species in the Coma}. Comets II, University of Arizona Press, p 425

\bibitem[{Fink(2009)}]{Fink2009}
Fink U (2009) {A taxonomic survey of comet composition 1985-2004 using CCD spectroscopy}. Icarus 201(1):311 -- 334. \doi{10.1016/j.icarus.2008.12.044}

\bibitem[{{Fraser} et~al(2022){Fraser}, {Dones}, {Volk}, {Womack}, and {Nesvorn{\'y}}}]{Fraser2022}
{Fraser} WC, {Dones} L, {Volk} K, et~al (2022) {The Transition from the Kuiper Belt to the Jupiter-Family (Comets)}. arXiv e-prints arXiv:2210.16354. \doi{10.48550/arXiv.2210.16354}, {\href{https://arxiv.org/abs/2210.16354}{{arXiv:2210.16354}}} {[astro-ph.EP]}

\bibitem[{{Fraser} et~al(2023){Fraser}, {Pike}, {Marsset}, {Schwamb}, {Bannister}, {Buchanan}, {Kavelaars}, {Benecchi}, {Tan}, {Peixinho}, {Gwyn}, {Alexandersen}, {Chen}, {Gladman}, and {Volk}}]{Fraser2023}
{Fraser} WC, {Pike} RE, {Marsset} M, et~al (2023) {Col-OSSOS: The Two Types of Kuiper Belt Surfaces}. \psj 4(5):80. \doi{10.3847/PSJ/acc844}, {\href{https://arxiv.org/abs/2206.04068}{{arXiv:2206.04068}}} {[astro-ph.EP]}

\bibitem[{{Friebe} et~al(2022){Friebe}, {Pearce}, and {L{\"o}hne}}]{Friebe2022}
{Friebe} MF, {Pearce} TD, {L{\"o}hne} T (2022) {Gap carving by a migrating planet embedded in a massive debris disc}. \mnras 512(3):4441--4454. \doi{10.1093/mnras/stac664}, {\href{https://arxiv.org/abs/2203.03611}{{arXiv:2203.03611}}} {[astro-ph.EP]}

\bibitem[{{Furuya} et~al(2022){Furuya}, {Lee}, and {Nomura}}]{Furuya2022}
{Furuya} K, {Lee} S, {Nomura} H (2022) {Different Degrees of Nitrogen and Carbon Depletion in the Warm Molecular Layers of Protoplanetary Disks}. \apj 938(1):29. \doi{10.3847/1538-4357/ac9233}, {\href{https://arxiv.org/abs/2209.07197}{{arXiv:2209.07197}}} {[astro-ph.EP]}

\bibitem[{{Garaud} et~al(2013){Garaud}, {Meru}, {Galvagni}, and {Olczak}}]{Garaud2013}
{Garaud} P, {Meru} F, {Galvagni} M, et~al (2013) {From Dust to Planetesimals: An Improved Model for Collisional Growth in Protoplanetary Disks}. \apj 764(2):146. \doi{10.1088/0004-637X/764/2/146}, {\href{https://arxiv.org/abs/1209.0013}{{arXiv:1209.0013}}} {[astro-ph.EP]}

\bibitem[{{Gerbig} et~al(2020){Gerbig}, {Murray-Clay}, {Klahr}, and {Baehr}}]{Gerbig2020}
{Gerbig} K, {Murray-Clay} RA, {Klahr} H, et~al (2020) {Requirements for Gravitational Collapse in Planetesimal Formation{\textemdash}The Impact of Scales Set by Kelvin-Helmholtz and Nonlinear Streaming Instability}. \apj 895(2):91. \doi{10.3847/1538-4357/ab8d37}, {\href{https://arxiv.org/abs/2001.10552}{{arXiv:2001.10552}}} {[astro-ph.EP]}

\bibitem[{{Gkotsinas} et~al(2024){Gkotsinas}, {Nesvorn{\'y}}, {Guilbert-Lepoutre}, {Raymond}, and {Kaib}}]{Gkotsinas:2024}
{Gkotsinas} A, {Nesvorn{\'y}} D, {Guilbert-Lepoutre} A, et~al (2024) {On the Early Thermal Processing of Planetesimals during and after the Giant Planet Instability}. \psj 5(11):243. \doi{10.3847/PSJ/ad7f4e}, {\href{https://arxiv.org/abs/2410.01923}{{arXiv:2410.01923}}} {[astro-ph.EP]}

\bibitem[{{Gladman} and {Volk}(2021)}]{Gladman2021}
{Gladman} B, {Volk} K (2021) {Transneptunian Space}. \araa 59:203--246. \doi{10.1146/annurev-astro-120920-010005}

\bibitem[{{Gladman} et~al(2008){Gladman}, {Marsden}, and {Vanlaerhoven}}]{Gladman2008}
{Gladman} B, {Marsden} BG, {Vanlaerhoven} C (2008) {Nomenclature in the Outer Solar System}. In: {Barucci} MA, {Boehnhardt} H, {Cruikshank} DP, et~al (eds) The Solar System Beyond Neptune. p 43--57

\bibitem[{Goldreich and Ward(1973)}]{Goldreich1973}
Goldreich P, Ward WR (1973) {The Formation of Planetesimals}. The Astrophysical Journal 183:1051--1061

\bibitem[{{Golimowski} et~al(2006){Golimowski}, {Ardila}, {Krist}, {Clampin}, {Ford}, {Illingworth}, {Bartko}, {Ben{\'\i}tez}, {Blakeslee}, {Bouwens}, {Bradley}, {Broadhurst}, {Brown}, {Burrows}, {Cheng}, {Cross}, {Demarco}, {Feldman}, {Franx}, {Goto}, {Gronwall}, {Hartig}, {Holden}, {Homeier}, {Infante}, {Jee}, {Kimble}, {Lesser}, {Martel}, {Mei}, {Menanteau}, {Meurer}, {Miley}, {Motta}, {Postman}, {Rosati}, {Sirianni}, {Sparks}, {Tran}, {Tsvetanov}, {White}, {Zheng}, and {Zirm}}]{Golimowski2006}
{Golimowski} DA, {Ardila} DR, {Krist} JE, et~al (2006) {Hubble Space Telescope ACS Multiband Coronagraphic Imaging of the Debris Disk around {\ensuremath{\beta}} Pictoris}. \aj 131(6):3109--3130. \doi{10.1086/503801}, {\href{https://arxiv.org/abs/astro-ph/0602292}{{arXiv:astro-ph/0602292}}} {[astro-ph]}

\bibitem[{{Goodman} et~al(1993){Goodman}, {Benson}, {Fuller}, and {Myers}}]{Goodman1993}
{Goodman} AA, {Benson} PJ, {Fuller} GA, et~al (1993) {Dense Cores in Dark Clouds. VIII. Velocity Gradients}. \apj 406:528. \doi{10.1086/172465}

\bibitem[{{Granvik} et~al(2012){Granvik}, {Vaubaillon}, and {Jedicke}}]{Granvik:2012}
{Granvik} M, {Vaubaillon} J, {Jedicke} R (2012) {The population of natural Earth satellites}. \icarus 218(1):262--277. \doi{10.1016/j.icarus.2011.12.003}, {\href{https://arxiv.org/abs/1112.3781}{{arXiv:1112.3781}}} {[astro-ph.EP]}

\bibitem[{{Greenstreet} et~al(2024){Greenstreet}, {Gladman}, and {Juri{\'c}}}]{Greenstreet:2024}
{Greenstreet} S, {Gladman} B, {Juri{\'c}} M (2024) {Jupiter's Metastable Companions}. \apjl 963(2):L40. \doi{10.3847/2041-8213/ad28c5}, {\href{https://arxiv.org/abs/2309.06609}{{arXiv:2309.06609}}} {[astro-ph.EP]}

\bibitem[{{Gundlach} and {Blum}(2015)}]{Gundlach2015}
{Gundlach} B, {Blum} J (2015) {The Stickiness of Micrometer-sized Water-ice Particles}. \apj 798(1):34. \doi{10.1088/0004-637X/798/1/34}, {\href{https://arxiv.org/abs/1410.7199}{{arXiv:1410.7199}}} {[astro-ph.EP]}

\bibitem[{{Haisch} et~al(2001){Haisch}, {Lada}, and {Lada}}]{Haisch2001}
{Haisch} JKarl~E., {Lada} EA, {Lada} CJ (2001) {Disk Frequencies and Lifetimes in Young Clusters}. \apjl 553(2):L153--L156. \doi{10.1086/320685}, {\href{https://arxiv.org/abs/astro-ph/0104347}{{arXiv:astro-ph/0104347}}} {[astro-ph]}

\bibitem[{{Hands} and {Dehnen}(2020)}]{Hands2020}
{Hands} TO, {Dehnen} W (2020) {Capture of interstellar objects: a source of long-period comets}. \mnras 493(1):L59--L64. \doi{10.1093/mnrasl/slz186}, {\href{https://arxiv.org/abs/1910.06338}{{arXiv:1910.06338}}} {[astro-ph.EP]}

\bibitem[{{Hardegree-Ullman} et~al(2019){Hardegree-Ullman}, {Cushing}, {Muirhead}, and {Christiansen}}]{Hardegree-Ullman2019}
{Hardegree-Ullman} KK, {Cushing} MC, {Muirhead} PS, et~al (2019) {Kepler Planet Occurrence Rates for Mid-type M Dwarfs as a Function of Spectral Type}. \aj 158(2):75. \doi{10.3847/1538-3881/ab21d2}, {\href{https://arxiv.org/abs/1905.05900}{{arXiv:1905.05900}}} {[astro-ph.EP]}

\bibitem[{{Harrington Pinto} et~al(2022){Harrington Pinto}, {Womack}, {Fernandez}, and {Bauer}}]{Pinto2022}
{Harrington Pinto} O, {Womack} M, {Fernandez} Y, et~al (2022) {A Survey of CO, CO$_{2}$, and H$_{2}$O in Comets and Centaurs}. \psj 3(11):247. \doi{10.3847/PSJ/ac960d}, {\href{https://arxiv.org/abs/2209.09985}{{arXiv:2209.09985}}} {[astro-ph.EP]}

\bibitem[{{Harrington Pinto} et~al(2023){Harrington Pinto}, {Kelley}, {Villanueva}, {Womack}, {Faggi}, {McKay}, {DiSanti}, {Schambeau}, {Fernandez}, {Bauer}, {Feaga}, and {Wierzchos}}]{Harrington2023}
{Harrington Pinto} O, {Kelley} MSP, {Villanueva} GL, et~al (2023) {First Detection of CO$_{2}$ Emission in a Centaur: JWST NIRSpec Observations of 39P/Oterma}. \psj 4(11):208. \doi{10.3847/PSJ/acf928}, {\href{https://arxiv.org/abs/2309.11486}{{arXiv:2309.11486}}} {[astro-ph.EP]}

\bibitem[{{Harsono} et~al(2015){Harsono}, {Bruderer}, and {van Dishoeck}}]{Harsono2015}
{Harsono} D, {Bruderer} S, {van Dishoeck} EF (2015) {Volatile snowlines in embedded disks around low-mass protostars}. \aap 582:A41. \doi{10.1051/0004-6361/201525966}, {\href{https://arxiv.org/abs/1507.07480}{{arXiv:1507.07480}}} {[astro-ph.SR]}

\bibitem[{{Hartlep} and {Cuzzi}(2020)}]{Hartlep2020}
{Hartlep} T, {Cuzzi} JN (2020) {Cascade Model for Planetesimal Formation by Turbulent Clustering}. \apj 892(2):120. \doi{10.3847/1538-4357/ab76c3}, {\href{https://arxiv.org/abs/2002.06321}{{arXiv:2002.06321}}} {[astro-ph.EP]}

\bibitem[{{He} et~al(2021){He}, {Ford}, and {Ragozzine}}]{He2021}
{He} MY, {Ford} EB, {Ragozzine} D (2021) {Architectures of Exoplanetary Systems. II. An Increase in Inner Planetary System Occurrence toward Later Spectral Types for Kepler's FGK Dwarfs}. \aj 161(1):16. \doi{10.3847/1538-3881/abc68b}, {\href{https://arxiv.org/abs/2003.04348}{{arXiv:2003.04348}}} {[astro-ph.EP]}

\bibitem[{{Hollands} et~al(2021){Hollands}, {Tremblay}, {G{\"a}nsicke}, {Koester}, and {Gentile-Fusillo}}]{Hollands2021}
{Hollands} MA, {Tremblay} PE, {G{\"a}nsicke} BT, et~al (2021) {Alkali metals in white dwarf atmospheres as tracers of ancient planetary crusts}. Nature Astronomy 5:451--459. \doi{10.1038/s41550-020-01296-7}, {\href{https://arxiv.org/abs/2101.01225}{{arXiv:2101.01225}}} {[astro-ph.EP]}

\bibitem[{{Holman}(1997)}]{Holman:1997}
{Holman} MJ (1997) {A possible long-lived belt of objects between Uranus and Neptune}. \nat 387(6635):785--788. \doi{10.1038/42890}

\bibitem[{{Hsieh} and {Jewitt}(2006)}]{Hsieh2006}
{Hsieh} HH, {Jewitt} D (2006) {A Population of Comets in the Main Asteroid Belt}. Science 312(5773):561--563. \doi{10.1126/science.1125150}

\bibitem[{{Hsieh} et~al(2025){Hsieh}, {Noonan}, {Kelley}, {Bodewits}, {Pittichov{\'a}}, {Thirouin}, {Micheli}, {Knight}, {Bannister}, {Chandler}, {Holt}, {Hopkins}, {Kim}, {Moskovitz}, {Oldroyd}, {Patterson}, {Sheppard}, {Tan}, {Trujillo}, and {Ye}}]{Hsieh2025}
{Hsieh} HH, {Noonan} JW, {Kelley} MSP, et~al (2025) {The Volatile Composition and Activity Evolution of Main-belt Comet 358P/PANSTARRS}. \psj 6(1):3. \doi{10.3847/PSJ/ad9199}, {\href{https://arxiv.org/abs/2411.07435}{{arXiv:2411.07435}}} {[astro-ph.EP]}

\bibitem[{{Hughes} et~al(2018){Hughes}, {Duch{\^e}ne}, and {Matthews}}]{Hughes2018}
{Hughes} AM, {Duch{\^e}ne} G, {Matthews} BC (2018) {Debris Disks: Structure, Composition, and Variability}. \araa 56:541--591. \doi{10.1146/annurev-astro-081817-052035}, {\href{https://arxiv.org/abs/1802.04313}{{arXiv:1802.04313}}} {[astro-ph.EP]}

\bibitem[{{Imaz Blanco} et~al(2023){Imaz Blanco}, {Marino}, {Matr{\`a}}, {Booth}, {Carpenter}, {Faramaz}, {Henning}, {Hughes}, {Kennedy}, {P{\'e}rez}, {Ricci}, and {Wyatt}}]{ImazBlanco2023}
{Imaz Blanco} A, {Marino} S, {Matr{\`a}} L, et~al (2023) {Inner edges of planetesimal belts: collisionally eroded or truncated?} \mnras 522(4):6150--6169. \doi{10.1093/mnras/stad1221}, {\href{https://arxiv.org/abs/2304.12337}{{arXiv:2304.12337}}} {[astro-ph.EP]}

\bibitem[{{Jacquet} and {Robert}(2013)}]{Jacquet2013}
{Jacquet} E, {Robert} F (2013) {Water transport in protoplanetary disks and the hydrogen isotopic composition of chondrites}. \icarus 223(2):722--732. \doi{10.1016/j.icarus.2013.01.022}, {\href{https://arxiv.org/abs/1301.5665}{{arXiv:1301.5665}}} {[astro-ph.EP]}

\bibitem[{{Janson} et~al(2021){Janson}, {Gratton}, {Rodet}, {Vigan}, {Bonnefoy}, {Delorme}, {Mamajek}, {Reffert}, {Stock}, {Marleau}, {Langlois}, {Chauvin}, {Desidera}, {Ringqvist}, {Mayer}, {Viswanath}, {Squicciarini}, {Meyer}, {Samland}, {Petrus}, {Helled}, {Kenworthy}, {Quanz}, {Biller}, {Henning}, {Mesa}, {Engler}, and {Carson}}]{Janson2021}
{Janson} M, {Gratton} R, {Rodet} L, et~al (2021) {A wide-orbit giant planet in the high-mass b Centauri binary system}. \nat 600(7888):231--234. \doi{10.1038/s41586-021-04124-8}, {\href{https://arxiv.org/abs/2112.04833}{{arXiv:2112.04833}}} {[astro-ph.EP]}

\bibitem[{{Jedicke} et~al(2015){Jedicke}, {Granvik}, {Micheli}, {Ryan}, {Spahr}, and {Yeomans}}]{Jedicke2015}
{Jedicke} R, {Granvik} M, {Micheli} M, et~al (2015) {Surveys, Astrometric Follow-Up, and Population Statistics}. In: {Michel} P, {DeMeo} FE, {Bottke} WF (eds) Asteroids IV. p 795--813, \doi{10.2458/azu_uapress_9780816532131-ch040}

\bibitem[{{Jewitt} and {Haghighipour}(2007)}]{Jewitt2007}
{Jewitt} D, {Haghighipour} N (2007) {Irregular Satellites of the Planets: Products of Capture in the Early Solar System}. \araa 45(1):261--295. \doi{10.1146/annurev.astro.44.051905.092459}, {\href{https://arxiv.org/abs/astro-ph/0703059}{{arXiv:astro-ph/0703059}}} {[astro-ph]}

\bibitem[{{Johansen} and {Klahr}(2005)}]{Johansen2005}
{Johansen} A, {Klahr} H (2005) {Dust Diffusion in Protoplanetary Disks by Magnetorotational Turbulence}. \apj 634(2):1353--1371. \doi{10.1086/497118}, {\href{https://arxiv.org/abs/astro-ph/0501641}{{arXiv:astro-ph/0501641}}} {[astro-ph]}

\bibitem[{{Johansen} et~al(2006{\natexlab{a}}){Johansen}, {Henning}, and {Klahr}}]{Johansen2006b}
{Johansen} A, {Henning} T, {Klahr} H (2006{\natexlab{a}}) {Dust Sedimentation and Self-sustained Kelvin-Helmholtz Turbulence in Protoplanetary Disk Midplanes}. \apj 643(2):1219--1232. \doi{10.1086/502968}, {\href{https://arxiv.org/abs/astro-ph/0512272}{{arXiv:astro-ph/0512272}}} {[astro-ph]}

\bibitem[{{Johansen} et~al(2006{\natexlab{b}}){Johansen}, {Klahr}, and {Henning}}]{Johansen2006}
{Johansen} A, {Klahr} H, {Henning} T (2006{\natexlab{b}}) {Gravoturbulent Formation of Planetesimals}. \apj 636:1121--1134. \doi{10.1086/498078}

\bibitem[{{Johansen} et~al(2007){Johansen}, {Oishi}, {Mac Low}, {Klahr}, {Henning}, and {Youdin}}]{2007Natur.448.1022J}
{Johansen} A, {Oishi} JS, {Mac Low} MM, et~al (2007) {Rapid planetesimal formation in turbulent circumstellar disks}. \nat 448:1022--1025. \doi{10.1038/nature06086}, {\href{https://arxiv.org/abs/0708.3890}{{arXiv:0708.3890}}}

\bibitem[{{Johansen} et~al(2009){Johansen}, {Youdin}, and {Mac Low}}]{Johansen2009b}
{Johansen} A, {Youdin} A, {Mac Low} MM (2009) {Particle Clumping and Planetesimal Formation Depend Strongly on Metallicity}. \apjl 704(2):L75--L79. \doi{10.1088/0004-637X/704/2/L75}, {\href{https://arxiv.org/abs/0909.0259}{{arXiv:0909.0259}}} {[astro-ph.EP]}

\bibitem[{{Johansen} et~al(2015){Johansen}, {Mac Low}, {Lacerda}, and {Bizzarro}}]{Johansen2015}
{Johansen} A, {Mac Low} MM, {Lacerda} P, et~al (2015) {Growth of asteroids, planetary embryos, and Kuiper belt objects by chondrule accretion}. Science Advances 1:1500109. \doi{10.1126/sciadv.1500109}, {\href{https://arxiv.org/abs/1503.07347}{{arXiv:1503.07347}}} {[astro-ph.EP]}

\bibitem[{{Johnston} et~al(2024){Johnston}, {Pani{\'c}}, and {Liu}}]{Johnston2024}
{Johnston} HF, {Pani{\'c}} O, {Liu} B (2024) {Formation of giant planets around intermediate-mass stars}. \mnras 527(2):2303--2322. \doi{10.1093/mnras/stad3254}, {\href{https://arxiv.org/abs/2310.17767}{{arXiv:2310.17767}}} {[astro-ph.EP]}

\bibitem[{{Kaib} and {Volk}(2024)}]{Kaib:2024}
{Kaib} NA, {Volk} K (2024) {Dynamical Population of Comet Reservoirs}. In: {Meech} KJ, {Combi} MR, {Bockel{\'e}e-Morvan} D, et~al (eds) Comets III. p 97--120

\bibitem[{{Kavelaars} et~al(2020){Kavelaars}, {Lawler}, {Bannister}, and {Shankman}}]{Kavelaars2020}
{Kavelaars} JJ, {Lawler} SM, {Bannister} MT, et~al (2020) {Perspectives on the distribution of orbits of distant Trans-Neptunian objects}. In: {Prialnik} D, {Barucci} MA, {Young} L (eds) The Trans-Neptunian Solar System. p 61--77, \doi{10.1016/B978-0-12-816490-7.00003-5}

\bibitem[{{Kelley} et~al(2023){Kelley}, {Hsieh}, {Bodewits}, {Saki}, {Villanueva}, {Milam}, and {Hammel}}]{Kelley2023}
{Kelley} MSP, {Hsieh} HH, {Bodewits} D, et~al (2023) {Spectroscopic identification of water emission from a main-belt comet}. \nat 619(7971):720--723. \doi{10.1038/s41586-023-06152-y}

\bibitem[{{Kempf} et~al(1999){Kempf}, {Pfalzner}, and {Henning}}]{Kempf:1999}
{Kempf} S, {Pfalzner} S, {Henning} TK (1999) {N-Particle-Simulations of Dust Growth. I. Growth Driven by Brownian Motion}. \icarus 141(2):388--398. \doi{10.1006/icar.1999.6171}

\bibitem[{{Klahr} and {Schreiber}(2020)}]{Klahr2020}
{Klahr} H, {Schreiber} A (2020) {Turbulence Sets the Length Scale for Planetesimal Formation: Local 2D Simulations of Streaming Instability and Planetesimal Formation}. \apj 901(1):54. \doi{10.3847/1538-4357/abac58}, {\href{https://arxiv.org/abs/2007.10696}{{arXiv:2007.10696}}} {[astro-ph.EP]}

\bibitem[{{Klahr} and {Schreiber}(2021)}]{Klahr2021}
{Klahr} H, {Schreiber} A (2021) {Testing the Jeans, Toomre, and Bonnor-Ebert Concepts for Planetesimal Formation: 3D Streaming-instability Simulations of Diffusion-regulated Formation of Planetesimals}. \apj 911(1):9. \doi{10.3847/1538-4357/abca9b}, {\href{https://arxiv.org/abs/2011.07849}{{arXiv:2011.07849}}} {[astro-ph.EP]}

\bibitem[{{Klahr} and {Henning}(1997)}]{Klahr1997}
{Klahr} HH, {Henning} T (1997) {Particle-Trapping Eddies in Protoplanetary Accretion Disks}. \icarus 128(1):213--229. \doi{10.1006/icar.1997.5720}

\bibitem[{{Kral} et~al(2017){Kral}, {Matr{\`a}}, {Wyatt}, and {Kennedy}}]{Kral2017}
{Kral} Q, {Matr{\`a}} L, {Wyatt} MC, et~al (2017) {Predictions for the secondary CO, C and O gas content of debris discs from the destruction of volatile-rich planetesimals}. \mnras 469(1):521--550. \doi{10.1093/mnras/stx730}, {\href{https://arxiv.org/abs/1703.10693}{{arXiv:1703.10693}}} {[astro-ph.EP]}

\bibitem[{{Krapp} et~al(2019){Krapp}, {Ben{\'\i}tez-Llambay}, {Gressel}, and {Pessah}}]{Krapp2019}
{Krapp} L, {Ben{\'\i}tez-Llambay} P, {Gressel} O, et~al (2019) {Streaming Instability for Particle-size Distributions}. \apjl 878(2):L30. \doi{10.3847/2041-8213/ab2596}, {\href{https://arxiv.org/abs/1905.13139}{{arXiv:1905.13139}}} {[astro-ph.EP]}

\bibitem[{{Krijt} et~al(2020){Krijt}, {Bosman}, {Zhang}, {Schwarz}, {Ciesla}, and {Bergin}}]{Krijt2020}
{Krijt} S, {Bosman} AD, {Zhang} K, et~al (2020) {CO Depletion in Protoplanetary Disks: A Unified Picture Combining Physical Sequestration and Chemical Processing}. \apj 899(2):134. \doi{10.3847/1538-4357/aba75d}, {\href{https://arxiv.org/abs/2007.09517}{{arXiv:2007.09517}}} {[astro-ph.SR]}

\bibitem[{{Krijt} et~al(2023){Krijt}, {Kama}, {McClure}, {Teske}, {Bergin}, {Shorttle}, {Walsh}, and {Raymond}}]{Krijt2023}
{Krijt} S, {Kama} M, {McClure} M, et~al (2023) {Chemical Habitability: Supply and Retention of Life's Essential Elements During Planet Formation}. In: {Inutsuka} S, {Aikawa} Y, {Muto} T, et~al (eds) Protostars and Planets VII, p 1031, \doi{10.48550/arXiv.2203.10056}, \eprint{2203.10056}

\bibitem[{{Krivov}(2010)}]{Krivov2010}
{Krivov} AV (2010) {Debris disks: seeing dust, thinking of planetesimals and planets}. Research in Astronomy and Astrophysics 10(5):383--414. \doi{10.1088/1674-4527/10/5/001}, {\href{https://arxiv.org/abs/1003.5229}{{arXiv:1003.5229}}} {[astro-ph.EP]}

\bibitem[{{Krivov} and {Wyatt}(2021)}]{Krivov2021}
{Krivov} AV, {Wyatt} MC (2021) {Solution to the debris disc mass problem: planetesimals are born small?} \mnras 500(1):718--735. \doi{10.1093/mnras/staa2385}, {\href{https://arxiv.org/abs/2008.07406}{{arXiv:2008.07406}}} {[astro-ph.EP]}

\bibitem[{{Kunitomo} et~al(2021){Kunitomo}, {Ida}, {Takeuchi}, {Pani{\'c}}, {Miley}, and {Suzuki}}]{Kunitomo2021}
{Kunitomo} M, {Ida} S, {Takeuchi} T, et~al (2021) {Photoevaporative Dispersal of Protoplanetary Disks around Evolving Intermediate-mass Stars}. \apj 909(2):109. \doi{10.3847/1538-4357/abdb2a}, {\href{https://arxiv.org/abs/2103.07673}{{arXiv:2103.07673}}} {[astro-ph.EP]}

\bibitem[{{Lada} and {Lada}(2003)}]{Lada2003}
{Lada} CJ, {Lada} EA (2003) {Embedded Clusters in Molecular Clouds}. \araa 41:57--115. \doi{10.1146/annurev.astro.41.011802.094844}, {\href{https://arxiv.org/abs/astro-ph/0301540}{{arXiv:astro-ph/0301540}}} {[astro-ph]}

\bibitem[{{Lauretta} et~al(2024){Lauretta}, {Connolly}, {Aebersold}, {Alexander}, {Ballouz}, {Barnes}, {Bates}, {Bennett}, {Blanche}, {Blumenfeld}, {Clemett}, {Cody}, {DellaGiustina}, {Dworkin}, {Eckley}, {Foustoukos}, {Franchi}, {Glavin}, {Greenwood}, {Haenecour}, {Hamilton}, {Hill}, {Hiroi}, {Ishimaru}, {Jourdan}, {Kaplan}, {Keller}, {King}, {Koefoed}, {Kontogiannis}, {Le}, {Macke}, {McCoy}, {Milliken}, {Najorka}, {Nguyen}, {Pajola}, {Polit}, {Righter}, {Roper}, {Russell}, {Ryan}, {Sandford}, {Schofield}, {Schultz}, {Seifert}, {Tachibana}, {Thomas-Keprta}, {Thompson}, {Tu}, {Tusberti}, {Wang}, {Zega}, and {Wolner}}]{Lauretta2024}
{Lauretta} DS, {Connolly} HC, {Aebersold} JE, et~al (2024) {Asteroid (101955) Bennu in the laboratory: Properties of the sample collected by OSIRIS-REx}. MAPS 59(9):2453--2486. \doi{10.1111/maps.14227}, {\href{https://arxiv.org/abs/2404.12536}{{arXiv:2404.12536}}} {[astro-ph.EP]}

\bibitem[{{Lawler} et~al(2018){Lawler}, {Shankman}, {Kavelaars}, {Alexandersen}, {Bannister}, {Chen}, {Gladman}, {Fraser}, {Gwyn}, {Kaib}, {Petit}, and {Volk}}]{Lawler2018}
{Lawler} SM, {Shankman} C, {Kavelaars} JJ, et~al (2018) {OSSOS. VIII. The Transition between Two Size Distribution Slopes in the Scattering Disk}. \aj 155(5):197. \doi{10.3847/1538-3881/aab8ff}, {\href{https://arxiv.org/abs/1803.07521}{{arXiv:1803.07521}}} {[astro-ph.EP]}

\bibitem[{{Le Roy} et~al(2015){Le Roy}, {Altwegg}, {Balsiger}, {Berthelier}, {Bieler}, {Briois}, {Calmonte}, {Combi}, {De Keyser}, {Dhooghe}, {Fiethe}, {Fuselier}, {Gasc}, {Gombosi}, {H{\"a}ssig}, {J{\"a}ckel}, {Rubin}, and {Tzou}}]{LeRoy2015}
{Le Roy} L, {Altwegg} K, {Balsiger} H, et~al (2015) {Inventory of the volatiles on comet 67P/Churyumov-Gerasimenko from Rosetta/ROSINA}. \aap 583:A1. \doi{10.1051/0004-6361/201526450}

\bibitem[{{Lenz} et~al(2019{\natexlab{a}}){Lenz}, {Klahr}, and {Birnstiel}}]{2019ApJ...874...36L}
{Lenz} CT, {Klahr} H, {Birnstiel} T (2019{\natexlab{a}}) {Planetesimal Population Synthesis: Pebble Flux-regulated Planetesimal Formation}. \apj 874(1):36. \doi{10.3847/1538-4357/ab05d9}, {\href{https://arxiv.org/abs/1902.07089}{{arXiv:1902.07089}}} {[astro-ph.EP]}

\bibitem[{{Lenz} et~al(2019{\natexlab{b}}){Lenz}, {Klahr}, and {Birnstiel}}]{Lenz_2019}
{Lenz} CT, {Klahr} H, {Birnstiel} T (2019{\natexlab{b}}) {Planetesimal Population Synthesis: Pebble Flux-regulated Planetesimal Formation}. \apj 874(1):36. \doi{10.3847/1538-4357/ab05d9}, {\href{https://arxiv.org/abs/1902.07089}{{arXiv:1902.07089}}} {[astro-ph.EP]}

\bibitem[{{Lenz} et~al(2020){Lenz}, {Klahr}, {Birnstiel}, {Kretke}, and {Stammler}}]{Lenz2020}
{Lenz} CT, {Klahr} H, {Birnstiel} T, et~al (2020) {Constraining the parameter space for the solar nebula. The effect of disk properties on planetesimal formation}. \aap 640:A61. \doi{10.1051/0004-6361/202037878}, {\href{https://arxiv.org/abs/2006.08799}{{arXiv:2006.08799}}} {[astro-ph.EP]}

\bibitem[{{Lesur} et~al(2023){Lesur}, {Flock}, {Ercolano}, {Lin}, {Yang}, {Barranco}, {Benitez-Llambay}, {Goodman}, {Johansen}, {Klahr}, {Laibe}, {Lyra}, {Marcus}, {Nelson}, {Squire}, {Simon}, {Turner}, {Umurhan}, and {Youdin}}]{Lesur2023}
{Lesur} G, {Flock} M, {Ercolano} B, et~al (2023) {Hydro-, Magnetohydro-, and Dust-Gas Dynamics of Protoplanetary Disks}. In: {Inutsuka} S, {Aikawa} Y, {Muto} T, et~al (eds) Astronomical Society of the Pacific Conference Series, p 465

\bibitem[{{Levison} et~al(2010){Levison}, {Duncan}, {Brasser}, and {Kaufmann}}]{Levison2010}
{Levison} HF, {Duncan} MJ, {Brasser} R, et~al (2010) {Capture of the Sun's Oort Cloud from Stars in Its Birth Cluster}. Science 329(5988):187--190. \doi{10.1126/science.1187535}

\bibitem[{{Li} and {Youdin}(2021)}]{Li2021}
{Li} R, {Youdin} AN (2021) {Thresholds for Particle Clumping by the Streaming Instability}. \apj 919(2):107. \doi{10.3847/1538-4357/ac0e9f}, {\href{https://arxiv.org/abs/2105.06042}{{arXiv:2105.06042}}} {[astro-ph.EP]}

\bibitem[{{Li} et~al(2018){Li}, {Youdin}, and {Simon}}]{Li2018}
{Li} R, {Youdin} AN, {Simon} JB (2018) {On the Numerical Robustness of the Streaming Instability: Particle Concentration and Gas Dynamics in Protoplanetary Disks}. \apj 862(1):14. \doi{10.3847/1538-4357/aaca99}, {\href{https://arxiv.org/abs/1803.03638}{{arXiv:1803.03638}}} {[astro-ph.EP]}

\bibitem[{{Li} et~al(2019){Li}, {Youdin}, and {Simon}}]{Li2019}
{Li} R, {Youdin} AN, {Simon} JB (2019) {Demographics of Planetesimals Formed by the Streaming Instability}. \apj 885(1):69. \doi{10.3847/1538-4357/ab480d}, {\href{https://arxiv.org/abs/1906.09261}{{arXiv:1906.09261}}} {[astro-ph.EP]}

\bibitem[{{Licandro} et~al(2025){Licandro}, {Pinilla-Alonso}, {Holler}, {De Pr{\'a}}, {Melita}, {de Souza Feliciano}, {Brunetto}, {Guilbert-Lepoutre}, {H{\'e}nault}, {Lorenzi}, {Stansberry}, {Schambeau}, {Harvison}, {Pendleton}, {Cruikshank}, {M{\"u}ller}, {McClure}, {Emery}, {Peixinho}, {Bannister}, and {Wong}}]{Licandro2025}
{Licandro} J, {Pinilla-Alonso} N, {Holler} BJ, et~al (2025) {Thermal evolution of trans-Neptunian objects through observations of Centaurs with JWST}. Nature Astronomy 9:245--251. \doi{10.1038/s41550-024-02417-2}

\bibitem[{{Lim} et~al(2024){Lim}, {Simon}, {Li}, {Armitage}, {Carrera}, {Lyra}, {Rea}, {Yang}, and {Youdin}}]{Lim2024}
{Lim} J, {Simon} JB, {Li} R, et~al (2024) {Streaming Instability and Turbulence: Conditions for Planetesimal Formation}. \apj 969(2):130. \doi{10.3847/1538-4357/ad47a2}, {\href{https://arxiv.org/abs/2312.12508}{{arXiv:2312.12508}}} {[astro-ph.EP]}

\bibitem[{{Lin} et~al(2021){Lin}, {Chen}, {Volk}, {Gladman}, {Murray-Clay}, {Alexandersen}, {Bannister}, {Lawler}, {Ip}, {Lykawka}, {Kavelaars}, {Gwyn}, and {Petit}}]{Lin2021}
{Lin} HW, {Chen} YT, {Volk} K, et~al (2021) {OSSOS: The eccentricity and inclination distributions of the stable Neptunian Trojans}. \icarus 361:114391. \doi{10.1016/j.icarus.2021.114391}, {\href{https://arxiv.org/abs/2006.10674}{{arXiv:2006.10674}}} {[astro-ph.EP]}

\bibitem[{{Lis} et~al(2019){Lis}, {Bockel{\'e}e-Morvan}, {G{\"u}sten}, {Biver}, {Stutzki}, {Delorme}, {Dur{\'a}n}, {Wiesemeyer}, and {Okada}}]{Lis2019}
{Lis} DC, {Bockel{\'e}e-Morvan} D, {G{\"u}sten} R, et~al (2019) {Terrestrial deuterium-to-hydrogen ratio in water in hyperactive comets}. \aap 625:L5. \doi{10.1051/0004-6361/201935554}, {\href{https://arxiv.org/abs/1904.09175}{{arXiv:1904.09175}}} {[astro-ph.EP]}

\bibitem[{{Liu} et~al(2019){Liu}, {Lambrechts}, {Johansen}, and {Liu}}]{Liu2019}
{Liu} B, {Lambrechts} M, {Johansen} A, et~al (2019) {Super-Earth masses sculpted by pebble isolation around stars of different masses}. \aap 632:A7. \doi{10.1051/0004-6361/201936309}, {\href{https://arxiv.org/abs/1909.00759}{{arXiv:1909.00759}}} {[astro-ph.EP]}

\bibitem[{{Liu} et~al(2020){Liu}, {Lambrechts}, {Johansen}, {Pascucci}, and {Henning}}]{Liu2020}
{Liu} B, {Lambrechts} M, {Johansen} A, et~al (2020) {Pebble-driven planet formation around very low-mass stars and brown dwarfs}. \aap 638:A88. \doi{10.1051/0004-6361/202037720}, {\href{https://arxiv.org/abs/2004.07239}{{arXiv:2004.07239}}} {[astro-ph.EP]}

\bibitem[{{Lorek} et~al(2018){Lorek}, {Lacerda}, and {Blum}}]{Lorek:2018}
{Lorek} S, {Lacerda} P, {Blum} J (2018) {Local growth of dust- and ice-mixed aggregates as cometary building blocks in the solar nebula}. \aap 611:A18. \doi{10.1051/0004-6361/201630175}

\bibitem[{{Malhotra}(1993)}]{Malhotra1993}
{Malhotra} R (1993) {The origin of Pluto's peculiar orbit}. \nat 365(6449):819--821. \doi{10.1038/365819a0}

\bibitem[{{Manara} et~al(2018){Manara}, {Morbidelli}, and {Guillot}}]{Manara2018}
{Manara} CF, {Morbidelli} A, {Guillot} T (2018) {Why do protoplanetary disks appear not massive enough to form the known exoplanet population?} \aap 618:L3. \doi{10.1051/0004-6361/201834076}, {\href{https://arxiv.org/abs/1809.07374}{{arXiv:1809.07374}}} {[astro-ph.EP]}

\bibitem[{{Manara} et~al(2023){Manara}, {Ansdell}, {Rosotti}, {Hughes}, {Armitage}, {Lodato}, and {Williams}}]{Manara2023}
{Manara} CF, {Ansdell} M, {Rosotti} GP, et~al (2023) {Demographics of Young Stars and their Protoplanetary Disks: Lessons Learned on Disk Evolution and its Connection to Planet Formation}. In: {Inutsuka} S, {Aikawa} Y, {Muto} T, et~al (eds) Protostars and Planets VII, p 539, \doi{10.48550/arXiv.2203.09930}, \eprint{2203.09930}

\bibitem[{{Mandt} et~al(2024){Mandt}, {Lustig-Yaeger}, {Luspay-Kuti}, {Wurz}, {Bodewits}, {Fuselier}, {Mousis}, {Petrinec}, and {Trattner}}]{Mandt2024}
{Mandt} KE, {Lustig-Yaeger} J, {Luspay-Kuti} A, et~al (2024) {A nearly terrestrial D/H for comet 67P/Churyumov-Gerasimenko}. Science Advances 10(46):eadp2191. \doi{10.1126/sciadv.adp2191}

\bibitem[{{Manger} and {Klahr}(2018)}]{Manger2018}
{Manger} N, {Klahr} H (2018) {Vortex formation and survival in protoplanetary discs subject to vertical shear instability}. \mnras 480(2):2125--2136. \doi{10.1093/mnras/sty1909}, {\href{https://arxiv.org/abs/1807.06492}{{arXiv:1807.06492}}} {[astro-ph.EP]}

\bibitem[{{Marino}(2021)}]{Marino2021}
{Marino} S (2021) {Constraining planetesimal stirring: how sharp are debris disc edges?} \mnras 503(4):5100--5114. \doi{10.1093/mnras/stab771}, {\href{https://arxiv.org/abs/2104.02072}{{arXiv:2104.02072}}} {[astro-ph.EP]}

\bibitem[{{Marino}(2022)}]{Marino2022}
{Marino} S (2022) {Planetesimal/Debris discs}. arXiv e-prints arXiv:2202.03053. \doi{10.48550/arXiv.2202.03053}, {\href{https://arxiv.org/abs/2202.03053}{{arXiv:2202.03053}}} {[astro-ph.EP]}

\bibitem[{{Marino} et~al(2018){Marino}, {Carpenter}, {Wyatt}, {Booth}, {Casassus}, {Faramaz}, {Guzman}, {Hughes}, {Isella}, {Kennedy}, {Matr{\`a}}, {Ricci}, and {Corder}}]{Marino2018_HD107146}
{Marino} S, {Carpenter} J, {Wyatt} MC, et~al (2018) {A gap in the planetesimal disc around HD 107146 and asymmetric warm dust emission revealed by ALMA}. \mnras 479(4):5423--5439. \doi{10.1093/mnras/sty1790}, {\href{https://arxiv.org/abs/1805.01915}{{arXiv:1805.01915}}} {[astro-ph.EP]}

\bibitem[{{Markwardt} et~al(2023){Markwardt}, {Wen Lin}, {Gerdes}, and {Adams}}]{Markwardt2023}
{Markwardt} L, {Wen Lin} H, {Gerdes} D, et~al (2023) {Photometric Survey of Neptune's Trojan Asteroids. I. The Color Distribution}. \psj 4(8):135. \doi{10.3847/PSJ/ace528}, {\href{https://arxiv.org/abs/2307.10542}{{arXiv:2307.10542}}} {[astro-ph.EP]}

\bibitem[{{Marschall} et~al(2025){Marschall}, {Morbidelli}, and {Marrocchi}}]{Marschall2025}
{Marschall} R, {Morbidelli} A, {Marrocchi} Y (2025) {The refractory-to-ice ratio in comet 67P: implications on the composition of the comet-forming region of the protoplanetary disk}. arXiv e-prints arXiv:2501.17864. \doi{10.48550/arXiv.2501.17864}, {\href{https://arxiv.org/abs/2501.17864}{{arXiv:2501.17864}}} {[astro-ph.EP]}

\bibitem[{{Marsset} et~al(2019){Marsset}, {Fraser}, {Pike}, {Bannister}, {Schwamb}, {Volk}, {Kavelaars}, {Alexandersen}, {Chen}, {Gladman}, {Gwyn}, {Lehner}, {Peixinho}, {Petit}, and {Wang}}]{Marsset2019}
{Marsset} M, {Fraser} WC, {Pike} RE, et~al (2019) {Col-OSSOS: Color and Inclination Are Correlated throughout the Kuiper Belt}. \aj 157(3):94. \doi{10.3847/1538-3881/aaf72e}, {\href{https://arxiv.org/abs/1812.02190}{{arXiv:1812.02190}}} {[astro-ph.EP]}

\bibitem[{{Marsset} et~al(2020){Marsset}, {Fraser}, {Bannister}, {Schwamb}, {Pike}, {Benecchi}, {Kavelaars}, {Alexandersen}, {Chen}, {Gladman}, {Gwyn}, {Petit}, and {Volk}}]{Marsset2020}
{Marsset} M, {Fraser} WC, {Bannister} MT, et~al (2020) {Col-OSSOS: Compositional Homogeneity of Three Kuiper Belt Binaries}. \psj 1(1):16. \doi{10.3847/PSJ/ab8cc0}, {\href{https://arxiv.org/abs/2004.12996}{{arXiv:2004.12996}}} {[astro-ph.EP]}

\bibitem[{{Marsset} et~al(2023){Marsset}, {Fraser}, {Schwamb}, {Buchanan}, {Pike}, {Volk}, {Peixinho}, {Benecchi}, {Bannister}, {Tan}, and {Kavelaars}}]{Marsset2023}
{Marsset} M, {Fraser} WC, {Schwamb} ME, et~al (2023) {Col-OSSOS: Evidence for a Compositional Gradient Inherited from the Protoplanetary Disk?} \psj 4(9):160. \doi{10.3847/PSJ/ace7d0}, {\href{https://arxiv.org/abs/2206.04096}{{arXiv:2206.04096}}} {[astro-ph.EP]}

\bibitem[{{Matr{\`a}} et~al(2017){Matr{\`a}}, {MacGregor}, {Kalas}, {Wyatt}, {Kennedy}, {Wilner}, {Duchene}, {Hughes}, {Pan}, {Shannon}, {Clampin}, {Fitzgerald}, {Graham}, {Holland}, {Pani{\'c}}, and {Su}}]{Matra2017}
{Matr{\`a}} L, {MacGregor} MA, {Kalas} P, et~al (2017) {Detection of Exocometary CO within the 440 Myr Old Fomalhaut Belt: A Similar CO+CO$_{2}$ Ice Abundance in Exocomets and Solar System Comets}. \apj 842(1):9. \doi{10.3847/1538-4357/aa71b4}, {\href{https://arxiv.org/abs/1705.05868}{{arXiv:1705.05868}}} {[astro-ph.EP]}

\bibitem[{{Matr{\`a}} et~al(2025){Matr{\`a}}, {Marino}, {Wilner}, {Kennedy}, {Booth}, {Krivov}, {Williams}, {Hughes}, {del Burgo}, {Carpenter}, {Davies}, {Ertel}, {Kral}, {Lestrade}, {Marshall}, {Milli}, {{\"O}berg}, {Pawellek}, {Sepulveda}, {Wyatt}, {Matthews}, and {MacGregor}}]{Matra2025}
{Matr{\`a}} L, {Marino} S, {Wilner} DJ, et~al (2025) {REsolved ALMA and SMA Observations of Nearby Stars (REASONS): A population of 74 resolved planetesimal belts at millimetre wavelengths}. \aap 693:A151. \doi{10.1051/0004-6361/202451397}, {\href{https://arxiv.org/abs/2501.09058}{{arXiv:2501.09058}}} {[astro-ph.EP]}

\bibitem[{{Matthews} et~al(2014){Matthews}, {Krivov}, {Wyatt}, {Bryden}, and {Eiroa}}]{Matthews2014}
{Matthews} BC, {Krivov} AV, {Wyatt} MC, et~al (2014) {Observations, Modeling, and Theory of Debris Disks}. In: {Beuther} H, {Klessen} RS, {Dullemond} CP, et~al (eds) Protostars and Planets VI, pp 521--544, \doi{10.2458/azu_uapress_9780816531240-ch023}, \eprint{1401.0743}

\bibitem[{{Michel} et~al(2021){Michel}, {van der Marel}, and {Matthews}}]{Michel2021}
{Michel} A, {van der Marel} N, {Matthews} BC (2021) {Bridging the Gap between Protoplanetary and Debris Disks: Separate Evolution of Millimeter and Micrometer-sized Dust}. \apj 921(1):72. \doi{10.3847/1538-4357/ac1bbb}, {\href{https://arxiv.org/abs/2104.05894}{{arXiv:2104.05894}}} {[astro-ph.EP]}

\bibitem[{{Miller} et~al(2021){Miller}, {Marino}, {Stammler}, {Pinilla}, {Lenz}, {Birnstiel}, and {Henning}}]{Miller2021}
{Miller} E, {Marino} S, {Stammler} SM, et~al (2021) {The formation of wide exoKuiper belts from migrating dust traps}. \mnras 508(4):5638--5656. \doi{10.1093/mnras/stab2935}, {\href{https://arxiv.org/abs/2110.04007}{{arXiv:2110.04007}}} {[astro-ph.EP]}

\bibitem[{{Miotello} et~al(2023){Miotello}, {Kamp}, {Birnstiel}, {Cleeves}, and {Kataoka}}]{Miotello2023}
{Miotello} A, {Kamp} I, {Birnstiel} T, et~al (2023) {Setting the Stage for Planet Formation: Measurements and Implications of the Fundamental Disk Properties}. In: {Inutsuka} S, {Aikawa} Y, {Muto} T, et~al (eds) Protostars and Planets VII, p 501, \doi{10.48550/arXiv.2203.09818}, \eprint{2203.09818}

\bibitem[{{Mo{\'o}r} et~al(2016){Mo{\'o}r}, {K{\'o}sp{\'a}l}, {{\'A}brah{\'a}m}, {Balog}, {Csengeri}, {Henning}, {Juh{\'a}sz}, and {Kiss}}]{Moor2016}
{Mo{\'o}r} A, {K{\'o}sp{\'a}l} {\'A}, {{\'A}brah{\'a}m} P, et~al (2016) {New Debris Disks in Nearby Young Moving Groups}. \apj 826(2):123. \doi{10.3847/0004-637X/826/2/123}, {\href{https://arxiv.org/abs/1606.09179}{{arXiv:1606.09179}}} {[astro-ph.SR]}

\bibitem[{{Morais} and {Morbidelli}(2006)}]{Morais:2006}
{Morais} MHM, {Morbidelli} A (2006) {The population of Near Earth Asteroids in coorbital motion with Venus}. \icarus 185(1):29--38. \doi{10.1016/j.icarus.2006.06.009}

\bibitem[{{Morbidelli} et~al(2009){Morbidelli}, {Bottke}, {Nesvorn{\'y}}, and {Levison}}]{Morbidelli2009}
{Morbidelli} A, {Bottke} WF, {Nesvorn{\'y}} D, et~al (2009) {Asteroids were born big}. \icarus 204(2):558--573. \doi{10.1016/j.icarus.2009.07.011}, {\href{https://arxiv.org/abs/0907.2512}{{arXiv:0907.2512}}} {[astro-ph.EP]}

\bibitem[{{Morbidelli} et~al(2020){Morbidelli}, {Batygin}, {Brasser}, and {Raymond}}]{Morbidelli2020}
{Morbidelli} A, {Batygin} K, {Brasser} R, et~al (2020) {No evidence for interstellar planetesimals trapped in the Solar system}. \mnras 497(1):L46--L49. \doi{10.1093/mnrasl/slaa111}, {\href{https://arxiv.org/abs/2006.04534}{{arXiv:2006.04534}}} {[astro-ph.EP]}

\bibitem[{{Mouillet} et~al(1997){Mouillet}, {Larwood}, {Papaloizou}, and {Lagrange}}]{Mouillet1997}
{Mouillet} D, {Larwood} JD, {Papaloizou} JCB, et~al (1997) {A planet on an inclined orbit as an explanation of the warp in the beta Pictoris disc}. \mnras 292(4):896--904. \doi{10.1093/mnras/292.4.896}, {\href{https://arxiv.org/abs/astro-ph/9705100}{{arXiv:astro-ph/9705100}}} {[astro-ph]}

\bibitem[{{Mulders} et~al(2015){Mulders}, {Pascucci}, and {Apai}}]{Mulders2015}
{Mulders} GD, {Pascucci} I, {Apai} D (2015) {A Stellar-mass-dependent Drop in Planet Occurrence Rates}. \apj 798(2):112. \doi{10.1088/0004-637X/798/2/112}, {\href{https://arxiv.org/abs/1406.7356}{{arXiv:1406.7356}}} {[astro-ph.EP]}

\bibitem[{{Mulders} et~al(2021){Mulders}, {Drazkowska}, {van der Marel}, {Ciesla}, and {Pascucci}}]{Mulders2021}
{Mulders} GD, {Drazkowska} J, {van der Marel} N, et~al (2021) {Why Do M Dwarfs Have More Transiting Planets?} \apjl 920(1):L1. \doi{10.3847/2041-8213/ac2947}, {\href{https://arxiv.org/abs/2110.02971}{{arXiv:2110.02971}}} {[astro-ph.EP]}

\bibitem[{Mumma and Charnley(2011)}]{Mumma2011}
Mumma MJ, Charnley SB (2011) {The Chemical Composition of Comets—Emerging Taxonomies and Natal Heritage}. Annual Review of Astronomy and Astrophysics 49:471 -- 524. \doi{10.1146/annurev-astro-081309-130811}, although the OPRs for individual primary volatiles often show different values in a given comet, the spin temperatures derived from these ratios agree within confidence limits. The underlying meaning of that temperature, however, is uncertain.

\bibitem[{{Musiolik} et~al(2016){Musiolik}, {Teiser}, {Jankowski}, and {Wurm}}]{Musiolik2016}
{Musiolik} G, {Teiser} J, {Jankowski} T, et~al (2016) {Ice Grain Collisions in Comparison: CO2, H2O, and Their Mixtures}. \apj 827(1):63. \doi{10.3847/0004-637X/827/1/63}, {\href{https://arxiv.org/abs/1608.05017}{{arXiv:1608.05017}}} {[astro-ph.EP]}

\bibitem[{{Myers} and {Benson}(1983)}]{Myers1983}
{Myers} PC, {Benson} PJ (1983) {Dense cores in dark clouds. II. NH3 observations and star formation.} \apj 266:309--320. \doi{10.1086/160780}

\bibitem[{{Nakagawa} et~al(1986){Nakagawa}, {Sekiya}, and {Hayashi}}]{Nakagawa1986}
{Nakagawa} Y, {Sekiya} M, {Hayashi} C (1986) {Settling and growth of dust particles in a laminar phase of a low-mass solar nebula}. \icarus 67(3):375--390. \doi{10.1016/0019-1035(86)90121-1}

\bibitem[{{Napier} et~al(2021){Napier}, {Adams}, and {Batygin}}]{Napier2021}
{Napier} KJ, {Adams} FC, {Batygin} K (2021) {On the Fate of Interstellar Objects Captured by Our Solar System}. \psj 2(6):217. \doi{10.3847/PSJ/ac29bb}, {\href{https://arxiv.org/abs/2109.11017}{{arXiv:2109.11017}}} {[astro-ph.EP]}

\bibitem[{{Naraoka} et~al(2023){Naraoka}, {Takano}, {Dworkin}, {Oba}, {Hamase}, {Furusho}, {Ogawa}, {Hashiguchi}, {Fukushima}, {Aoki}, {Schmitt-Kopplin}, {Aponte}, {Parker}, {Glavin}, {McLain}, {Elsila}, {Graham}, {Eiler}, {Orthous-Daunay}, {Wolters}, {Isa}, {Vuitton}, {Thissen}, {Sakai}, {Yoshimura}, {Koga}, {Ohkouchi}, {Chikaraishi}, {Sugahara}, {Mita}, {Furukawa}, {Hertkorn}, {Ruf}, {Yurimoto}, {Nakamura}, {Noguchi}, {Okazaki}, {Yabuta}, {Sakamoto}, {Tachibana}, {Connolly}, {Lauretta}, {Abe}, {Yada}, {Nishimura}, {Yogata}, {Nakato}, {Yoshitake}, {Suzuki}, {Miyazaki}, {Furuya}, {Hatakeda}, {Soejima}, {Hitomi}, {Kumagai}, {Usui}, {Hayashi}, {Yamamoto}, {Fukai}, {Kitazato}, {Sugita}, {Namiki}, {Arakawa}, {Ikeda}, {Ishiguro}, {Hirata}, {Wada}, {Ishihara}, {Noguchi}, {Morota}, {Sakatani}, {Matsumoto}, {Senshu}, {Honda}, {Tatsumi}, {Yokota}, {Honda}, {Michikami}, {Matsuoka}, {Miura}, {Noda}, {Yamada}, {Yoshihara}, {Kawahara}, {Ozaki}, {Iijima}, {Yano}, {Hayakawa}, {Iwata}, {Tsukizaki}, {Sawada}, {Hosoda},
  {Ogawa}, {Okamoto}, {Hirata}, {Shirai}, {Shimaki}, {Yamada}, {Okada}, {Yamamoto}, {Takeuchi}, {Fujii}, {Takei}, {Yoshikawa}, {Mimasu}, {Ono}, {Ogawa}, {Kikuchi}, {Nakazawa}, {Terui}, {Tanaka}, {Saiki}, {Yoshikawa}, {Watanabe}, and {Tsuda}}]{Naraoka2023}
{Naraoka} H, {Takano} Y, {Dworkin} JP, et~al (2023) {Soluble organic molecules in samples of the carbonaceous asteroid (162173) Ryugu}. Science 379(6634):abn9033. \doi{10.1126/science.abn9033}

\bibitem[{{Nesvorn{\'y}}(2018)}]{Nesvorny2018}
{Nesvorn{\'y}} D (2018) {Dynamical Evolution of the Early Solar System}. \araa 56:137--174. \doi{10.1146/annurev-astro-081817-052028}, {\href{https://arxiv.org/abs/1807.06647}{{arXiv:1807.06647}}} {[astro-ph.EP]}

\bibitem[{{Nesvorn{\'y}} et~al(2019){Nesvorn{\'y}}, {Li}, {Youdin}, {Simon}, and {Grundy}}]{Nesvorny2019}
{Nesvorn{\'y}} D, {Li} R, {Youdin} AN, et~al (2019) {Trans-Neptunian binaries as evidence for planetesimal formation by the streaming instability}. Nature Astronomy 3:808--812. \doi{10.1038/s41550-019-0806-z}, {\href{https://arxiv.org/abs/1906.11344}{{arXiv:1906.11344}}} {[astro-ph.EP]}

\bibitem[{{Nesvorn{\'y}} et~al(2020){Nesvorn{\'y}}, {Vokrouhlick{\'y}}, {Alexandersen}, {Bannister}, {Buchanan}, {Chen}, {Gladman}, {Gwyn}, {Kavelaars}, {Petit}, {Schwamb}, and {Volk}}]{Nesvorny2020}
{Nesvorn{\'y}} D, {Vokrouhlick{\'y}} D, {Alexandersen} M, et~al (2020) {OSSOS XX: The Meaning of Kuiper Belt Colors}. \aj 160(1):46. \doi{10.3847/1538-3881/ab98fb}, {\href{https://arxiv.org/abs/2006.01806}{{arXiv:2006.01806}}} {[astro-ph.EP]}

\bibitem[{{Nesvorn{\'y}} et~al(2023){Nesvorn{\'y}}, {Bernardinelli}, {Vokrouhlick{\'y}}, and {Batygin}}]{Nesvorny2023b}
{Nesvorn{\'y}} D, {Bernardinelli} P, {Vokrouhlick{\'y}} D, et~al (2023) {Radial distribution of distant trans-Neptunian objects points to Sun's formation in a stellar cluster}. \icarus 406:115738. \doi{10.1016/j.icarus.2023.115738}, {\href{https://arxiv.org/abs/2308.11059}{{arXiv:2308.11059}}} {[astro-ph.EP]}

\bibitem[{{Nesvorn{\'y}} et~al(2024){Nesvorn{\'y}}, {Vokrouhlick{\'y}}, {Shelly}, {Deienno}, {Bottke}, {Fuls}, {Jedicke}, {Naidu}, {Chesley}, {Chodas}, {Farnocchia}, and {Delbo}}]{Nesvorny:2024}
{Nesvorn{\'y}} D, {Vokrouhlick{\'y}} D, {Shelly} F, et~al (2024) {NEOMOD 3: The debiased size distribution of Near Earth Objects}. \icarus 417:116110. \doi{10.1016/j.icarus.2024.116110}, {\href{https://arxiv.org/abs/2404.18805}{{arXiv:2404.18805}}} {[astro-ph.EP]}

\bibitem[{{Nielsen} et~al(2019){Nielsen}, {De Rosa}, {Macintosh}, {Wang}, {Ruffio}, {Chiang}, {Marley}, {Saumon}, {Savransky}, {Ammons}, {Bailey}, {Barman}, {Blain}, {Bulger}, {Burrows}, {Chilcote}, {Cotten}, {Czekala}, {Doyon}, {Duch{\^e}ne}, {Esposito}, {Fabrycky}, {Fitzgerald}, {Follette}, {Fortney}, {Gerard}, {Goodsell}, {Graham}, {Greenbaum}, {Hibon}, {Hinkley}, {Hirsch}, {Hom}, {Hung}, {Dawson}, {Ingraham}, {Kalas}, {Konopacky}, {Larkin}, {Lee}, {Lin}, {Maire}, {Marchis}, {Marois}, {Metchev}, {Millar-Blanchaer}, {Morzinski}, {Oppenheimer}, {Palmer}, {Patience}, {Perrin}, {Poyneer}, {Pueyo}, {Rafikov}, {Rajan}, {Rameau}, {Rantakyr{\"o}}, {Ren}, {Schneider}, {Sivaramakrishnan}, {Song}, {Soummer}, {Tallis}, {Thomas}, {Ward-Duong}, and {Wolff}}]{Nielsen2019}
{Nielsen} EL, {De Rosa} RJ, {Macintosh} B, et~al (2019) {The Gemini Planet Imager Exoplanet Survey: Giant Planet and Brown Dwarf Demographics from 10 to 100 au}. \aj 158(1):13. \doi{10.3847/1538-3881/ab16e9}, {\href{https://arxiv.org/abs/1904.05358}{{arXiv:1904.05358}}} {[astro-ph.EP]}

\bibitem[{{Oba} et~al(2023){Oba}, {Takano}, {Dworkin}, and {Naraoka}}]{Oba2023}
{Oba} Y, {Takano} Y, {Dworkin} JP, et~al (2023) {Ryugu asteroid sample return provides a natural laboratory for primordial chemical evolution}. Nature Communications 14:3107. \doi{10.1038/s41467-023-38518-1}

\bibitem[{{{\"O}berg} et~al(2023){{\"O}berg}, {Facchini}, and {Anderson}}]{Oberg2023}
{{\"O}berg} KI, {Facchini} S, {Anderson} DE (2023) {Protoplanetary Disk Chemistry}. \araa 61:287--328. \doi{10.1146/annurev-astro-022823-040820}, {\href{https://arxiv.org/abs/2309.05685}{{arXiv:2309.05685}}} {[astro-ph.EP]}

\bibitem[{{Ofek} et~al(2024){Ofek}, {Spitzer}, and {Nir}}]{Ofek2024A}
{Ofek} EO, {Spitzer} SA, {Nir} G (2024) {An Efficient Observational Strategy for the Detection of the Oort Cloud}. \aj 168(4):147. \doi{10.3847/1538-3881/ad644b}, {\href{https://arxiv.org/abs/2409.02170}{{arXiv:2409.02170}}} {[astro-ph.IM]}

\bibitem[{{Ohashi} et~al(1997){Ohashi}, {Hayashi}, {Ho}, {Momose}, {Tamura}, {Hirano}, and {Sargent}}]{Ohashi1997}
{Ohashi} N, {Hayashi} M, {Ho} PTP, et~al (1997) {Rotation in the Protostellar Envelopes around IRAS 04169+2702 and IRAS 04365+2535: The Size Scale for Dynamical Collapse}. \apj 488(1):317--329. \doi{10.1086/304685}

\bibitem[{{Ohashi} et~al(2014){Ohashi}, {Saigo}, {Aso}, {Aikawa}, {Koyamatsu}, {Machida}, {Saito}, {Takahashi}, {Takakuwa}, {Tomida}, {Tomisaka}, and {Yen}}]{Ohashi2014}
{Ohashi} N, {Saigo} K, {Aso} Y, et~al (2014) {Formation of a Keplerian Disk in the Infalling Envelope around L1527 IRS: Transformation from Infalling Motions to Kepler Motions}. \apj 796(2):131. \doi{10.1088/0004-637X/796/2/131}, {\href{https://arxiv.org/abs/1410.0172}{{arXiv:1410.0172}}} {[astro-ph.GA]}

\bibitem[{{Ohashi} et~al(2023){Ohashi}, {Tobin}, {J{\o}rgensen}, {Takakuwa}, {Sheehan}, {Aikawa}, {Li}, {Looney}, {Williams}, {Aso}, {Sharma}, {Sai}, {Yamato}, {Lee}, {Tomida}, {Yen}, {Encalada}, {Flores}, {Gavino}, {Kido}, {Han}, {Lin}, {Narayanan}, {Phuong}, {Santamar{\'\i}a-Miranda}, {Thieme}, {van't Hoff}, {de Gregorio-Monsalvo}, {Koch}, {Kwon}, {Lai}, {Lee}, {Plunkett}, {Saigo}, {Hirano}, {Lam}, and {Mori}}]{Ohashi2023}
{Ohashi} N, {Tobin} JJ, {J{\o}rgensen} JK, et~al (2023) {Early Planet Formation in Embedded Disks (eDisk). I. Overview of the Program and First Results}. \apj 951(1):8. \doi{10.3847/1538-4357/acd384}, {\href{https://arxiv.org/abs/2306.15406}{{arXiv:2306.15406}}} {[astro-ph.EP]}

\bibitem[{{Oka} et~al(2011){Oka}, {Nakamoto}, and {Ida}}]{Oka2011}
{Oka} A, {Nakamoto} T, {Ida} S (2011) {Evolution of Snow Line in Optically Thick Protoplanetary Disks: Effects of Water Ice Opacity and Dust Grain Size}. \apj 738(2):141. \doi{10.1088/0004-637X/738/2/141}, {\href{https://arxiv.org/abs/1106.2682}{{arXiv:1106.2682}}} {[astro-ph.EP]}

\bibitem[{{Okuzumi} et~al(2012){Okuzumi}, {Tanaka}, {Kobayashi}, and {Wada}}]{Okuzumi2012}
{Okuzumi} S, {Tanaka} H, {Kobayashi} H, et~al (2012) {Rapid Coagulation of Porous Dust Aggregates outside the Snow Line: A Pathway to Successful Icy Planetesimal Formation}. \apj 752(2):106. \doi{10.1088/0004-637X/752/2/106}, {\href{https://arxiv.org/abs/1204.5035}{{arXiv:1204.5035}}} {[astro-ph.EP]}

\bibitem[{{Okuzumi} et~al(2016){Okuzumi}, {Momose}, {Sirono}, {Kobayashi}, and {Tanaka}}]{Okuzumi2016}
{Okuzumi} S, {Momose} M, {Sirono} Si, et~al (2016) {Sintering-induced Dust Ring Formation in Protoplanetary Disks: Application to the HL Tau Disk}. \apj 821(2):82. \doi{10.3847/0004-637X/821/2/82}, {\href{https://arxiv.org/abs/1510.03556}{{arXiv:1510.03556}}} {[astro-ph.SR]}

\bibitem[{{Oort}(1950)}]{Oort1950}
{Oort} JH (1950) {The structure of the cloud of comets surrounding the Solar System and a hypothesis concerning its origin}. \bain 11:91--110

\bibitem[{{Ormel} and {Huang}(2025)}]{Ormel2025}
{Ormel} CW, {Huang} Y (2025) {From Planetesimals to Dwarf Planets by Pebble Accretion}. arXiv e-prints arXiv:2502.04016. \doi{10.48550/arXiv.2502.04016}, {\href{https://arxiv.org/abs/2502.04016}{{arXiv:2502.04016}}} {[astro-ph.EP]}

\bibitem[{{Ould Rouis} et~al(2024){Ould Rouis}, {Hermes}, {G{\"a}nsicke}, {Sahu}, {Koester}, {Tremblay}, {Veras}, {Farihi}, {Heintz}, {Gentile Fusillo}, and {Redfield}}]{OuldRouis2024}
{Ould Rouis} LB, {Hermes} JJ, {G{\"a}nsicke} BT, et~al (2024) {Constraints on Remnant Planetary Systems as a Function of Main-Sequence Mass with HST/COS}. arXiv e-prints arXiv:2410.06335. \doi{10.48550/arXiv.2410.06335}, {\href{https://arxiv.org/abs/2410.06335}{{arXiv:2410.06335}}} {[astro-ph.SR]}

\bibitem[{{Paganini} et~al(2017){Paganini}, {Mumma}, {Gibb}, and {Villanueva}}]{Paganini2017}
{Paganini} L, {Mumma} MJ, {Gibb} EL, et~al (2017) {Ground-based Detection of Deuterated Water in Comet C/2014 Q2 (Lovejoy) at IR Wavelengths}. \apjl 836(2):L25. \doi{10.3847/2041-8213/aa5cb3}

\bibitem[{{Pan} et~al(2011){Pan}, {Padoan}, {Scalo}, {Kritsuk}, and {Norman}}]{Pan2011}
{Pan} L, {Padoan} P, {Scalo} J, et~al (2011) {Turbulent Clustering of Protoplanetary Dust and Planetesimal Formation}. \apj 740(1):6. \doi{10.1088/0004-637X/740/1/6}, {\href{https://arxiv.org/abs/1106.3695}{{arXiv:1106.3695}}} {[astro-ph.EP]}

\bibitem[{{Pan} and {Gallardo}(2025)}]{Pan:2025}
{Pan} N, {Gallardo} T (2025) {An attempt to build a dynamical catalog of present-day solar system co-orbitals: An attempt to build a dynamical catalog}. Celestial Mechanics and Dynamical Astronomy 137(1):2. \doi{10.1007/s10569-024-10234-y}, {\href{https://arxiv.org/abs/2410.20015}{{arXiv:2410.20015}}} {[astro-ph.EP]}

\bibitem[{{Parker}(2015)}]{Parker2015}
{Parker} AH (2015) {The intrinsic Neptune Trojan orbit distribution: Implications for the primordial disk and planet migration}. \icarus 247:112--125. \doi{10.1016/j.icarus.2014.09.043}, {\href{https://arxiv.org/abs/1409.6735}{{arXiv:1409.6735}}} {[astro-ph.EP]}

\bibitem[{{Pe{\~n}arrubia}(2023)}]{Penarubbia2023}
{Pe{\~n}arrubia} J (2023) {A halo of trapped interstellar matter surrounding the Solar system}. \mnras 519(2):1955--1980. \doi{10.1093/mnras/stac3642}, {\href{https://arxiv.org/abs/2206.08535}{{arXiv:2206.08535}}} {[astro-ph.GA]}

\bibitem[{{Pearce}(2024)}]{Pearce2024Review}
{Pearce} TD (2024) {Debris disks around main-sequence stars}. arXiv e-prints arXiv:2403.11804. {\href{https://arxiv.org/abs/2403.11804}{{arXiv:2403.11804}}} {[astro-ph.EP]}

\bibitem[{{Pearce} and {Wyatt}(2014)}]{Pearce2014}
{Pearce} TD, {Wyatt} MC (2014) {Dynamical evolution of an eccentric planet and a less massive debris disc}. \mnras 443(3):2541--2560. \doi{10.1093/mnras/stu1302}, {\href{https://arxiv.org/abs/1406.7294}{{arXiv:1406.7294}}} {[astro-ph.EP]}

\bibitem[{{Pearce} and {Wyatt}(2015)}]{Pearce2015}
{Pearce} TD, {Wyatt} MC (2015) {Double-ringed debris discs could be the work of eccentric planets: explaining the strange morphology of HD 107146}. \mnras 453(3):3329--3340. \doi{10.1093/mnras/stv1847}, {\href{https://arxiv.org/abs/1507.04367}{{arXiv:1507.04367}}} {[astro-ph.EP]}

\bibitem[{{Pearce} et~al(2020){Pearce}, {Krivov}, and {Booth}}]{Pearce2020}
{Pearce} TD, {Krivov} AV, {Booth} M (2020) {Gas trapping of hot dust around main-sequence stars}. \mnras 498(2):2798--2813. \doi{10.1093/mnras/staa2514}, {\href{https://arxiv.org/abs/2008.07505}{{arXiv:2008.07505}}} {[astro-ph.EP]}

\bibitem[{{Pearce} et~al(2022{\natexlab{a}}){Pearce}, {Launhardt}, {Ostermann}, {Kennedy}, {Gennaro}, {Booth}, {Krivov}, {Cugno}, {Henning}, {Quirrenbach}, {Barcucci}, {Matthews}, {Ruh}, and {Stone}}]{Pearce2022ISPY}
{Pearce} TD, {Launhardt} R, {Ostermann} R, et~al (2022{\natexlab{a}}) {Planet populations inferred from debris discs. Insights from 178 debris systems in the ISPY, LEECH, and LIStEN planet-hunting surveys}. \aap 659:A135. \doi{10.1051/0004-6361/202142720}, {\href{https://arxiv.org/abs/2201.08369}{{arXiv:2201.08369}}} {[astro-ph.EP]}

\bibitem[{{Pearce} et~al(2024){Pearce}, {Krivov}, {Sefilian}, {Jankovic}, {L{\"o}hne}, {Morgner}, {Wyatt}, {Booth}, and {Marino}}]{Pearce2024Edges}
{Pearce} TD, {Krivov} AV, {Sefilian} AA, et~al (2024) {The effect of sculpting planets on the steepness of debris-disc inner edges}. \mnras 527(2):3876--3899. \doi{10.1093/mnras/stad3462}, {\href{https://arxiv.org/abs/2311.04265}{{arXiv:2311.04265}}} {[astro-ph.EP]}

\bibitem[{{Pearce} et~al(2022{\natexlab{b}})}]{Pearce2022Comets}
{Pearce} TD, et~al (2022{\natexlab{b}}) {Hot exozodis: cometary supply without trapping is unlikely to be the mechanism}. \mnras 517(1):1436--1451. \doi{10.1093/mnras/stac2773}, {\href{https://arxiv.org/abs/2209.11219}{{arXiv:2209.11219}}} {[astro-ph.EP]}

\bibitem[{{Peixinho} et~al(2015){Peixinho}, {Delsanti}, and {Doressoundiram}}]{Peixinho2015}
{Peixinho} N, {Delsanti} A, {Doressoundiram} A (2015) {Reanalyzing the visible colors of Centaurs and KBOs: what is there and what we might be missing}. \aap 577:A35. \doi{10.1051/0004-6361/201425436}, {\href{https://arxiv.org/abs/1502.04145}{{arXiv:1502.04145}}} {[astro-ph.EP]}

\bibitem[{{Perrine}(1902)}]{Perrine1902}
{Perrine} CD (1902) {The Search for an Intra-Mercurial Planet at the Total Solar Eclipse of 1901, May 18}. \pasp 14(86):160. \doi{10.1086/121490}

\bibitem[{{Petit} et~al(2001){Petit}, {Morbidelli}, and {Chambers}}]{Petit2001}
{Petit} JM, {Morbidelli} A, {Chambers} J (2001) {The Primordial Excitation and Clearing of the Asteroid Belt}. \icarus 153(2):338--347. \doi{10.1006/icar.2001.6702}

\bibitem[{{Pfalzner}(2009)}]{Pfalzner2009}
{Pfalzner} S (2009) {Universality of young cluster sequences}. \aap 498(2):L37--L40. \doi{10.1051/0004-6361/200912056}, {\href{https://arxiv.org/abs/0904.0523}{{arXiv:0904.0523}}} {[astro-ph.GA]}

\bibitem[{{Pfalzner} and {Govind}(2021)}]{Pfalzner:2021}
{Pfalzner} S, {Govind} A (2021) {Close Stellar Flybys Common in Low-mass Clusters}. \apj 921(1):90. \doi{10.3847/1538-4357/ac19aa}, {\href{https://arxiv.org/abs/2108.10296}{{arXiv:2108.10296}}} {[astro-ph.SR]}

\bibitem[{{Pfalzner} and {Kaczmarek}(2013)}]{Pfalzner2013}
{Pfalzner} S, {Kaczmarek} T (2013) {The expansion of massive young star clusters - observation meets theory}. \aap 559:A38. \doi{10.1051/0004-6361/201322134}, {\href{https://arxiv.org/abs/1309.0315}{{arXiv:1309.0315}}} {[astro-ph.GA]}

\bibitem[{{Pfalzner} et~al(2022){Pfalzner}, {Dehghani}, and {Michel}}]{Pfalzner:2022}
{Pfalzner} S, {Dehghani} S, {Michel} A (2022) {Most Planets Might Have More than 5 Myr of Time to Form}. \apjl 939(1):L10. \doi{10.3847/2041-8213/ac9839}, {\href{https://arxiv.org/abs/2210.02420}{{arXiv:2210.02420}}} {[astro-ph.GA]}

\bibitem[{{Pfalzner} et~al(2024{\natexlab{a}}){Pfalzner}, {Govind}, and {Portegies Zwart}}]{Pfalzner2024}
{Pfalzner} S, {Govind} A, {Portegies Zwart} S (2024{\natexlab{a}}) {Trajectory of the stellar flyby that shaped the outer Solar System}. Nature Astronomy 8:1380--1386. \doi{10.1038/s41550-024-02349-x}, {\href{https://arxiv.org/abs/2409.03342}{{arXiv:2409.03342}}} {[astro-ph.EP]}

\bibitem[{{Pfalzner} et~al(2024{\natexlab{b}}){Pfalzner}, {Govind}, and {Portegies Zwart}}]{Pfalzner:2024}
{Pfalzner} S, {Govind} A, {Portegies Zwart} S (2024{\natexlab{b}}) {Trajectory of the stellar flyby that shaped the outer Solar System}. Nature Astronomy 8:1380--1386. \doi{10.1038/s41550-024-02349-x}, {\href{https://arxiv.org/abs/2409.03342}{{arXiv:2409.03342}}} {[astro-ph.EP]}

\bibitem[{{Pfalzner} et~al(2024{\natexlab{c}}){Pfalzner}, {Govind}, and {Wagner}}]{Pfalzner2024b}
{Pfalzner} S, {Govind} A, {Wagner} FW (2024{\natexlab{c}}) {Irregular Moons Possibly Injected from the Outer Solar System by a Stellar Flyby}. \apjl 972(2):L21. \doi{10.3847/2041-8213/ad63a6}, {\href{https://arxiv.org/abs/2409.03529}{{arXiv:2409.03529}}} {[astro-ph.EP]}

\bibitem[{{Pfalzner} et~al(2025){Pfalzner}, {Wagner}, and {Gibbon}}]{Pfalzner:2025}
{Pfalzner} S, {Wagner} FW, {Gibbon} P (2025) {TNO colours provide new evidence for a past close flyby of another star to the Solar System}. arXiv e-prints arXiv:2507.06693. \doi{10.48550/arXiv.2507.06693}, {\href{https://arxiv.org/abs/2507.06693}{{arXiv:2507.06693}}} {[astro-ph.EP]}

\bibitem[{{Pinilla-Alonso} et~al(2025){Pinilla-Alonso}, {Brunetto}, {De Pr{\'a}}, {Holler}, {H{\'e}nault}, {Feliciano}, {Lorenzi}, {Pendleton}, {Cruikshank}, {M{\"u}ller}, {Stansberry}, {Emery}, {Schambeau}, {Licandro}, {Harvison}, {McClure}, {Guilbert-Lepoutre}, {Peixinho}, {Bannister}, and {Wong}}]{Pinilla-Alonso2025}
{Pinilla-Alonso} N, {Brunetto} R, {De Pr{\'a}} MN, et~al (2025) {A JWST/DiSCo-TNOs portrait of the primordial Solar System through its trans-Neptunian objects}. Nature Astronomy 9:230--244. \doi{10.1038/s41550-024-02433-2}

\bibitem[{{Pirani} et~al(2019){Pirani}, {Johansen}, {Bitsch}, {Mustill}, and {Turrini}}]{Pirani2019}
{Pirani} S, {Johansen} A, {Bitsch} B, et~al (2019) {Consequences of planetary migration on the minor bodies of the early solar system}. \aap 623:A169. \doi{10.1051/0004-6361/201833713}, {\href{https://arxiv.org/abs/1902.04591}{{arXiv:1902.04591}}} {[astro-ph.EP]}

\bibitem[{{Poleski} et~al(2021){Poleski}, {Skowron}, {Mr{\'o}z}, {Udalski}, {Szyma{\'n}ski}, {Pietrukowicz}, {Ulaczyk}, {Rybicki}, {Iwanek}, {Wrona}, and {Gromadzki}}]{Poleski2021}
{Poleski} R, {Skowron} J, {Mr{\'o}z} P, et~al (2021) {Wide-Orbit Exoplanets are Common. Analysis of Nearly 20 Years of OGLE Microlensing Survey Data}. Acta Astronomica 71(1):1--23. \doi{10.32023/0001-5237/71.1.1}, {\href{https://arxiv.org/abs/2104.02079}{{arXiv:2104.02079}}} {[astro-ph.EP]}

\bibitem[{{Portegies Zwart} et~al(2021){Portegies Zwart}, {Torres}, {Cai}, and {Brown}}]{Portegies2021}
{Portegies Zwart} S, {Torres} S, {Cai} MX, et~al (2021) {Oort cloud Ecology. II. the chronology of the formation of the Oort cloud}. \aap 652:A144. \doi{10.1051/0004-6361/202040096}, {\href{https://arxiv.org/abs/2105.12816}{{arXiv:2105.12816}}} {[astro-ph.EP]}

\bibitem[{{Raettig} et~al(2015){Raettig}, {Klahr}, and {Lyra}}]{Raettig2015}
{Raettig} N, {Klahr} H, {Lyra} W (2015) {Particle Trapping and Streaming Instability in Vortices in Protoplanetary Disks}. \apj 804(1):35. \doi{10.1088/0004-637X/804/1/35}, {\href{https://arxiv.org/abs/1501.05364}{{arXiv:1501.05364}}} {[astro-ph.EP]}

\bibitem[{{Rafikov}(2023)}]{Rafikov2023}
{Rafikov} RR (2023) {Radial profiles of surface density in debris discs}. \mnras 519(4):5607--5622. \doi{10.1093/mnras/stac3411}, {\href{https://arxiv.org/abs/2207.07678}{{arXiv:2207.07678}}} {[astro-ph.EP]}

\bibitem[{{Reffert} et~al(2015){Reffert}, {Bergmann}, {Quirrenbach}, {Trifonov}, and {K{\"u}nstler}}]{Reffert2015}
{Reffert} S, {Bergmann} C, {Quirrenbach} A, et~al (2015) {Precise radial velocities of giant stars. VII. Occurrence rate of giant extrasolar planets as a function of mass and metallicity}. \aap 574:A116. \doi{10.1051/0004-6361/201322360}, {\href{https://arxiv.org/abs/1412.4634}{{arXiv:1412.4634}}} {[astro-ph.EP]}

\bibitem[{{Ribas} et~al(2015){Ribas}, {Bouy}, and {Mer{\'\i}n}}]{Ribas2015}
{Ribas} {\'A}, {Bouy} H, {Mer{\'\i}n} B (2015) {Protoplanetary disk lifetimes vs. stellar mass and possible implications for giant planet populations}. \aap 576:A52. \doi{10.1051/0004-6361/201424846}, {\href{https://arxiv.org/abs/1502.00631}{{arXiv:1502.00631}}} {[astro-ph.SR]}

\bibitem[{Russo et~al(2016)Russo, Kawakita, Vervack, and Weaver}]{DelloRusso2016}
Russo ND, Kawakita H, Vervack RJ, et~al (2016) {Emerging trends and a comet taxonomy based on the volatile chemistry measured in thirty comets with high-resolution infrared spectroscopy between 1997 and 2013}. Icarus 278:301--332. \doi{10.1016/j.icarus.2016.05.039}

\bibitem[{Safronov(1972)}]{Safronov1972}
Safronov V (1972) {Evolution of the Protoplanetary Cloud and Formation of the Earth and the Planets}. Israel Program for Scientific Translations, Jerusalem p~11

\bibitem[{Saki et~al(2024)Saki, Bodewits, Bonev, Russo, Luspay-Kuti, Noonan, Combi, and Shou}]{Saki2024}
Saki M, Bodewits D, Bonev BP, et~al (2024) {Parent Volatile Outgassing Associations in Cometary Nuclei: Synthesizing Rosetta Measurements and Ground-based Observations}. The Planetary Science Journal 5(3):70. \doi{10.3847/psj/ad118f}

\bibitem[{{Sch{\"a}fer} et~al(2017){Sch{\"a}fer}, {Yang}, and {Johansen}}]{Schaefer2017}
{Sch{\"a}fer} U, {Yang} CC, {Johansen} A (2017) {Initial mass function of planetesimals formed by the streaming instability}. \aap 597:A69. \doi{10.1051/0004-6361/201629561}, {\href{https://arxiv.org/abs/1611.02285}{{arXiv:1611.02285}}} {[astro-ph.EP]}

\bibitem[{{Schaffer} et~al(2021){Schaffer}, {Johansen}, and {Lambrechts}}]{Schaffer2021}
{Schaffer} N, {Johansen} A, {Lambrechts} M (2021) {Streaming instability of multiple particle species. II. Numerical convergence with increasing particle number}. \aap 653:A14. \doi{10.1051/0004-6361/202140690}, {\href{https://arxiv.org/abs/2106.15302}{{arXiv:2106.15302}}} {[astro-ph.EP]}

\bibitem[{Schleicher et~al(2007)Schleicher, Farnham, and Bair}]{Schleicher2007}
Schleicher DG, Farnham TL, Bair AN (2007) {The Chemical and Physical Properties of Comets: Uniform Analyses of Narrowband Photometry}. Workshop on Planetary Atmospheres p 103. \urlprefix\url{http://adsabs.harvard.edu/cgi-bin/nph-data\_query?bibcode=2007plat.work..103S\&link\_type=ABSTRACT}

\bibitem[{{Schlichting} et~al(2012){Schlichting}, {Ofek}, {Sari}, {Nelan}, {Gal-Yam}, {Wenz}, {Muirhead}, {Javanfar}, and {Livio}}]{Schlichting2012}
{Schlichting} HE, {Ofek} EO, {Sari} R, et~al (2012) {Measuring the Abundance of Sub-kilometer-sized Kuiper Belt Objects Using Stellar Occultations}. \apj 761(2):150. \doi{10.1088/0004-637X/761/2/150}, {\href{https://arxiv.org/abs/1210.8155}{{arXiv:1210.8155}}} {[astro-ph.EP]}

\bibitem[{{Schwamb} et~al(2010){Schwamb}, {Brown}, {Rabinowitz}, and {Ragozzine}}]{Schwamb2010}
{Schwamb} ME, {Brown} ME, {Rabinowitz} DL, et~al (2010) {Properties of the Distant Kuiper Belt: Results from the Palomar Distant Solar System Survey}. \apj 720(2):1691--1707. \doi{10.1088/0004-637X/720/2/1691}, {\href{https://arxiv.org/abs/1007.2954}{{arXiv:1007.2954}}} {[astro-ph.EP]}

\bibitem[{{Sefilian} et~al(2021){Sefilian}, {Rafikov}, and {Wyatt}}]{Sefilian2021}
{Sefilian} AA, {Rafikov} RR, {Wyatt} MC (2021) {Formation of Gaps in Self-gravitating Debris Disks by Secular Resonance in a Single-planet System. I. A Simplified Model}. \apj 910(1):13. \doi{10.3847/1538-4357/abda46}, {\href{https://arxiv.org/abs/2010.15617}{{arXiv:2010.15617}}} {[astro-ph.EP]}

\bibitem[{{Seligman} et~al(2022){Seligman}, {Rogers}, {Cabot}, {Noonan}, {Kareta}, {Mandt}, {Ciesla}, {McKay}, {Feinstein}, {Levine}, {Bean}, {Nordlander}, {Krumholz}, {Mansfield}, {Hoover}, and {Van Clepper}}]{Seligman2022}
{Seligman} DZ, {Rogers} LA, {Cabot} SHC, et~al (2022) {The Volatile Carbon-to-oxygen Ratio as a Tracer for the Formation Locations of Interstellar Comets}. \psj 3(7):150. \doi{10.3847/PSJ/ac75b5}, {\href{https://arxiv.org/abs/2204.13211}{{arXiv:2204.13211}}} {[astro-ph.EP]}

\bibitem[{{Sezestre} et~al(2019){Sezestre}, {Augereau}, and {Th{\'e}bault}}]{Sezestre2019}
{Sezestre} {\'E}, {Augereau} JC, {Th{\'e}bault} P (2019) {Hot exozodiacal dust: an exocometary origin?} \aap 626:A2. \doi{10.1051/0004-6361/201935250}, {\href{https://arxiv.org/abs/1903.06130}{{arXiv:1903.06130}}} {[astro-ph.EP]}

\bibitem[{{Sheehan} et~al(2022){Sheehan}, {Tobin}, {Looney}, and {Megeath}}]{Sheehan2022}
{Sheehan} PD, {Tobin} JJ, {Looney} LW, et~al (2022) {The VLA/ALMA Nascent Disk and Multiplicity (VANDAM) Survey of Orion Protostars. VI. Insights from Radiative Transfer Modeling}. \apj 929(1):76. \doi{10.3847/1538-4357/ac574d}, {\href{https://arxiv.org/abs/2203.00029}{{arXiv:2203.00029}}} {[astro-ph.SR]}

\bibitem[{{Shu} et~al(1987){Shu}, {Adams}, and {Lizano}}]{Shu1987}
{Shu} FH, {Adams} FC, {Lizano} S (1987) {Star formation in molecular clouds: observation and theory.} \araa 25:23--81. \doi{10.1146/annurev.aa.25.090187.000323}

\bibitem[{{Shvartzvald} et~al(2016){Shvartzvald}, {Maoz}, {Udalski}, {Sumi}, {Friedmann}, {Kaspi}, {Poleski}, {Szyma{\'n}ski}, {Skowron}, {Koz{\l}owski}, {Wyrzykowski}, {Mr{\'o}z}, {Pietrukowicz}, {Pietrzy{\'n}ski}, {Soszy{\'n}ski}, {Ulaczyk}, {Abe}, {Barry}, {Bennett}, {Bhattacharya}, {Bond}, {Freeman}, {Inayama}, {Itow}, {Koshimoto}, {Ling}, {Masuda}, {Fukui}, {Matsubara}, {Muraki}, {Ohnishi}, {Rattenbury}, {Saito}, {Sullivan}, {Suzuki}, {Tristram}, {Wakiyama}, and {Yonehara}}]{Shvartzvald2016}
{Shvartzvald} Y, {Maoz} D, {Udalski} A, et~al (2016) {The frequency of snowline-region planets from four years of OGLE-MOA-Wise second-generation microlensing}. \mnras 457(4):4089--4113. \doi{10.1093/mnras/stw191}, {\href{https://arxiv.org/abs/1510.04297}{{arXiv:1510.04297}}} {[astro-ph.EP]}

\bibitem[{{Sibthorpe} et~al(2018){Sibthorpe}, {Kennedy}, {Wyatt}, {Lestrade}, {Greaves}, {Matthews}, and {Duch{\^e}ne}}]{Sibthorpe2018}
{Sibthorpe} B, {Kennedy} GM, {Wyatt} MC, et~al (2018) {Analysis of the Herschel DEBRIS Sun-like star sample}. \mnras 475(3):3046--3064. \doi{10.1093/mnras/stx3188}, {\href{https://arxiv.org/abs/1803.00072}{{arXiv:1803.00072}}} {[astro-ph.EP]}

\bibitem[{{Silverberg} et~al(2020){Silverberg}, {Wisniewski}, {Kuchner}, {Lawson}, {Bans}, {Debes}, {Biggs}, {Bosch}, {Doll}, {Luca}, {Enachioaie}, {Hamilton}, {Holden}, and {Hyogo}}]{Silverberg:2020}
{Silverberg} SM, {Wisniewski} JP, {Kuchner} MJ, et~al (2020) {Peter Pan Disks: Long-lived Accretion Disks Around Young M Stars}. \apj 890(2):106. \doi{10.3847/1538-4357/ab68e6}, {\href{https://arxiv.org/abs/2001.05030}{{arXiv:2001.05030}}} {[astro-ph.SR]}

\bibitem[{{Simon} et~al(2016){Simon}, {Armitage}, {Li}, and {Youdin}}]{Simon2016}
{Simon} JB, {Armitage} PJ, {Li} R, et~al (2016) {The Mass and Size Distribution of Planetesimals Formed by the Streaming Instability. I. The Role of Self-gravity}. \apj 822(1):55. \doi{10.3847/0004-637X/822/1/55}, {\href{https://arxiv.org/abs/1512.00009}{{arXiv:1512.00009}}} {[astro-ph.SR]}

\bibitem[{{Squire} and {Hopkins}(2018)}]{Squire2018SI}
{Squire} J, {Hopkins} PF (2018) {Resonant Drag Instabilities in protoplanetary disks: the streaming instability and new, faster-growing instabilities}. \mnras p 823. \doi{10.1093/mnras/sty854}

\bibitem[{{Steffl} et~al(2013){Steffl}, {Cunningham}, {Shinn}, {Durda}, and {Stern}}]{Steffl2013}
{Steffl} AJ, {Cunningham} NJ, {Shinn} AB, et~al (2013) {A search for Vulcanoids with the STEREO Heliospheric Imager}. \icarus 223(1):48--56. \doi{10.1016/j.icarus.2012.11.031}, {\href{https://arxiv.org/abs/1301.3804}{{arXiv:1301.3804}}} {[astro-ph.SR]}

\bibitem[{{Stern} et~al(2019){Stern}, {Weaver}, {Spencer}, {Olkin}, {Gladstone}, {Grundy}, {Moore}, {Cruikshank}, {Elliott}, {McKinnon}, {Parker}, {Verbiscer}, {Young}, {Aguilar}, {Albers}, {Andert}, {Andrews}, {Bagenal}, {Banks}, {Bauer}, {Bauman}, {Bechtold}, {Beddingfield}, {Behrooz}, {Beisser}, {Benecchi}, {Bernardoni}, {Beyer}, {Bhaskaran}, {Bierson}, {Binzel}, {Birath}, {Bird}, {Boone}, {Bowman}, {Bray}, {Britt}, {Brown}, {Buckley}, {Buie}, {Buratti}, {Burke}, {Bushman}, {Carcich}, {Chaikin}, {Chavez}, {Cheng}, {Colwell}, {Conard}, {Conner}, {Conrad}, {Cook}, {Cooper}, {Custodio}, {Dalle Ore}, {Deboy}, {Dharmavaram}, {Dhingra}, {Dunn}, {Earle}, {Egan}, {Eisig}, {El-Maarry}, {Engelbrecht}, {Enke}, {Ercol}, {Fattig}, {Ferrell}, {Finley}, {Firer}, {Fischetti}, {Folkner}, {Fosbury}, {Fountain}, {Freeze}, {Gabasova}, {Glaze}, {Green}, {Griffith}, {Guo}, {Hahn}, {Hals}, {Hamilton}, {Hamilton}, {Hanley}, {Harch}, {Harmon}, {Hart}, {Hayes}, {Hersman}, {Hill}, {Hill}, {Hofgartner}, {Holdridge}, {Hor{\'a}nyi},
  {Hosadurga}, {Howard}, {Howett}, {Jaskulek}, {Jennings}, {Jensen}, {Jones}, {Kang}, {Katz}, {Kaufmann}, {Kavelaars}, {Keane}, {Keleher}, {Kinczyk}, {Kochte}, {Kollmann}, {Krimigis}, {Kruizinga}, {Kusnierkiewicz}, {Lahr}, {Lauer}, {Lawrence}, {Lee}, {Lessac-Chenen}, {Linscott}, {Lisse}, {Lunsford}, {Mages}, {Mallder}, {Martin}, {May}, {McComas}, {McNutt}, {Mehoke}, {Mehoke}, {Nelson}, {Nguyen}, {N{\'u}{\~n}ez}, {Ocampo}, {Owen}, {Oxton}, {Parker}, {P{\"a}tzold}, {Pelgrift}, {Pelletier}, {Pineau}, {Piquette}, {Porter}, {Protopapa}, {Quirico}, {Redfern}, {Regiec}, {Reitsema}, {Reuter}, {Richardson}, {Riedel}, {Ritterbush}, {Robbins}, {Rodgers}, {Rogers}, {Rose}, {Rosendall}, {Runyon}, {Ryschkewitsch}, {Saina}, {Salinas}, {Schenk}, {Scherrer}, {Schlei}, {Schmitt}, {Schultz}, {Schurr}, {Scipioni}, {Sepan}, {Shelton}, {Showalter}, {Simon}, {Singer}, {Stahlheber}, {Stanbridge}, {Stansberry}, {Steffl}, {Strobel}, {Stothoff}, {Stryk}, {Stuart}, {Summers}, {Tapley}, {Taylor}, {Taylor}, {Tedford}, {Throop}, {Turner},
  {Umurhan}, {Van Eck}, {Velez}, {Versteeg}, {Vincent}, {Webbert}, {Weidner}, {Weigle}, {Wendel}, {White}, {Whittenburg}, and {Williams}}]{Stern2019}
{Stern} SA, {Weaver} HA, {Spencer} JR, et~al (2019) {Initial results from the New Horizons exploration of 2014 MU$_{69}$, a small Kuiper Belt object}. Science 364(6441):aaw9771. \doi{10.1126/science.aaw9771}, {\href{https://arxiv.org/abs/2004.01017}{{arXiv:2004.01017}}} {[astro-ph.EP]}

\bibitem[{{Su} et~al(2006){Su}, {Rieke}, {Stansberry}, {Bryden}, {Stapelfeldt}, {Trilling}, {Muzerolle}, {Beichman}, {Moro-Martin}, {Hines}, and {Werner}}]{Su2006}
{Su} KYL, {Rieke} GH, {Stansberry} JA, et~al (2006) {Debris Disk Evolution around A Stars}. \apj 653(1):675--689. \doi{10.1086/508649}, {\href{https://arxiv.org/abs/astro-ph/0608563}{{arXiv:astro-ph/0608563}}} {[astro-ph]}

\bibitem[{{Szab{\'o}} et~al(2007){Szab{\'o}}, {Ivezi{\'c}}, {Juri{\'c}}, and {Lupton}}]{Szabo2007}
{Szab{\'o}} GM, {Ivezi{\'c}} {\v{Z}}, {Juri{\'c}} M, et~al (2007) {The properties of Jovian Trojan asteroids listed in SDSS Moving Object Catalogue 3}. \mnras 377(4):1393--1406. \doi{10.1111/j.1365-2966.2007.11687.x}, {\href{https://arxiv.org/abs/astro-ph/0703026}{{arXiv:astro-ph/0703026}}} {[astro-ph]}

\bibitem[{{Tabachnik} and {Evans}(1999)}]{Tabachnik:1999}
{Tabachnik} S, {Evans} NW (1999) {Cartography for Martian Trojans}. \apjl 517(1):L63--L66. \doi{10.1086/312019}, {\href{https://arxiv.org/abs/astro-ph/9904085}{{arXiv:astro-ph/9904085}}} {[astro-ph]}

\bibitem[{{Takahashi} and {Muto}(2018)}]{Takahashi2018}
{Takahashi} SZ, {Muto} T (2018) {Structure Formation in a Young Protoplanetary Disk by a Magnetic Disk Wind}. \apj 865(2):102. \doi{10.3847/1538-4357/aadda0}, {\href{https://arxiv.org/abs/1808.10220}{{arXiv:1808.10220}}} {[astro-ph.SR]}

\bibitem[{{Takakuwa} et~al(2024){Takakuwa}, {Saigo}, {Kido}, {Ohashi}, {Tobin}, {J{\o}rgensen}, {Aikawa}, {Aso}, {Gavino}, {Han}, {Koch}, {Kwon}, {Lee}, {Lee}, {Li}, {Lin}, {Looney}, {Mori}, {Sai}, {Sharma}, {Sheehan}, {Tomida}, {Williams}, {Yamato}, and {Yen}}]{Takakuwa2024}
{Takakuwa} S, {Saigo} K, {Kido} M, et~al (2024) {Early Planet Formation in Embedded Disks (eDisk). XIV. Flared Dust Distribution and Viscous Accretion Heating of the Disk around R CrA IRS 7B-a}. \apj 964(1):24. \doi{10.3847/1538-4357/ad1f57}, {\href{https://arxiv.org/abs/2401.08722}{{arXiv:2401.08722}}} {[astro-ph.EP]}

\bibitem[{{Tegler} et~al(2016){Tegler}, {Romanishin}, {Consolmagno}, and {J.}}]{Tegler2016}
{Tegler} SC, {Romanishin} W, {Consolmagno} GJ, et~al (2016) {Two Color Populations of Kuiper Belt and Centaur Objects and the Smaller Orbital Inclinations of Red Centaur Objects}. \aj 152(6):210. \doi{10.3847/0004-6256/152/6/210}

\bibitem[{{Terebey} et~al(1984){Terebey}, {Shu}, and {Cassen}}]{Terebey1984}
{Terebey} S, {Shu} FH, {Cassen} P (1984) {The collapse of the cores of slowly rotating isothermal clouds}. \apj 286:529--551. \doi{10.1086/162628}

\bibitem[{{Tsuchiyama} et~al(2011){Tsuchiyama}, {Uesugi}, {Matsushima}, {Michikami}, {Kadono}, {Nakamura}, {Uesugi}, {Nakano}, {Sandford}, {Noguchi}, {Matsumoto}, {Matsuno}, {Nagano}, {Imai}, {Takeuchi}, {Suzuki}, {Ogami}, {Katagiri}, {Ebihara}, {Ireland}, {Kitajima}, {Nagao}, {Naraoka}, {Noguchi}, {Okazaki}, {Yurimoto}, {Zolensky}, {Mukai}, {Abe}, {Yada}, {Fujimura}, {Yoshikawa}, and {Kawaguchi}}]{Tsuchiyama2011}
{Tsuchiyama} A, {Uesugi} M, {Matsushima} T, et~al (2011) {Three-Dimensional Structure of Hayabusa Samples: Origin and Evolution of Itokawa Regolith}. Science 333(6046):1125. \doi{10.1126/science.1207807}

\bibitem[{{Tychoniec} et~al(2020){Tychoniec}, {Manara}, {Rosotti}, {van Dishoeck}, {Cridland}, {Hsieh}, {Murillo}, {Segura-Cox}, {van Terwisga}, and {Tobin}}]{Tychoniec2020}
{Tychoniec} {\L}, {Manara} CF, {Rosotti} GP, et~al (2020) {Dust masses of young disks: constraining the initial solid reservoir for planet formation}. \aap 640:A19. \doi{10.1051/0004-6361/202037851}, {\href{https://arxiv.org/abs/2006.02812}{{arXiv:2006.02812}}} {[astro-ph.EP]}

\bibitem[{{Veras} et~al(2020){Veras}, {Tremblay}, {Hermes}, {McDonald}, {Kennedy}, {Meru}, and {G{\"a}nsicke}}]{Veras2020HighMass}
{Veras} D, {Tremblay} PE, {Hermes} JJ, et~al (2020) {Constraining planet formation around 6-8 M$_{{\ensuremath{\odot}}}$ stars}. \mnras 493(1):765--775. \doi{10.1093/mnras/staa241}, {\href{https://arxiv.org/abs/2001.08757}{{arXiv:2001.08757}}} {[astro-ph.EP]}

\bibitem[{{Vigan} et~al(2021){Vigan}, {Fontanive}, {Meyer}, {Biller}, {Bonavita}, {Feldt}, {Desidera}, {Marleau}, {Emsenhuber}, {Galicher}, {Rice}, {Forgan}, {Mordasini}, {Gratton}, {Le Coroller}, {Maire}, {Cantalloube}, {Chauvin}, {Cheetham}, {Hagelberg}, {Lagrange}, {Langlois}, {Bonnefoy}, {Beuzit}, {Boccaletti}, {D'Orazi}, {Delorme}, {Dominik}, {Henning}, {Janson}, {Lagadec}, {Lazzoni}, {Ligi}, {Menard}, {Mesa}, {Messina}, {Moutou}, {M{\"u}ller}, {Perrot}, {Samland}, {Schmid}, {Schmidt}, {Sissa}, {Turatto}, {Udry}, {Zurlo}, {Abe}, {Antichi}, {Asensio-Torres}, {Baruffolo}, {Baudoz}, {Baudrand}, {Bazzon}, {Blanchard}, {Bohn}, {Brown Sevilla}, {Carbillet}, {Carle}, {Cascone}, {Charton}, {Claudi}, {Costille}, {De Caprio}, {Delboulb{\'e}}, {Dohlen}, {Engler}, {Fantinel}, {Feautrier}, {Fusco}, {Gigan}, {Girard}, {Giro}, {Gisler}, {Gluck}, {Gry}, {Hubin}, {Hugot}, {Jaquet}, {Kasper}, {Le Mignant}, {Llored}, {Madec}, {Magnard}, {Martinez}, {Maurel}, {M{\"o}ller-Nilsson}, {Mouillet}, {Moulin}, {Orign{\'e}},
  {Pavlov}, {Perret}, {Petit}, {Pragt}, {Puget}, {Rabou}, {Ramos}, {Rickman}, {Rigal}, {Rochat}, {Roelfsema}, {Rousset}, {Roux}, {Salasnich}, {Sauvage}, {Sevin}, {Soenke}, {Stadler}, {Suarez}, {Wahhaj}, {Weber}, and {Wildi}}]{Vigan2021}
{Vigan} A, {Fontanive} C, {Meyer} M, et~al (2021) {The SPHERE infrared survey for exoplanets (SHINE). III. The demographics of young giant exoplanets below 300 au with SPHERE}. \aap 651:A72. \doi{10.1051/0004-6361/202038107}, {\href{https://arxiv.org/abs/2007.06573}{{arXiv:2007.06573}}} {[astro-ph.EP]}

\bibitem[{{Vincke} and {Pfalzner}(2018)}]{Vincke:2018}
{Vincke} K, {Pfalzner} S (2018) {How Do Disks and Planetary Systems in High-mass Open Clusters Differ from Those around Field Stars?} \apj 868(1):1. \doi{10.3847/1538-4357/aae7d1}, {\href{https://arxiv.org/abs/1810.04453}{{arXiv:1810.04453}}} {[astro-ph.GA]}

\bibitem[{{Wahlberg Jansson} and {Johansen}(2014)}]{WahlbergJansson2014}
{Wahlberg Jansson} K, {Johansen} A (2014) {Formation of pebble-pile planetesimals}. \aap 570:A47. \doi{10.1051/0004-6361/201424369}, {\href{https://arxiv.org/abs/1408.2535}{{arXiv:1408.2535}}} {[astro-ph.EP]}

\bibitem[{{Weidenschilling}(1977)}]{Weidenschilling1977}
{Weidenschilling} SJ (1977) {Aerodynamics of solid bodies in the solar nebula.} \mnras 180:57--70. \doi{10.1093/mnras/180.2.57}

\bibitem[{{Weidenschilling}(1980)}]{Weidenschilling1980}
{Weidenschilling} SJ (1980) {Dust to planetesimals: Settling and coagulation in the solar nebula}. \icarus 44(1):172--189. \doi{10.1016/0019-1035(80)90064-0}

\bibitem[{{Weidenschilling} and {Cuzzi}(1993)}]{Weidenschilling1993}
{Weidenschilling} SJ, {Cuzzi} JN (1993) {Formation of Planetesimals in the Solar Nebula}. In: {Levy} EH, {Lunine} JI (eds) Protostars and Planets III, p 1031

\bibitem[{{Williams} and {Cieza}(2011)}]{Williams2011}
{Williams} JP, {Cieza} LA (2011) {Protoplanetary Disks and Their Evolution}. \araa 49(1):67--117. \doi{10.1146/annurev-astro-081710-102548}, {\href{https://arxiv.org/abs/1103.0556}{{arXiv:1103.0556}}} {[astro-ph.GA]}

\bibitem[{{Windmark} et~al(2012){Windmark}, {Birnstiel}, {Ormel}, and {Dullemond}}]{Windmark2012}
{Windmark} F, {Birnstiel} T, {Ormel} CW, et~al (2012) {Breaking through: The effects of a velocity distribution on barriers to dust growth}. \aap 544:L16. \doi{10.1051/0004-6361/201220004}, {\href{https://arxiv.org/abs/1208.0304}{{arXiv:1208.0304}}} {[astro-ph.EP]}

\bibitem[{{Winter} et~al(2018){Winter}, {Clarke}, {Rosotti}, {Ih}, {Facchini}, and {Haworth}}]{Winter2018}
{Winter} AJ, {Clarke} CJ, {Rosotti} G, et~al (2018) {Protoplanetary disc truncation mechanisms in stellar clusters: comparing external photoevaporation and tidal encounters}. \mnras 478(2):2700--2722. \doi{10.1093/mnras/sty984}, {\href{https://arxiv.org/abs/1804.00013}{{arXiv:1804.00013}}} {[astro-ph.SR]}

\bibitem[{{Wong} and {Brown}(2016)}]{Wong2016}
{Wong} I, {Brown} ME (2016) {A Hypothesis for the Color Bimodality of Jupiter Trojans}. \aj 152(4):90. \doi{10.3847/0004-6256/152/4/90}, {\href{https://arxiv.org/abs/1607.04133}{{arXiv:1607.04133}}} {[astro-ph.EP]}

\bibitem[{{Wong} and {Brown}(2017)}]{Wong2017}
{Wong} I, {Brown} ME (2017) {The Bimodal Color Distribution of Small Kuiper Belt Objects}. \aj 153(4):145. \doi{10.3847/1538-3881/aa60c3}, {\href{https://arxiv.org/abs/1702.02615}{{arXiv:1702.02615}}} {[astro-ph.EP]}

\bibitem[{{Wong} et~al(2024){Wong}, {Brown}, {Emery}, {Binzel}, {Grundy}, {Marchi}, {Martin}, {Noll}, and {Sunshine}}]{Wong2024}
{Wong} I, {Brown} ME, {Emery} JP, et~al (2024) {JWST Near-infrared Spectroscopy of the Lucy Jupiter Trojan Flyby Targets: Evidence for OH Absorption, Aliphatic Organics, and CO$_{2}$}. \psj 5(4):87. \doi{10.3847/PSJ/ad2fc3}, {\href{https://arxiv.org/abs/2311.11531}{{arXiv:2311.11531}}} {[astro-ph.EP]}

\bibitem[{Wurm and Teiser(2021)}]{Wurm2021}
Wurm G, Teiser J (2021) Understanding planet formation using microgravity experiments. Nature Reviews Physics 3(6):405--421

\bibitem[{{Wyatt}(2008)}]{Wyatt2008}
{Wyatt} MC (2008) {Evolution of debris disks.} \araa 46:339--383. \doi{10.1146/annurev.astro.45.051806.110525}

\bibitem[{{Wyatt} et~al(2017){Wyatt}, {Bonsor}, {Jackson}, {Marino}, and {Shannon}}]{Wyatt2017}
{Wyatt} MC, {Bonsor} A, {Jackson} AP, et~al (2017) {How to design a planetary system for different scattering outcomes: giant impact sweet spot, maximizing exocomets, scattered discs}. \mnras 464(3):3385--3407. \doi{10.1093/mnras/stw2633}, {\href{https://arxiv.org/abs/1610.00714}{{arXiv:1610.00714}}} {[astro-ph.EP]}

\bibitem[{{Yang} and {Zhu}(2021)}]{Yang2021}
{Yang} CC, {Zhu} Z (2021) {Streaming instability with multiple dust species - II. Turbulence and dust-gas dynamics at non-linear saturation}. \mnras 508(4):5538--5553. \doi{10.1093/mnras/stab2959}, {\href{https://arxiv.org/abs/2110.06248}{{arXiv:2110.06248}}} {[astro-ph.EP]}

\bibitem[{{Yang} et~al(2017){Yang}, {Johansen}, and {Carrera}}]{Yang2017}
{Yang} CC, {Johansen} A, {Carrera} D (2017) {Concentrating small particles in protoplanetary disks through the streaming instability}. \aap 606:A80. \doi{10.1051/0004-6361/201630106}, {\href{https://arxiv.org/abs/1611.07014}{{arXiv:1611.07014}}} {[astro-ph.EP]}

\bibitem[{{Yang} et~al(2020){Yang}, {Xie}, and {Zhou}}]{Yang2020}
{Yang} JY, {Xie} JW, {Zhou} JL (2020) {Occurrence and Architecture of Kepler Planetary Systems as Functions of Stellar Mass and Effective Temperature}. \aj 159(4):164. \doi{10.3847/1538-3881/ab7373}, {\href{https://arxiv.org/abs/2002.02840}{{arXiv:2002.02840}}} {[astro-ph.EP]}

\bibitem[{{Yen} et~al(2024){Yen}, {Williams}, {Sai}, {Koch}, {Han}, {J{\o}rgensen}, {Kwon}, {Lee}, {Li}, {Looney}, {Narang}, {Ohashi}, {Takakuwa}, {Tobin}, {de Gregorio-Monsalvo}, {Lai}, {Lee}, and {Tomida}}]{Yen2024}
{Yen} HW, {Williams} JP, {Sai} J, et~al (2024) {Early Planet Formation in Embedded Disks (eDisk). XV. Influence of Magnetic Field Morphology in Dense Cores on Sizes of Protostellar Disks}. \apj 969(2):125. \doi{10.3847/1538-4357/ad4c6b}, {\href{https://arxiv.org/abs/2405.09063}{{arXiv:2405.09063}}} {[astro-ph.SR]}

\bibitem[{{Yokoyama} et~al(2024){Yokoyama}, {Dauphas}, {Fukai}, {Usui}, {Tachibana}, {Sch{\"o}nb{\"a}chler}, {Busemann}, {Abe}, and {Yada}}]{Yokoyama2025}
{Yokoyama} T, {Dauphas} N, {Fukai} R, et~al (2024) {The Chemical Composition of Ryugu: Prospects as a Reference Material for Solar System Composition}. arXiv e-prints arXiv:2405.04500. \doi{10.48550/arXiv.2405.04500}, {\href{https://arxiv.org/abs/2405.04500}{{arXiv:2405.04500}}} {[astro-ph.EP]}

\bibitem[{{Yoshida} et~al(2022){Yoshida}, {Nomura}, {Tsukagoshi}, {Furuya}, and {Ueda}}]{Yoshida2022}
{Yoshida} TC, {Nomura} H, {Tsukagoshi} T, et~al (2022) {Discovery of Line Pressure Broadening and Direct Constraint on Gas Surface Density in a Protoplanetary Disk}. \apjl 937(1):L14. \doi{10.3847/2041-8213/ac903a}, {\href{https://arxiv.org/abs/2209.03367}{{arXiv:2209.03367}}} {[astro-ph.EP]}

\bibitem[{{Youdin} and {Goodman}(2005)}]{Youdin2005}
{Youdin} AN, {Goodman} J (2005) {Streaming Instabilities in Protoplanetary Disks}. \apj 620(1):459--469. \doi{10.1086/426895}, {\href{https://arxiv.org/abs/astro-ph/0409263}{{arXiv:astro-ph/0409263}}} {[astro-ph]}

\bibitem[{{Zhang} and {Gladman}(2022)}]{Zhang2022}
{Zhang} K, {Gladman} BJ (2022) {GLISSE: A GPU-optimized planetary system integrator with application to orbital stability calculations.} \na 90:101659. \doi{10.1016/j.newast.2021.101659}

\bibitem[{{Zhang} et~al(2019){Zhang}, {Bergin}, {Schwarz}, {Krijt}, and {Ciesla}}]{Zhang2019}
{Zhang} K, {Bergin} EA, {Schwarz} K, et~al (2019) {Systematic Variations of CO Gas Abundance with Radius in Gas-rich Protoplanetary Disks}. \apj 883(1):98. \doi{10.3847/1538-4357/ab38b9}, {\href{https://arxiv.org/abs/1908.03267}{{arXiv:1908.03267}}} {[astro-ph.EP]}

\bibitem[{{Zhang} et~al(2020){Zhang}, {Schwarz}, and {Bergin}}]{Zhang2020}
{Zhang} K, {Schwarz} KR, {Bergin} EA (2020) {Rapid Evolution of Volatile CO from the Protostellar Disk Stage to the Protoplanetary Disk Stage}. \apjl 891(1):L17. \doi{10.3847/2041-8213/ab7823}, {\href{https://arxiv.org/abs/2002.08522}{{arXiv:2002.08522}}} {[astro-ph.EP]}

\bibitem[{{Zhu} and {Dong}(2021)}]{Zhu2021}
{Zhu} W, {Dong} S (2021) {Exoplanet Statistics and Theoretical Implications}. \araa 59:291--336. \doi{10.1146/annurev-astro-112420-020055}, {\href{https://arxiv.org/abs/2103.02127}{{arXiv:2103.02127}}} {[astro-ph.EP]}

\bibitem[{{Zhu} and {Yang}(2021)}]{ZhuYang2021}
{Zhu} Z, {Yang} CC (2021) {Streaming instability with multiple dust species - I. Favourable conditions for the linear growth}. \mnras 501(1):467--482. \doi{10.1093/mnras/staa3628}, {\href{https://arxiv.org/abs/2008.01119}{{arXiv:2008.01119}}} {[astro-ph.EP]}

\bibitem[{{Zsom} et~al(2010){Zsom}, {Ormel}, {G{\"u}ttler}, {Blum}, and {Dullemond}}]{Zsom2010}
{Zsom} A, {Ormel} CW, {G{\"u}ttler} C, et~al (2010) {The outcome of protoplanetary dust growth: pebbles, boulders, or planetesimals? II. Introducing the bouncing barrier}. \aap 513:A57. \doi{10.1051/0004-6361/200912976}, {\href{https://arxiv.org/abs/1001.0488}{{arXiv:1001.0488}}} {[astro-ph.EP]}

\end{thebibliography}

\end{document}